\pdfoutput=1
\documentclass[twocolumn]{aastex7}

\usepackage{amsmath}
\usepackage{graphicx}
\usepackage{url}
\usepackage{tabularx}
\usepackage{footmisc}
\usepackage{verbatim}
\usepackage{appendix}
\usepackage{enumerate}
\usepackage{xcolor}
\usepackage[normalem]{ulem}

\definecolor{SECTIONRED}{HTML}{FF0000}

\graphicspath{{figures/}} 

\submitjournal{PASP}

\shorttitle{Why Estimating $\eta_\oplus$ is Difficult}
\shortauthors{Bryson et al.}

\begin{document}

\title{Why Estimating $\eta_\oplus$ is Difficult: A Kepler-Centric Perspective}

\correspondingauthor{Michelle Kunimoto}
\email{mkuni@phas.ubc.ca}

\author[0000-0003-0081-1797]{Steve Bryson}
\affiliation{NASA Ames Research Center, Moffett Field, CA 94035, USA}
\email{steve.bryson@nasa.gov}

\author[0000-0001-9269-8060]{Michelle Kunimoto}
\affiliation{Department of Physics and Astronomy, University of British Columbia, 6224 Agricultural Road, Vancouver, BC V6T 1Z1, Canada}
\email{mkuni@phas.ubc.ca}

\author{Ruslan Belikov}
\affiliation{NASA Ames Research Center, Moffett Field, CA 94035, USA}
\email{ruslan.belikov-1@nasa.gov}

\author[0000-0003-4500-8850]{Galen J. Bergsten}
\affil{Lunar and Planetary Laboratory, The University of Arizona, Tucson, AZ 85721, USA}
\email{gbergsten@arizona.edu}

\author[0000-0002-6673-8206]{Sakhee Bhure}
\affil{Centre for Astrophysics, University of Southern Queensland, Toowoomba, QLD 4350, Australia}
\email{sakhee.bhure@unisq.edu.au}

\author[0000-0003-2139-2405]{William J. Borucki}
\affil{Associate, NASA Ames Research Center, Moffett Field, CA 94035, USA}
\email{william.j.borucki@nasa.gov}

\author[0000-0003-1963-9616]{Douglas A. Caldwell}
\affiliation{SETI Institute, Mountain View, CA 94043 USA}
\affiliation{NASA Ames Research Center, Moffett Field, CA 94035, USA}
\email{dcaldwell@seti.org}

\author[0000-0001-6703-0798]{Aritra Chakrabarty}
\altaffiliation{NASA Postdoctoral Program Fellow}
\affil{NASA Ames Research Center, Moffett Field, CA 94035, USA}
\email{aritra.astrophysics@gmail.com}

\author[0000-0002-3853-7327]{Rachel B. Fernandes}
\altaffiliation{President's Postdoctoral Fellow}
\affil{Department of Astronomy \& Astrophysics, 525 Davey Laboratory, The Pennsylvania State University, University Park, PA 16802, USA}
\affil{Center for Exoplanets and Habitable Worlds, 525 Davey Laboratory, The Pennsylvania State University, University Park, PA 16802, USA}
\email{rbf5378@psu.edu}

\author[0000-0002-5223-7945]{Matthias Y. He}
\altaffiliation{NASA Postdoctoral Program Fellow}
\affil{NASA Ames Research Center, Moffett Field, CA 94035, USA}
\email{matthias.y.he@nasa.gov}

\author[0000-0002-4715-9460]{Jon M. Jenkins}
\affiliation{NASA Ames Research Center, Moffett Field, CA 94035, USA}
\email{jon.jenkins@nasa.gov}

\author[0000-0001-5847-9147]{Kristo Ment}
\affil{Department of Astronomy \& Astrophysics, 525 Davey Laboratory, The Pennsylvania State University, University Park, PA 16802, USA}
\affil{Center for Exoplanets and Habitable Worlds, 525 Davey Laboratory, The Pennsylvania State University, University Park, PA 16802, USA}
\email{kxm821@psu.edu}

\author[0000-0003-1227-3084]{Michael R. Meyer}
\affiliation{Department of Astronomy, The University of Michigan, Ann Arbor, MI 48109}
\email{mrmeyer@umich.edu}

\author[0000-0002-1078-9493]{Gijs D. Mulders}
\affil{Instituto de Astrof\'isica, Pontificia Universidad Cat\'olica de Chile, Av. Vicu\~na Mackenna 4860, 7820436 Macul, Santiago, Chile}
\email{gijs.mulders@uc.cl}

\author[0000-0001-7962-1683]{Ilaria Pascucci}
\affiliation{Lunar and Planetary Laboratory, The University of Arizona, Tucson, AZ 85721}
\email{pascucci@arizona.edu}

\author[0000-0002-8864-1667]{Peter Plavchan}
\affiliation{George Mason University, 4400 University Dr, MS 6D5, Fairfax, VA 22030, USA}
\email{pplavcha@gmu.edu}

\begin{abstract}
$\eta_{\oplus}$, the occurrence rate of rocky habitable zone exoplanets orbiting Sun-like stars, is of great interest to both the astronomical community and the general public.  The Kepler space telescope has made it possible to estimate $\eta_{\oplus}$, but estimates by different groups vary by more than an order of magnitude.  We identify several causes for this range of estimates.  We first review why, despite being designed to estimate $\eta_{\oplus}$, Kepler's observations are not sufficient for a high-confidence estimate, due to Kepler's detection limit coinciding with the $\eta_{\oplus}$ regime.  This results in a need to infer $\eta_{\oplus}$, for example extrapolating from a regime of non-habitable zone, non-rocky exoplanets.  We examine two broad classes of causes that can account for the large discrepancy in $\eta_\oplus$ found in the literature: a) differences in definitions and input data between studies, and b) fundamental limits in Kepler data that lead to large uncertainties and poor accuracy.  We highlight the risk of large biases when using extrapolation to describe small exoplanet populations in the habitable zone.  We discuss how $\eta_{\oplus}$ estimates based on Kepler data can be improved, such as reprocessing Kepler data for more complete, higher-reliability detections and better exoplanet catalog characterization.  We briefly survey upcoming space telescopes capable of measuring $\eta_{\oplus}$, and how they can be used to supplement Kepler data.

 \end{abstract}



\section{Introduction} \label{section:introduction}
The Kepler mission \citep{Borucki2010,Koch2010,Borucki2016} was designed  to estimate $\eta_{\oplus}$ $-$ defined by the mission as the occurrence of habitable-zone rocky exoplanets around Sun-like stars.  Understanding $\eta_{\oplus}$ is of great interest to the exoplanet community and the general public, and is an important input to mission design for instruments designed to detect and characterize habitable-zone exoplanets such as the Habitable Worlds Observatory \citep{Feinberg2024,Stark2024}.

Kepler's strategy to estimate $\eta_\oplus$ was to continuously observe $>$150,000 Solar-like main-sequence dwarf stars (primarily spectral types F, G, and K) for at least four years, with a highly sensitive photometer in Solar orbit, identifying exoplanets through the detection of transits. Launched in March 2009, Kepler's continuous period of observation was from May 2009 to August 2013, when it was interrupted by the loss of ability to maintain stable pointing at the original star field. The Kepler telescope was then repurposed as the K2 mission \citep{howell2014} to observe ecliptic fields for $\sim 90$ days at a time, delivering many wonderful and diverse science results.  In this work, we focus on the long-stare observations of the same stars from 2009 to 2013, hereafter referred to as Kepler's prime mission.

The data collected from the prime mission was processed by the Kepler Science Operation Center pipeline \citep{Jenkins2002, KSOCoverview2010, Jenkins2020KDPH}, resulting in several data releases, many of which included exoplanet candidate catalogs.  The final Kepler data release 25 (DR25) \citep{Thompson2018} included a uniformly created exoplanet candidate catalog with several products that can be used to measure the catalog's completeness and reliability, enabling bias-corrected calculations of intrinsic exoplanet occurrence rates from the Kepler sample.  A sampling of values of $\eta_\oplus$ from Kepler data is shown in Figure~\ref{fig:etaEarthHistory}, exhibiting the large range that have been published in the literature.

In this work, we attempt to identify common difficulties underlying the computation of $\eta_\oplus$, and discuss whether or not the range of values in Figure~\ref{fig:etaEarthHistory} can be resolved without more observational data.  We do not advocate that any method for estimating $\eta_{\oplus}$ is better than any other, nor do we present our ``best estimate'' of $\eta_{\oplus}$.  We do not consider issues of confirmation or statistical validation, because the confirmation/validation process is not sufficiently well characterized to support demographic studies (see the discussion in \S\ref{section:doingOurBest}).  Beyond generally considering false positives, we do not address astrophysical issues -- such as stellar and planetary multiplicity -- and how they impact $\eta_{\oplus}$ estimates.   This work is not a review of the $\eta_{\oplus}$ in general -- in particular we do not comprehensively discuss $\eta_{\oplus}$ measurements in the literature. Rather, we seek to raise awareness of issues with Kepler data that will be encountered in any attempt to estimate $\eta_{\oplus}$ from that data.

For a broader review of $\eta_{\oplus}$ beyond issues with Kepler data, see \citet{Fernandes2025b}.

\subsection{What is $\eta_\oplus$?} \label{section:whatisEA}


In this work we adopt the broad sense of $\eta_\oplus$ as the number of exoplanets per star that may be potentially habitable for life as we know it.  This choice is motivated by the goal of the Habitable Worlds Observatory (HWO) to detect biosignatures around ``Earth-sized'' exoplanets.  We assume that life as we know it requires Earth-like conditions: a rocky world with an atmosphere, and the presence of liquid water.  In this section we briefly survey the conceptual issues with turning this characterization of $\eta_\oplus$ into data useful for $\eta_\oplus$ estimates, and present a more detailed discussion in \S\ref{section:spread}.

Relating these criteria to Kepler observables such as exoplanet size and orbital period is challenging.  A consensus has emerged that the rocky exoplanet radius upper limit is about $1.64 \pm 0.05 R_\oplus$\citep{Rogers2015,Muller2024}, though, for example, \citet{Stark2024} adapts $1.4 R_\oplus$ and \citet{Otegi2020} suggests that some exoplanets with radii up to $2.5 R_\oplus$ may be rocky.  The lower limit on the size of a rocky exoplanet in the habitable zone that can retain an atmosphere is less well constrained, partly because there are currently no exoplanets with a radius less than $1 R_\oplus$ with observed atmospheres.  This lower radius limit likely depends on the stellar instellation on the exoplanet.  \citet{Stark2024} adopts the minimum radius of $0.8 (a/\mathrm{EEID})^{-0.5} R_\oplus$, where $a$ is the exoplanet's semi-major axis in AU and EEID is the distance in AU at which the exoplanet would have the same instellation flux as the Earth.  

Similarly, the relationship between the habitable zone and orbital period is ambiguous for a collection of stars.  The habitable zone of a specific star can be well-characterized in terms of that star's effective temperature as a range of instellation fluxes \citep{Kopparapu2013}, which corresponds to a range of orbital periods.  Therefore, for a collection of stars with a range of effective temperatures, like that required to statistically estimate $\eta_\oplus$, there is no single range of orbital periods that exactly corresponds to the habitable zones of those stars.  Further, the magnitude of the atmospheric greenhouse effect will depend on the composition of the atmosphere, which in turn may be impacted significantly by surface--atmosphere interactions over geologic time.  It is worth remembering that there would not be liquid water on the Earth's surface but for the magnitude of our atmospheric greenhouse.    

There is also the question of which stellar population supports potentially habitable exoplanets.  For example, the potential habitability of tidally locked exoplanets is an open issue, and exoplanets in the habitable zones of stars cooler than $\approx 4800$~K are likely tidally locked \citep{Yang2013, Yang2014b, Wolf2015a, Way2015, Godolt2015, Kopparapu2016, Kopparapu2017, Bin2018}.  

These ambiguities in how ``a rocky world with an atmosphere, and the presence of liquid water'' translated into Kepler observables allows different studies to make different choices in the definition of $\eta_\oplus$.  In this work we do not advocate for one choice over another.  In \S\ref{section:spread} we discuss how these choices impact the differing values of $\eta_\oplus$ in the literature and Figure~\ref{fig:etaEarthHistory}.

Throughout this paper, we refer to small (in the above sense) exoplanets orbiting FGKM stars in their host star's habitable zone as exoplanets in the {\it $\eta_\oplus$ regime}. 



\section{The Evolution of the Community's Estimates of $\eta_{\oplus}$} \label{section:etaEarthHistory}
\begin{figure*}
\begin{center}
\includegraphics[width=0.98\textwidth]{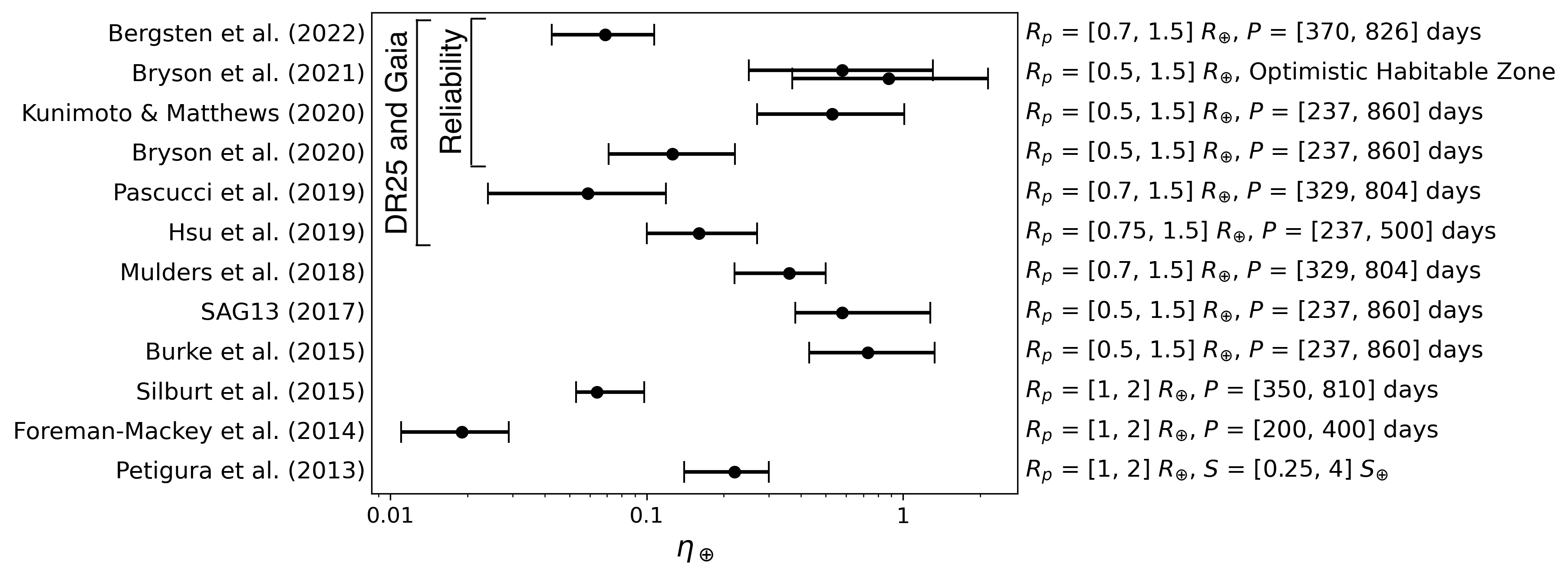} 
\caption{A comparison of representative $\eta_\oplus$ estimates (\citet{Petigura2013, ForemanMackey2014, Silburt2015, Burke2015}, SAG13, \citet{Mulders2018, Hsu2019, Pascucci2019, Bryson2020, Kunimoto2020a, Bryson2021, Bergsten2022}), with the definition of the $\eta_\oplus$ regime shown on the right. \cite{Bryson2021} integrated over the optimistic habitable zone \cite{Kopparapu2013} based on individual stars' stellar effective temperature ($237 - 860$ days for a Sun-like star), while all other works integrated over a single period or instellation flux range for all stars. We show the \cite{Bryson2021} estimates for two bounding completeness extrapolations. The bracket labeled "DR25 and Gaia" shows the studies using the DR25 exoplanet candidate catalog and Gaia stellar properties as discussed in \S\ref{section:catalogs}.  The bracket labeled "Reliability" shows the studies that applied reliability correction, discussed in \S\ref{section:reliability}.
}
\label{fig:etaEarthHistory}
\end{center}
\end{figure*}

Kepler launched a sustained effort within the exoplanet community to determine $\eta_{\oplus}$\footnote{\label{footnote:sag13}\url{https://exoplanetarchive.ipac.caltech.edu/docs/occurrence_rate_papers.html}}. Several recent papers summarize the literature \citep{Kunimoto2020a,Bryson2021,Bergsten2022} which contains a range of $\eta_{\oplus}$ studies from 2013 to the present day.  
These studies yielded a dramatically large range of values, shown in Figure~\ref{fig:etaEarthHistory}, spanning two orders of magnitude and often with very large uncertainties.  SAG13, the Study Analysis Group on Exoplanet Occurrence Rates and Distributions of the NASA Exoplanet Exploration Program Analysis Group, gathered much of the exoplanet community to find a consensus value for $\eta_{\oplus}$, released in 2017\footnote{\label{footnote:sag13}\url{https://exoplanets.nasa.gov/exep/exopag/sag/\#sag13}}. Comparisons between these $\eta_{\oplus}$ studies are challenging due to several factors:
\begin{itemize}
    \item Different studies used different Kepler-based exoplanet catalogs, some of which are based on only a fraction of the data used by others. 
    \item Different studies made different assumptions about how the population of larger, short-period exoplanets, where exoplanet detections are abundant, relates to the $\eta_{\oplus}$ regime, where there are fewer detections. 
    \item Different studies used different stellar catalogs, with changes in stellar properties, such as stellar brightness, that impact exoplanet properties necessary to estimate $\eta_{\oplus}$.
    \item Catalog completeness corrections varied between studies, with some assuming a simple analytic model of detection efficiency \citep[e.g.,] []{Youdin2011, Howard2012}, some empirically estimating detection efficiency with transit injection/recovery tests \citep[e.g.,][]{Petigura2013, Burke2015}, and others simply assuming a catalog is complete beyond some threshold \citep[e.g.,][]{CatanzariteShao2011}. 
    \item Catalog reliability correction (see \S\ref{section:doingOurBest}) was only possible after the DR25 data release in 2018.
\end{itemize}

The final Kepler data release (DR25) was based on the full set of Kepler prime mission observations. With its comprehensive data products for characterizing the resulting exoplanet catalog, DR25 has been essential for alleviating issues of completeness and reliability. As such, the DR25 catalog is now the standard used by occurrence rate studies \citep[e.g.,][]{Mulders2018, Hsu2018, zinkChristiansen2019, Bryson2021}. DR25 was the first data release to include products that allowed for the characterization of its exoplanet catalog reliability against false alarms due to noise and systematic instrumental artifacts, which are the most prevalent contaminants in the $\eta_{\oplus}$ regime.  Thus nearly all works prior to DR25 did not incorporate reliability against false alarms in their estimates. \citet{Mulders2018} attempted to mitigate the impact of false-alarm contamination by using a DR25 Disposition Score cut \citep[see \S7.3.4 of][]{Thompson2018}, as an alternative to reliability correction. \citet{Bryson2020} was the first to directly take into account reliability against both noise/systematics and astrophysical false positives, and in doing so found that occurrence rates for small exoplanets in long-period orbits dropped by about a factor of two after reliability correction. As discussed in \S\ref{section:reliability}, \citet{Bryson2020b} showed that fully accounting for reliability is necessary to produce $\eta_\oplus$ estimates that are robust against catalog variations.

Studies have also varied in the choice of stellar catalogs, and exoplanet occurrence rates have been shown to be sensitive to such choices because different catalogs have different stellar properties for the same star. For instance, the discovery of a gap in the radius distribution of small exoplanets -- first uncovered by \citet{Fulton2017} -- was enabled by improvements in stellar radius measurements by the California Kepler Survey \citep[CKS;][]{Petigura2017,Johnson2017}. 
One major advancement with important implications for $\eta_{\oplus}$ is the use of Gaia DR2 parallaxes, which reduced the median stellar radius uncertainty of Kepler stars from $\approx27\%$ to $\approx4\%$ \citep{Berger2020a}. \citet{Bryson2020} showed that occurrence rates of exoplanets near Earth's orbit and size are overestimated by a factor of two when using pre-Gaia Kepler Input Catalog stellar properties.

To summarize, Kepler exoplanet catalogs and host star stellar properties have been changing over time during the period shown in Figure~\ref{fig:etaEarthHistory}.  As discussed in more detail in \S\ref{section:underControl}, to some extent these changes can account for the range of eta-Earth values shown in Figure~\ref{fig:etaEarthHistory}. After 2019 both the exoplanet catalogs and stellar properties stabilized, yet subsequent studies still produced significantly different values of $\eta_{\oplus}$.  The reasons for this continued lack of agreement are discussed in \S\ref{section:needExtrapolate}.



\section{Kepler Observations Fall Short of Supporting Robust Estimates of $\eta_\oplus$} \label{section:currentSituation}
There is no question that the dominant underlying cause of the lack of agreement on the value $\eta_{\oplus}$ is that Kepler's detection limit falls just short of what is needed for an accurate estimate. Simply put, there are not enough detections of exoplanets in the $\eta_{\oplus}$ regime to support a precise estimate.  How this came to be is the topic of the this section.

\subsection{How We Got Here}\label{section:KeplerProblems}
The Kepler mission was designed with the very ambitious goal of estimating $\eta_{\oplus}$ \citep{Borucki2008}.  This design was based on the following assumptions:
\begin{itemize}
    \item {\bf Kepler would be a very low noise instrument.} Kepler would be placed in Solar orbit, have very stable pointing, and would be insulated from thermal effects so that Kepler's photometric performance would not prevent the detection of Earth-size exoplanets around sufficiently bright Solar-type stars \citep{Borucki2008}.
    \item {\bf Sun-like stars are typically as quiet as our Sun.}  Many main-sequence early K through late F stars would be as quiet as the Sun on the time scale of an Earth-analog transit, so that stellar variability would not prevent the detection of Earth-size exoplanets around sufficiently bright Solar-type stars \citep{Koch2010}.
\end{itemize}
Under these assumptions, simulations showed that a fast $\sim 1$-meter Schmidt camera would have sufficient photometric precision to detect an Earth-Sun analog from three or four transit events \citep{Borucki2008}.  This determined the nominal mission duration of four years.  The requirement to detect a few 10s of Earth-Sun analogs, assuming that every Sun-like star has such an Earth-size exoplanet, determined the wide-field design that, when pointed at an appropriate field, allowed the (nearly) continuous four-year observation of over 150,000 stars.  The capability to detect a few tens of Earth-Sun analogs provides sufficient statistics to derive good constraints on $\eta_{\oplus}$: no detections would be a significant null result.  For margin, Kepler was designed to allow ten years of observation, in case a longer baseline was required to detect enough Earth-Sun analog systems.

Unfortunately, many of these assumptions turned out to not be true:
\begin{itemize}
    \item {\bf The Sun is quieter than most similar main-sequence dwarfs.} The Sun's activity on the time scale of an Earth transit is in the lower third of the distribution of stars observed by Kepler (see Figures 7 and 8 in \citet{gilliland2011} and Figure 2 in \citet{Gilliland2015})
    Therefore, more transits are required for most stars to achieve the design S/N for an Earth-Sun analog detection.
    \item {\bf Thermally-dependent, spatial-temporal oscillations were present in the CCD signal.} A flaw in the design of the on-board electronics introduced spatial oscillations into the photometry on some CCD channels.  The location on the CCD of these oscillations was thermally dependent \citep{VanCleve2009}.  These oscillations were discovered during ground testing, but it was deemed too risky and expensive to open the hardware and fix the electronics.  This problem was made significantly worse by the next issue.
    \item {\bf Kepler photometry was more sensitive to thermal variations than expected.}  Despite careful design to thermally isolate the Kepler photometer from the rest of the spacecraft and the environment, thermal variations were apparent in the photometry.  The switching on and off of battery heaters was easily seen in the photometry, as well as dependence of Kepler's focus on where sunlight fell on the spacecraft \citep{VanCleve2009}.  This led to a strong coupling between thermal artifacts and Kepler's 372.5-day orbit.  This thermal dependence, combined with the aforementioned thermally dependent electronic oscillations, resulted in oscillations in the photometry with periods and amplitudes comparable to an Earth transit that repeated with a period of about 372.5 days.  This resulted in a very large number of instrumental transit-like false alarms smack in the middle of the $\eta_{\oplus}$ regime.  
\end{itemize}

The result of these problems was a reduction in Kepler's photometric precision.  This could be compensated for by extending Kepler's prime-mission observation time to detect more transits.  Indeed, Kepler was approved for an extended mission adding four years of observation.  Simulations indicated that a total of eight years of observations would, with the actual photometric performance of Kepler and the actual noise of the observed stars, be sufficient to recover the transit S/N planned in Kepler's original design \citep{Jenkins2012}.  But this was not to be:
\begin{itemize}
    \item {\bf The failure of a second (of four) reaction wheels in 2013 prevented Kepler from staying pointed at the original Kepler field.}  Kepler requires three reaction wheels for precise, stable pointing, and was built with four reaction wheels, providing margin against failure.  But one wheel failed in 2012 and the second in 2013, ending Kepler's long-stare high-precision observation of the same stars for several years.
\end{itemize}
Kepler collected data on the same star field for a time span of 1,420 days, though data was not collected on every day of this span for a variety of reasons such as quarterly rolls, data downlinks and occasional safe modes.  The Kepler team required three transits to consider an exoplanet ``detected''.  Therefore the longest orbital period that can possibly be detected is 710 days ($= 1420/2)$), in the fortuitous circumstance when the first transit is at the beginning and the third transit is just before the end of those 1,420 days.  Any orbital period longer than 710 days cannot be detected by Kepler because three transits cannot be observed.  An orbital period $<473$ days ($= 1420/3)$) will always have three transits falling within Kepler prime's observation period, though transits may be missed due to the aforementioned gaps in Kepler's observations.  

Kepler would live on as K2 and continue successfully taking highly precise (though not as precise as the Kepler prime mission) stellar observations. However, because K2 observed patches of the sky for considerably shorter periods ($\sim$85 days), Kepler's quest to estimate $\eta_{\oplus}$ for Sun-like stars is limited to the data collected by the Kepler prime mission between 2009 and 2013, with all that data's limitations and flaws.

\subsection{Making the Best of the Situation}\label{section:doingOurBest}
The Kepler team spent five years, from 2013 to 2018, understanding the Kepler prime mission data and devising strategies to overcome its problems, as reflected in increasingly sophisticated methods to create exoplanet catalogs \citep{Borucki2011, Batalha2013, Burke2014, Mullally2015, Coughlin2016, Thompson2018}.  This culminated in the DR25 data release. \citet{Thompson2018} does a thorough job describing DR25's exoplanet catalog and products supporting the strategy of characterizing completeness and reliability.  DR25 implemented a number of strategies to deal with the situation described in \S\ref{section:KeplerProblems}.  

\begin{figure}[ht]
  \centering
  \includegraphics[width=0.95\linewidth]{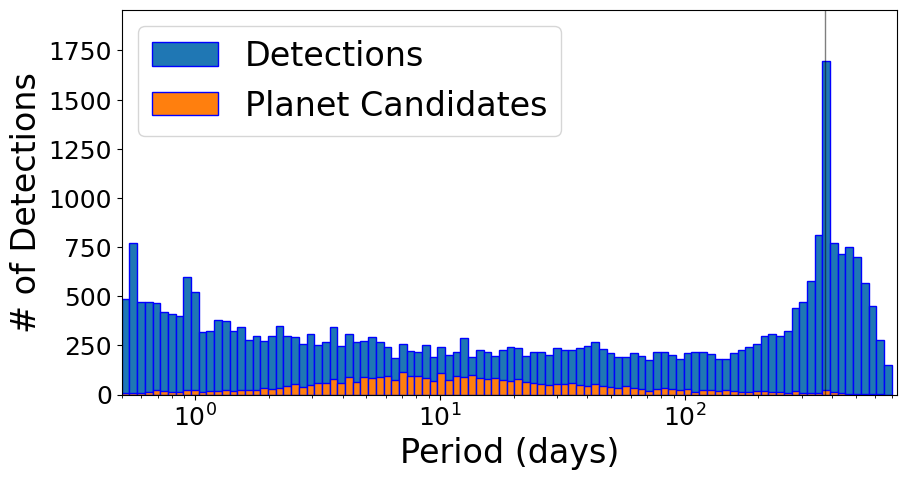} 
\caption{The period distribution of transit detections (blue) and exoplanet candidates (orange) in the Kepler DR25 catalog.  The vertical grey line at 372.5 days shows the orbital period of the Kepler telescope.  The large spike in detections near the Kepler orbital period is due to instrumental false alarms.} \label{figure:tceDist}
\end{figure}

The higher stellar noise and Kepler instrumental artifacts, combined with Kepler's goal of estimating $\eta_{\oplus}$, forced the Kepler exoplanet search to reach as far as possible into lower S/N transits.  This resulted in the enormous number of detections shown in Figure~\ref{figure:tceDist}, whose period distribution clearly has structure due to false positives that does not reflect the true exoplanet population.  These detections required aggressive vetting, most fully developed in the DR25 release, to remove false positives while keeping as many exoplanet candidates as possible.  But it is almost certainly the case that this aggressive vetting removed detected true exoplanets in the $\eta_{\oplus}$ regime.

DR25 vetting is based on several tests, described in \citet{Thompson2018}.  Some of these tests detect astrophysical false positives, such as tests for alternating transit depth or V-shaped transits that indicate an eclipsing binary false positive, and transit offset tests that check that the transit signal is due to a background binary.  Other tests compare the transit depth and shape to the light curve's noise characteristics to check for false alarms due to instrumental artifacts or stellar variability.  There are a total of 23 tests, see \citet{Thompson2018} for details.  These tests were implemented by a fully automated \texttt{Robovetter} \citep{Coughlin2017} in order to facilitate the catalog characterization described below.  Detections that passed all tests were elevated to exoplanet candidate status. Detections that were sufficiently transit-like were elevated to Kepler Objects of Interest (KOIs) even if they failed astrophysical false positive tests, but we do not address these in this work.

While the resulting DR25 exoplanet candidate population had a much more reasonable period distribution than the underlying transit detections (see Figure~\ref{figure:tceDist}), it was expected that this exoplanet candidate population was contaminated with undetected astrophysical false positives and false alarms.  \citet{Morton2011} estimated that, overall, $\sim 10\%$ of the exoplanet candidates were astrophysical false positives.  The ability to detect false positives and false alarms degrades with lower S/N, so we expect that lower S/N exoplanet candidates are more contaminated by undetected false positive/false alarms, so they are less {\it reliable} than higher S/N candidates. In addition, many exoplanet transits with lower S/N likely went undetected, even after accounting for geometric probability, resulting in the DR25 exoplanet candidate catalog being {\it incomplete}.  As described below, catalog reliability and completeness need to be balanced when creating an exoplanet catalog.

To estimate $\eta_{\oplus}$ we have to reconstruct the underlying true exoplanet population from the exoplanet catalog.  That reconstruction must correct for both the completeness and reliability of the catalog.  This requires that the catalog is generated in a uniform manner, detecting and vetting all entries with the same processes.  The  confirmed exoplanet table on the NASA Exoplanet Archive\footnote{\label{footnote:confirmedTable}\url{https://exoplanetarchive.ipac.caltech.edu/cgi-bin/TblView/nph-tblView?app=ExoTbls&config=PSCompPars}} \citep{Christiansen2025}, on the other hand, is not uniform, with different entries being confirmed or statistically validated with different approaches.  Therefore the completeness of the confirmed exoplanet table is very difficult, if not impossible, to measure (the reliability is likely very high). The DR25 catalog, on the other hand, was uniformly generated, and the DR25 release included extensive products designed to facilitate the measurement of the DR25 exoplanet candidate catalog's completeness and reliability, making DR25 suitable for exoplanet demographics studies, in particular estimating $\eta_{\oplus}$.  

Catalog completeness characterization is the problem of measuring the fraction of true exoplanets which are actually identified as exoplanet candidates in that catalog.  Completeness was measured by injecting synthetic transits into the observed Kepler data, running exactly the same detection pipeline and \texttt{Robovetter} software as is used for the exoplanet search and vetting, and seeing what injected synthetic exoplanets are recovered as exoplanet candidates.  The results were presented both as a function of S/N averaged over stars (completeness curves) \citep{Christiansen2013,Christiansen2015,Christiansen2016,Christiansen2020, Christiansen2017}, and as a function of S/N and orbital period for each star (completeness contours) \citep{BurkeJCat2017}.  Due to resource limitations, completeness was only measured for orbital periods $< 500$ days, which led to the issues discussed in \S\ref{section:completeness}.  

Catalog reliability characterization is the measurement of how many identified exoplanet candidates are true exoplanets.  There are two broad classes of false exoplanet candidate identifications: {\it astrophysical false positives,} true detections of repeating transit-like signals that are not exoplanets orbiting the target star, typically eclipsing binaries; and {\it false alarms,} false exoplanet detections due to stellar variability, instrument systematics or statistical fluctuations.

Reliability against undetected astrophysical false positives in the DR25 exoplanet catalog was estimated by the computation of Astrophysical False Positive Probability (FPP) \citep{Morton2012,Morton2016} and the Astrophysical Positional Probability (APP) \citep{Bryson2017}.  APP computed the probability that a transit signal is co-located with the presumed host star, and is sensitive to diluted resolved but nearby eclipsing binary false positives.  FPP is sensitive to eclipsing binaries not resolved from the presumed host star, which can include grazing binaries and dilution due to stellar multiplicity.  Both APP and FPP are very sensitive to stars in pixels near a presumed transit host star, including their stellar properties.  

Reliability against false alarms is more difficult to measure.  Examination of the Kepler data revealed that the large excess of detections seen in Figure~\ref{figure:tceDist} near the 372.5 day Kepler orbital period were due to two causes.  The large, narrow spike at the Kepler orbital period is due to the thermally-dependent electronic oscillations in the CCD pixels described in \S\ref{section:KeplerProblems}.  The broad shoulder between 200 and 700 days is due to pixel responses to cosmic ray impacts (so-called Sudden Pixel Sensitivity Dropouts or SPSDs) lining up with statistical fluctuations, resulting in an apparent transit-like signal in period-folded light curves.  For more details, see \citet{VanCleve2009} and \citet{Thompson2018}.  Both phenomena are difficult to identify directly, but their statistics can be measured using manipulations of Kepler observed data that removes real transit signals while maintaining the statistics of these false alarm sources.

Inverting Kepler light curves preserves the CCD oscillations while making real exoplanets undetectable, and scrambling seasonal Kepler observational periods breaks the periodic ephemeris of real exoplanets while preserving the statistics of the SPSD alignments that create false alarm detections.  These inverted and scrambled datasets provide, in a statistical sense, a ground truth that includes these types of false alarms and no detectable exoplanets, so all Kepler pipeline detections in these data can be treated as false alarms \citep{Thompson2018}.  Indeed the pipeline detections from these data sets reproduce the large excess detections seen in Figure~\ref{figure:tceDist}.  The inverted and scrambled data were used to develop effective \texttt{Robovetter} tests that can detect these false alarms.  The DR25 data products include the result of running the Kepler detection pipeline and \texttt{Robovetter} on the inverted and scrambled data, which can be used to measure the reliability against such false alarms of the DR25 exoplanet candidate catalog \citep{Thompson2018,Bryson2020} as a function of, say, orbital period and exoplanet radius.  We will discuss issues with this DR25 characterization of reliability in \S\ref{section:reliability}. 

When creating the DR25 exoplanet candidate catalog, the Kepler team had to decide how to balance completeness and reliability.  Should the catalog be more complete but less reliable, with more exoplanets but more false positives/alarms? Or should the catalog be more reliable but less complete, with fewer exoplanets but greater confidence in those that survive the vetting process?  This choice is implemented by tuning the strictness of vetting tests in the Robovetter.  \citet{Bryson2020b} compares $\eta_{\oplus}$ computations for various choices of \texttt{Robovetter} strictness. For the DR25 release, \cite{Thompson2018} chose a balance between completeness and reliability.  The result is more exoplanets in the $\eta_{\oplus}$ regime, but these exoplanets are less reliable.  In the next section we discuss the DR25 exoplanet catalog in more detail.

\subsection{Where We Are At}\label{section:whereWereAt}
\begin{figure*}[ht]
    \centering
    \includegraphics[width=0.95\textwidth]{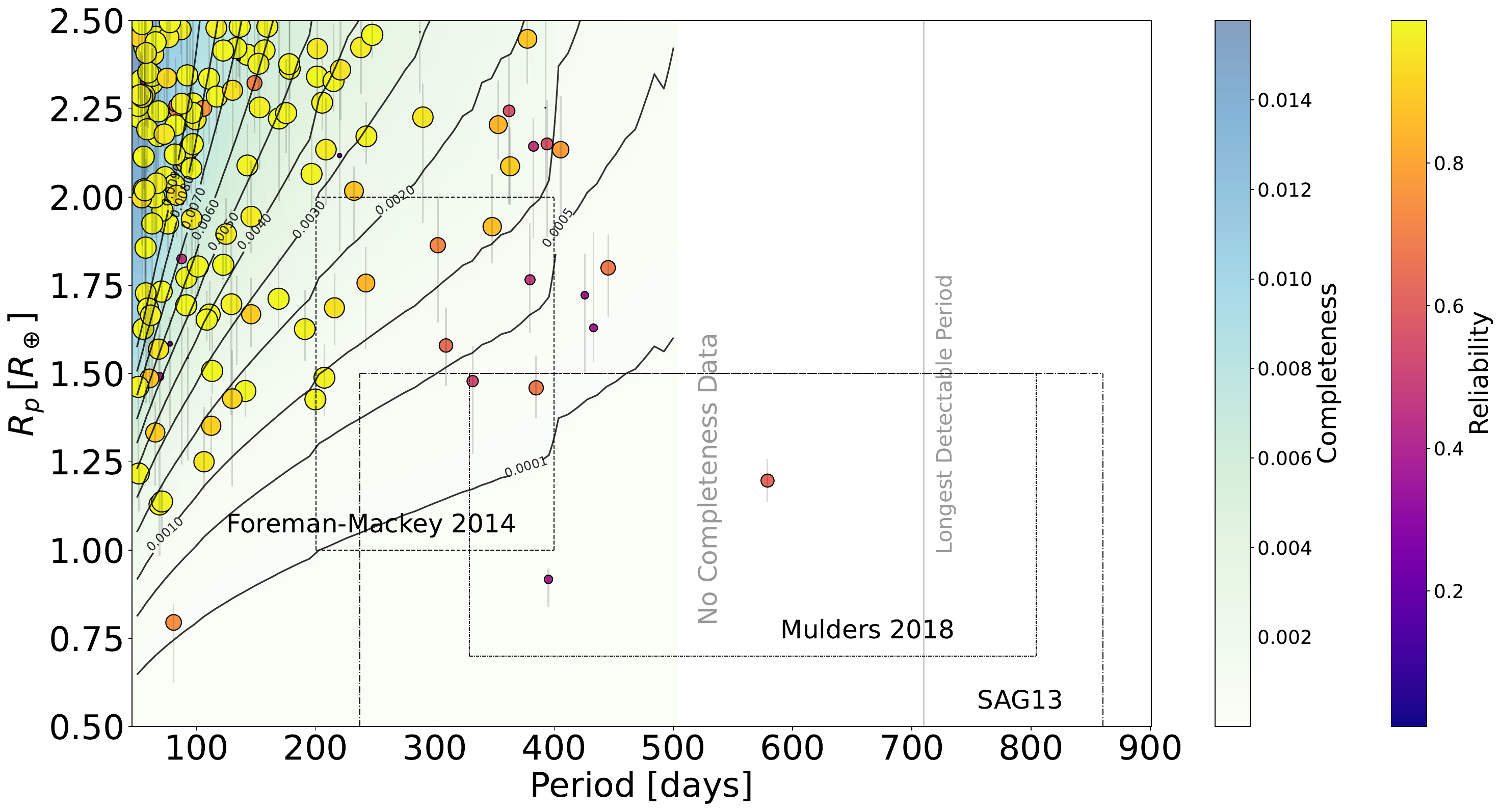}  \
    \includegraphics[width=0.95\textwidth]{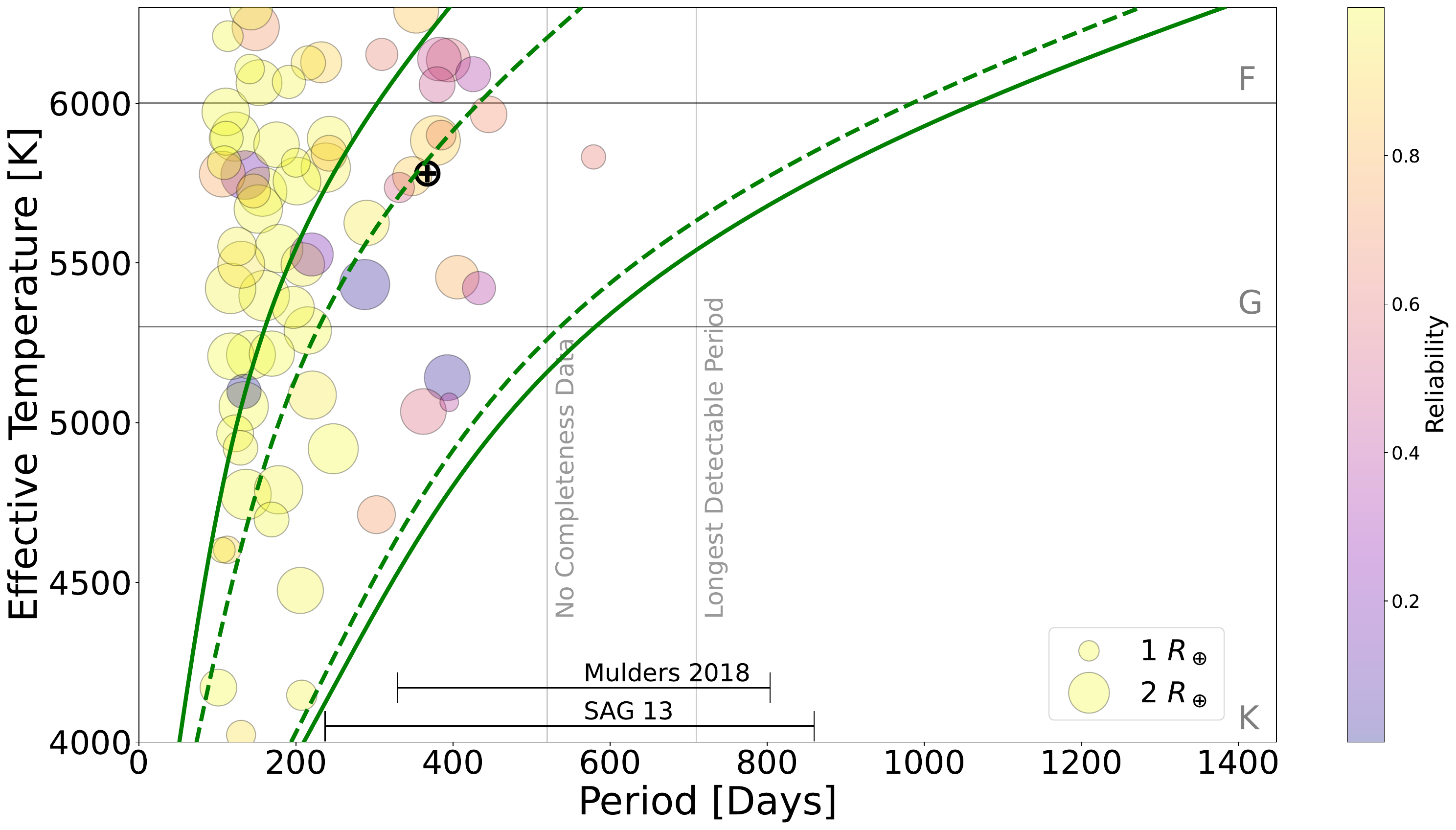}  
    \caption{Top: the Kepler exoplanet candidate population around FGK stars used in the analysis of \citet{Bryson2021}, shown in period and radius, both colored and sized by catalog reliability with exoplanet radius error bars.  The background color map and contours indicate detection completeness.  The rectangles show example $\eta_\oplus$ definitions from different authors. Adapted from~\cite{Bryson2020}. Bottom: The same exoplanet population shown in period and host star effective temperature, which allows comparison with and approximation of the habitable zone.  The green lines show period range of the optimistic (solid) and conservative (dashed) habitable zones for average dwarf stars given effective temperature. The period beyond which there is no completeness data and the longest detectable period are shown by vertical lines.  The $\oplus$ symbol shows the Earth's orbital period at the Sun's effective temperature.  Reliability values in both panels are from \citet{Bryson2020}.}\label{figure:populationPerRad}
\end{figure*}

\begin{figure*}[ht]
  \centering
  \includegraphics[width=0.9\linewidth]{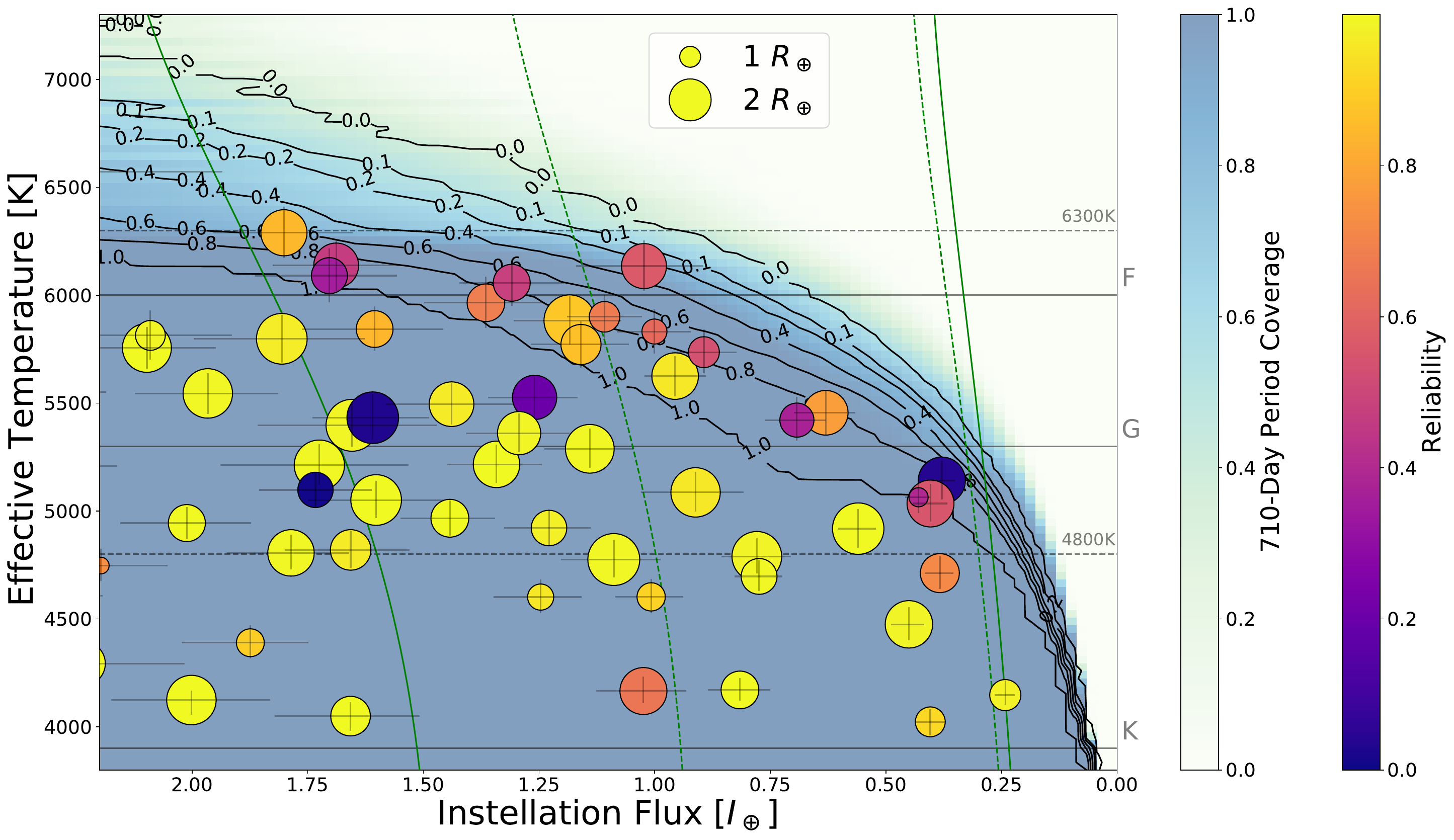} \\
  \includegraphics[width=0.91\linewidth]{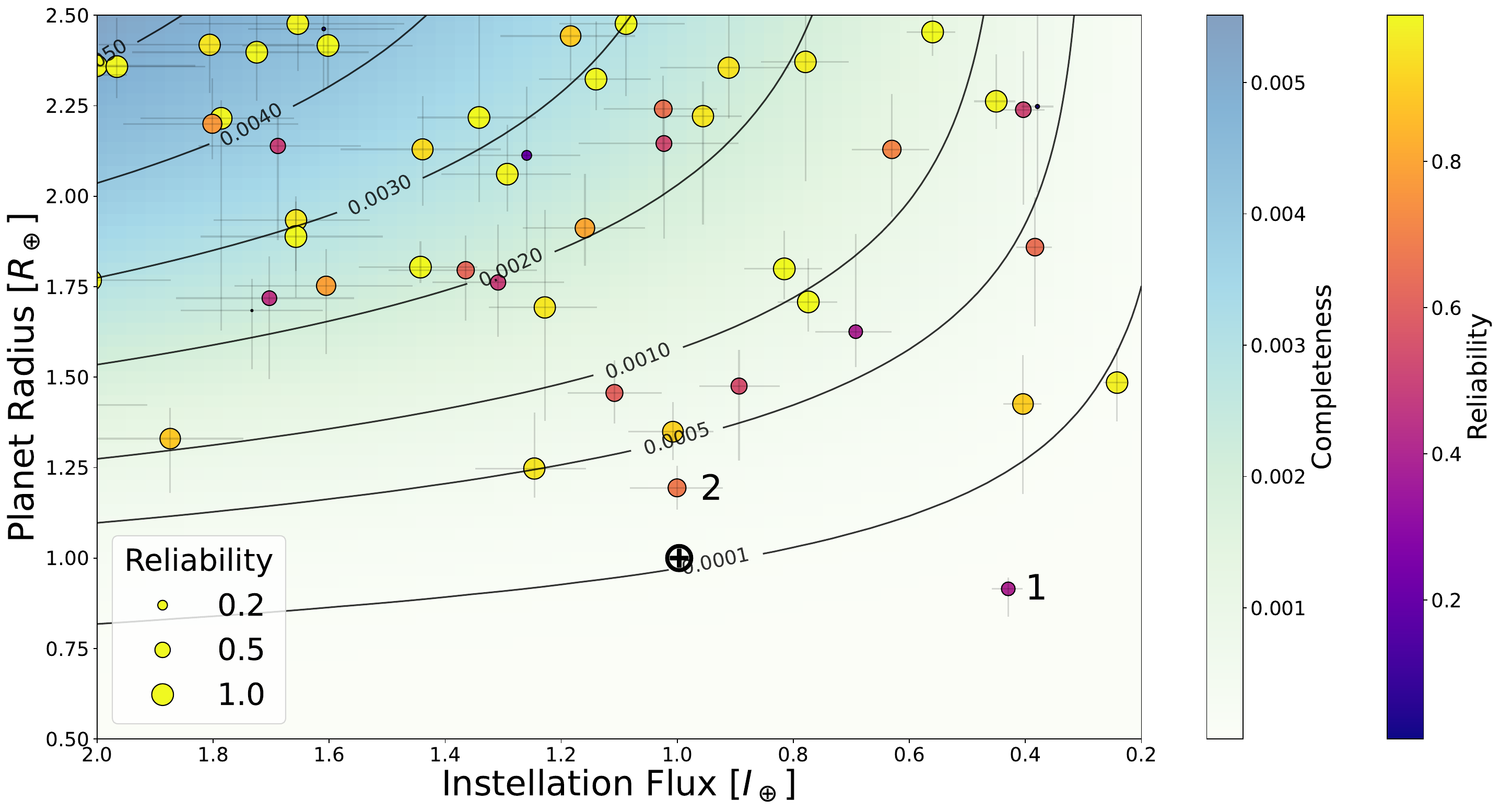} 
\caption{Two views of the DR25 PC population around FGK stars with radii smaller than $2.5~R_\oplus$ and instellation flux near their host star's habitable zone around main sequence dwarf stars. Top: Instellation flux vs. stellar effective temperature, showing the habitable zone and Kepler observational coverage.  The background color map gives, at each point, the fraction of Kepler target stars at that effective temperature whose exoplanets at that instellation flux would have orbital periods of 710 days or less, so it is possible to observe three transits.  The contours show the fraction of exoplanets with periods of 500 days or less, indicating available completeness measurements.  The solid green lines are the boundaries of the optimistic habitable zone, while the dashed green lines are the boundaries of the conservative habitable zone.  The exoplanets are sized by their radius and colored by their catalog reliability.  Bottom: Instellation flux vs. exoplanet radius.  The color map and contours show the average completeness for the stellar population (in the case of zero-completeness extrapolation, see \S\ref{section:completeness}).  The exoplanets are sized and colored by catalog reliability, with radius and instellation flux error bars.  In the lower panel the $\oplus$ symbol shows the Earth.  The two numbered exoplanets are discussed in \S\ref{section:needExtrapolate}. From~\cite{Bryson2021}. Reliability values in both panels are from \citet{Bryson2020}.}  \label{figure:populations}
\end{figure*}

Figures~\ref{figure:populationPerRad} and \ref{figure:populations} give four views of of the DR25 exoplanet candidate catalog in the $\eta_{\oplus}$ regime.  These exoplanet populations come from selections of exoplanet host stars appropriate for $\eta_{\oplus}$ calculations, as described in \citet{Bryson2020,Bryson2021}.

Figure~\ref{figure:populationPerRad} shows the exoplanet population around FGK stars in period-radius space, which is a very popular parameterization for $\eta_{\oplus}$ calculations.  The boxes in the top panel of Figure~\ref{figure:populationPerRad} show example regions used by various authors to define $\eta_{\oplus}$.  In this case $\eta_{\oplus}$ is defined as the average number of exoplanets in the underlying true population (after completeness correction) per star whose radius and orbital period are in these boxes. Figure~\ref{fig:etaEarthHistory} gives these regions for specific authors alongside their corresponding estimates for $\eta_{\oplus}$.  The bottom panel of Figure~\ref{figure:populationPerRad} shows the same exoplanet population in period-host star effective temperature in order to delineate the habitable zone of the host stars from \cite{Kopparapu2013} (green lines).  The habitable zone is computed as a function of period using a fitted average radius and log(g) for Kepler target stars as a function of effective temperature, and may not be accurate for a particular exoplanet's host star.  This panel also shows the 500-day period limit of the DR25 completeness measurement and the 710-day period limit, beyond which three transits cannot occur during Kepler observations.  The most salient features to note are:
\begin{itemize}
    \item {\bf Disagreement about what region of period-radius defines $\eta_{\oplus}$.}  This lack of consensus reflects the fact that, while a specific star's habitable zone can be defined by orbital period, different stars will have habitable zones with different period ranges. This forces an approximate period range that covers the habitable zone for some population of stars, and different authors make different approximations.  Also there is disagreement about how the size of an exoplanet impacts habitability, leading to different ranges of exoplanet radius.  Unless the average number of habitable zone exoplanets per star is independent of period and radius, choices of period and radius range to define $\eta_{\oplus}$ leads to differences in $\eta_{\oplus}$ values even when the analysis is otherwise the same.  We examine this issue in \S\ref{section:habZone}.
    \item {\bf Lack of detections in the $\eta_{\oplus}$ region, and $\eta_{\oplus}$ regions are far from  regions with many detections.}  In particular, significant portions of the habitable zone of solar-like and hotter stars have orbital periods that cannot be detected by Kepler.  We examine this issue in Sections \ref{section:lackDetections} and \ref{section:needExtrapolate}.
    \item {\bf Lack of completeness information for large portions of $\eta_{\oplus}$ regions.}  The $\eta_{\oplus}$ region chosen by most authors, as well as the habitable zone of G and F stars, extends beyond a 500-day orbital period.  But DR25 only provides completeness data out to 500 days.  We examine this issue in \ref{section:completeness}. 
\end{itemize}

Figure~\ref{figure:populations} shows the same exoplanet population as Figure~\ref{figure:populationPerRad}, but now replacing orbital period with planetary instellation flux.  The top panel of Figure~\ref{figure:populations} shows instellation flux vs. stellar host effective temperature, which has the major advantage of being the parameter space in which the habitable zone is defined by \cite{Kopparapu2013}.  As explained in the next paragraph, the color map shows the coverage of the Kepler observations and the contour show the completeness coverage.  The lower panel shows instellation flux vs. exoplanet radius, with the color map showing detection and vetting completeness.   In both panels the exoplanets are colored by their catalog reliability.

In the upper panel of Figure~\ref{figure:populations}, we see the exoplanet population relative to the habitable zone, which is shown by the green lines: the solid green lines enclose the so-called optimistic habitable zone, which would have habitable surface temperatures with just the right atmosphere, and the dashed green lines enclose the conservative habitable zone, which has habitable surface temperatures for a wide range of atmospheres \citep{Kopparapu2013}.  The exoplanets are sized by their radius.  We see that, contrary to a quick reading of Figure~\ref{figure:populationPerRad}, the habitable zone(s) are well-populated with Kepler exoplanet detections, albeit mostly with larger exoplanets in the $\sim 2 R_\oplus$ range.  

The upper panel of Figure~\ref{figure:populations} also shows how well Kepler observations cover the habitable zone, both in terms of where in the habitable zone exoplanets can be detected (the color map), and how well completeness is characterized (the contours). 
The color map gives, at each point, the fraction of Kepler target stars at that effective temperature $\mathrm{Teff}$ whose exoplanets at that instellation flux $I$ would have orbital periods of 710 days or less, so it is possible to observe three transits.
More precisely, at each $(I, \mathrm{Teff})$ in the top panel, there are $N(\mathrm{Teff})$ Kepler target stars with varying stellar properties (because the number of stars does not depend on $I$, which is a property of the exoplanets).  Of these $N(\mathrm{Teff})$ stars, there are $N_{<710}(I, \mathrm{Teff})$ stars for which the (circular) orbital period of an exoplanet with instellation $I$ is less than 710 days, so it is possible for three transits to occur during Kepler's observational time span.  The colormap in the top panel of Figure~\ref{figure:populations} shows the fraction $N_{<710}(I, \mathrm{Teff})/N(\mathrm{Teff})$.  When this fraction is 1, any exoplanet with that instellation flux orbiting any star with that effective temperature could have three transits during Kepler's observations.  If the fraction is 0.5, for half of those stars, exoplanets with that instellation flux would have orbital periods longer than 710 days, so such exoplanets could not have been detected by Kepler.  If the fraction is zero, there are no target stars with that effective temperature whose exoplanets with that instellation flux can be detected.  

For stars hotter than about 5500K, there are regions of both the conservative and optimistic habitable zones that have essentially no coverage, so exoplanets in those portions of the habitable zone could not have been detected by Kepler.  We also see that the coverage of the entire habitable zone drops off dramatically for stars hotter than 6300K.  The habitable zone of F stars is therefore poorly covered.  The lack of coverage of the habitable zone implies that extrapolation is necessary to estimate $\eta_\oplus$, which will be discussed in \S\ref{section:needExtrapolate}.

Similarly, the contours in the top panel of Figure~\ref{figure:populations} show how well completeness is characterized.  The analysis is the same as three-transit coverage, with the 710-day threshold replaced by a 500-day threshold, reflecting the longest orbital period with measured completeness in DR25.  The contours show the fraction $N_{<500}(I, \mathrm{Teff})/N(\mathrm{Teff})$.  Inside the 1.0 contour, completeness is well characterized for all the stars.  Near the 0.5 contour, half the stars have unknown completeness for exoplanets with the orbital period with that instellation flux.  Outside the 0.0 contour, none of the stars have characterized completeness for exoplanets with that instellation flux.  We discuss how this impacts completeness correction in \S\ref{section:completeness}.

The lower panel of Figure~\ref{figure:populations} shows the same exoplanet population in instellation flux and exoplanet radius, with the symbols both sized and colored by catalog reliability.  The color map and contours show the value of the completeness averaged over the target stars.  Here we see the dearth of detections below $1.25 R_\oplus$, implying that extrapolation is required to estimate the population of Earth-size, habitable-zone exoplanets.

In this section, we've seen that Kepler data, as presented in the DR25 exoplanet candidate catalog, presents several challenges in the estimation of $\eta_\oplus$:
\begin{itemize}
    \item Very few detections of Earth-size exoplanets in the habitable zone, due to very low Kepler completeness for small, habitable-zone exoplanets.
    \item No coverage of much of the habitable zone for main-sequence dwarf stars hotter than the Sun, due to the three-transit requirement.
\end{itemize}
Both of these challenges imply that, to describe exoplanet populations in the habitable zone based on Kepler data, some kind of extrapolation is required, both from larger exoplanets to smaller exoplanets, and from hotter (short-period) exoplanets to temperate (long-period) exoplanets.  The difficulty of such extrapolation is discussed more fully in \S\ref{section:needExtrapolate}.

\section{Causes of the Spread in $\eta_\oplus$ Estimates}\label{section:spread}

The previous sections describe a situation presenting several difficulties when trying to estimate $\eta_\oplus$: incomplete and unreliable data that requires extrapolation to cover the habitable zone, a lack of consensus on the definition of $\eta_\oplus$, and input data sets that change over time.  We believe that differences in the ways authors approach these difficulties, coupled with inherent limitations of the Kepler sample, are enough to account for the spread in values of $\eta_\oplus$ in Figure~\ref{fig:etaEarthHistory}.  It is our hope that understanding these difficulties will help reconcile $\eta_\oplus$ estimates by different studies.  

We classify these difficulties in two high-level categories, and address each in turn:
\begin{itemize}
    \item \textbf{Differences due to choices made by authors.} Much of the spread in Figure~\ref{fig:etaEarthHistory} is due to different authors making different choices, so that not all the values shown in Figure~\ref{fig:etaEarthHistory} refer to the same quantity, despite being given the same name.  In some cases these choices are in definitions, while in other cases the choice is what data to use and how that data are interpreted.  Older studies necessarily used earlier datasets with shorter observational duration.  Some studies used different vetting methods to compute their exoplanet catalogs than others.  
    \item \textbf{Inherent limitations of the Kepler data.}  Because the $\eta_\oplus$ regime is at the Kepler detection limit, and coincides with the occurrence of a large number of false alarm-generating systematics, there are inherent limitations of the data.  The lack of detections leads to large uncertainties, and the lack of coverage of the habitable zone (see Figure~\ref{figure:populationPerRad}) forces extrapolation using poorly constrained models.  
\end{itemize}


\subsection{The Easy Problem: Differing Choices}\label{section:underControl}
An immediate, and relatively easy to account for, cause of the spread of values in Figure~\ref{fig:etaEarthHistory} is the fact that different studies do not use the same definitions or input data.  Between the studies, there is variation on the definition of the habitable zone, what counts as a rocky exoplanet, which exoplanet and stellar properties are used as input and whether or not the catalogs are corrected for reliability.  In this section we survey the impacts of the differing choices.


\subsubsection{What is the Habitable Zone?}\label{section:habZone}
As discussed in \S\ref{section:whatisEA}, while it is common for $\eta_\oplus$ to mean the number of rocky exoplanets in a star's habitable zone, the definition of ``rocky'' and ``habitable zone'' varies between studies.  What qualifies as a ``rocky'' exoplanet is defined as a range of exoplanet radii. Historically, most studies defined the habitable zone as a range of orbital periods, with some studies defining the habitable zone in terms of a range of instellation flux. Figures~\ref{fig:etaEarthHistory} and \ref{figure:populationPerRad} show the ranges used by different studies. In typical studies, $\eta_\oplus$ is the integration of a (usually Poisson) rate over period (or instellation flux) and exoplanet radius ranges. Differences can arise if, for example, two studies find the same occurrence rate-period distribution but use different period ranges in their final integration, resulting in two different values of $\eta_{\oplus}$.  
\citet{Kunimoto2020a} showed that $\eta_\oplus$ can increase by more than a factor of 2 when changing the lower limit of exoplanet radius from $0.75$ to $0.5~R_{\oplus}$.

Careful definitions of the habitable zone have been provided in terms of an exoplanet's instellation flux and its host star's effective temperature \citep{Kopparapu2013, Kopparapu2014}.  The definition of \citet{Kopparapu2014} as parameterized curves in the stellar instellation-stellar effective temperature plane, shown in the top panel of Figure~\ref{figure:populations}.  \citet{Kopparapu2014} defines the conservative and optimistic habitable zone, where the conservative inner and outer edges are defined by the `runaway greenhouse' and `maximum greenhouse' limits, while the optimistic inner and outer boundaries are the `recent Venus' and `early Mars' limits.  It is important to remember that the magnitude of the greenhouse effect depends critically on the atmospheric composition.  It is usually assumed that in addition to a range of masses and radii comparable to the Earth, an Earth-centric greenhouse is also assumed, implying an Earth-like atmospheric composition.  The range of compositions implied by exoplanet formation models, as well as the lack of current constraints from bulk density and atmospheric characterization, introduce further uncertainty into the definition of the habitable zone. 

\begin{figure}[ht]
  \centering
  \includegraphics[width=0.99\linewidth]{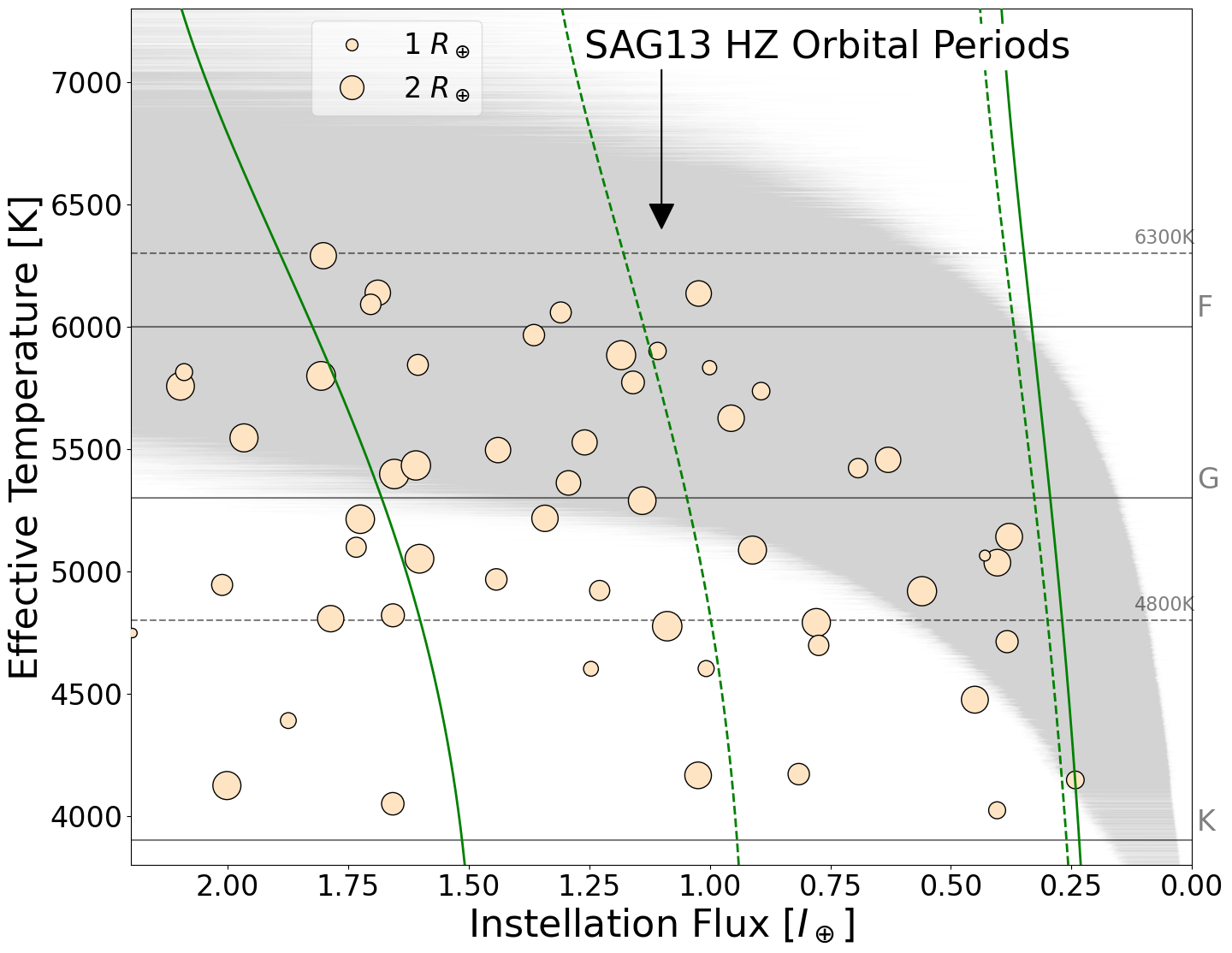} 
\caption{The habitable zone flux range compared with example orbital periods, previously used to estimate habitable zone occurrence for F, G, and K stars.  For each star in the stellar parent sample, we show the SAG13 instellation flux range of the SAG13 orbital period range as a horizontal grey line.  The solid green lines are the boundaries of the optimistic habitable zone, while the dashed green lines are the boundaries of the conservative habitable zone.  The exoplanet population is the same as in Figure~\ref{figure:populations}, and exoplanets are sized by their radius. Adapted from~\cite{Bryson2021}.} \label{figure:fluxPeriod}
\end{figure}

An exoplanet's instellation flux depends on its host star effective temperature and radius.  Therefore, the habitable zone as defined by \citet{Kopparapu2014} is not well defined in terms of orbital period: an exoplanet around one star may be in that star's habitable zone, but another exoplanet with the same orbital period around a different star may not be in this second star's habitable zone.  This is illustrated in Figure~\ref{figure:fluxPeriod}, which shows the range of instellation fluxes for for the SAG13 orbital period ranges from for FGK stars.  In this figure, we see that, while much of an interesting portion of the habitable zone is well covered, other portions are not.  Further, some ``in the habitable zone represented by SAG13'' periods are outside the optimistic habitable zone defined by \citet{Kopparapu2014}, so using the SAG13 orbital period range will include planets not in the habitable zone.  There is no range of stellar effective temperatures for which the example orbital periods closely match the habitable zone.  Therefore defining the habitable zone in terms of orbital period requires compromise and approximation, and different authors made different choices.

\paragraph{Is it helpful to use $\Gamma_\oplus$?}\label{section:etaVsGamma}
$\Gamma_\oplus$ is defined by  \citet{ForemanMackey2014} as $\Gamma_{\oplus} \equiv \mathrm{d}^2 f / \mathrm{d} \log p \, \mathrm{d} \log r$, evaluated at $p=$ one year and $r = 1 R_\oplus$, where $f$ is the number of exoplanets per star (this is called $C_{\oplus,\mathrm{yr}}$ in \citet{Youdin2011}).  When expressing the orbital period in years and radius in Earth radii, $\Gamma_\oplus$ is an approximation to $\frac{\eta_{\oplus}}{\Delta p \Delta r}$, with an error that is proportional to the derivative of $f$ and $(\Delta p \Delta r)^2$ \citep{Bryson2025b}.  By ``dividing out'' the range of period and radius with $\Delta p \Delta r$, $\Gamma_\oplus$ is often used as a proxy that allows direct comparison between different $\eta_\oplus$ estimates that are independent of the period and radius ranges used to define $\eta_\oplus$.  But when $f$ is not constant and for the large ranges of $\Delta p$ and $\Delta r$ that we see in Figure~\ref{fig:etaEarthHistory}, the approximation error of $\Gamma_\oplus$ can be large, a factor of two or more \citep{Bryson2025b}.  Because of this additional source of error, we do not see $\Gamma_\oplus$ as helpful in estimating $\eta_\oplus$ in a way that is useful for, for example, mission design for the Habitable Worlds Observatory.  We also do not believe it is useful to separate the problem of estimating $\eta_\oplus$ from the problem of finding the right range of parameters that define $\eta_\oplus$.

\subsubsection{Catalogs Changing Over Time} \label{section:catalogs}
$\eta_\oplus$ studies are critically dependent on both the input candidate exoplanet catalog and stellar properties from stellar catalogs.  Both types of catalogs changed dramatically from 2013 to 2019, much of the period covered in Figure~\ref{fig:etaEarthHistory}.  

\paragraph{Exoplanet Candidate Catalogs}  Though Kepler's prime mission ended in the late spring of 2013 (as described in \S\ref{section:KeplerProblems}), the Kepler team released several exoplanet catalogs after 2013: \citet{Batalha2013, Burke2014, Mullally2015, Coughlin2016, Thompson2018}.  Each catalog improved over time through the development of detection and vetting methods described in \S\ref{section:doingOurBest}.  Further, some of the $\eta_\oplus$ studies in Figure~\ref{fig:etaEarthHistory} -- such as \citet{Petigura2013} and \citet{Kunimoto2020a} -- performed their own exoplanet searches, generating their own exoplanet candidate catalogs (\citet{ForemanMackey2014} used the exoplanet catalog from \citet{Petigura2013}). While there was a large overlap between these catalogs, they contain different transit detections, particularly near Kepler's detection limit, near the $\eta_\oplus$ regime.  As we show in \S\ref{section:needExtrapolate}, a difference of a single exoplanet between catalogs can have a dramatic impact on $\eta_\oplus$ for common $\eta_\oplus$ estimation methods.  Further, as described in \S\ref{section:doingOurBest}, any exoplanet candidate catalog has completeness and reliability biases that must be corrected to infer the underlying exoplanet population.  We expect that completeness corrections for all these catalogs were of uniform quality (though see \S\ref{section:completeness}). Correction for catalog reliability bias is more difficult and is discussed in \S\ref{section:reliability}.  

\paragraph{Stellar Catalogs and Stellar Properties} In transit surveys, exoplanet candidate properties critical to $\eta_\oplus$ such as radius and instellation depend strongly on the properties of their host star.  $\eta_\oplus$ studies get these stellar properties from input stellar catalogs, whose accuracy has improved dramatically from 2013 to 2021.  Prior to 2017, stellar properties were obtained from the Kepler Input Catalog (KIC) \citep{Brown2011}, a comprehensive survey of about six million stars in the Kepler field that integrated multiple existing catalog sources with new ground-based observation.  Stellar radii were provided for many brighter stars through $\log(g)$, which had an uncertainty of about 0.4 dex for dwarfs, so stellar radii derived from KIC data had relatively low precision.  Improved characterization of 2,762 stars in the KIC was provided by \citet{Huber2014}. By 2017, exoplanet candidate stellar hosts were identified in the Kepler field, and became subject to careful ground-based follow-up resulting in improved stellar properties for these host stars \citep{Petigura2017}.  At the same time, \citet{Mathur2017ApJS} provided improved stellar properties for most Kepler target stars.

These improved stellar properties, however, were not available for stars that were not targeted by Kepler for study, such as background stars whose brightness is important for correcting exoplanet radii for background dilution.  In 2018, Gaia DR2 enabled improved characterization of Kepler target stars through Gaia parallax \citep{berger18}, which were combined with ground-based followup in 2020 to provide higher accuracy and precision exoplanet radii \citep{Berger2020a, Berger2020b}.  \citet{Bryson2020} studied the impact of stellar catalog changes on $\eta_\oplus$ by performing the same analysis based on a power law population model in orbital period and exoplanet radius using the KIC (with improvements from \citet{Mathur2017ApJS}) and the stellar and exoplanet properties from \citet{Berger2020a, Berger2020b}, using the SAG13 definition of $\eta_\oplus$.  They found that using the KIC produced a catalog reliability-corrected estimate of $\eta_{\oplus} = 0.223^{+0.136}_{-0.087}$, while using the 2020 Berger catalog gave $\eta_{\oplus} = 0.126^{+0.095}_{-0.055}$.  While these are superficially within one $\sigma$ of each other, this is a systematic change of $\eta_\oplus$ by nearly a factor of two.  Gaia also enables the identification of likely unidentified stellar host multiplicity, which can have a dramatic effect on exoplanet properties \citep{Furlan2020} for example, through the RUWE metric \citep{gaiaRuwe2018}.  Studies that don't use RUWE to filter out likely binary host stars are likely more contaminated by  stellar multiplicity \citep{Savel2020,Sullivan2022}, compared with studies that do use RUWE such as \citet{Bryson2020, Bryson2021}.

\paragraph{Differing Stellar Populations} Even when using the same stellar and exoplanet catalogs, different studies consider different stellar populations.  For example, \citet{Burke2015} and \citet{zink2019} considered GK stars, \citet{Bryson2021} concentrated on stars with effective temperatures between 4800 K and 6300 K, while \citet{Mulders2018} treated all host stars as having Solar stellar properties.  There is weak evidence of a dependence of $\eta_\oplus$ on stellar temperature \citep{Bergsten2022, Bryson2021}.  \citet{Bryson2021} computed the $\eta_\oplus$ by extrapolating powers laws over the conservative habitable zone using the same analysis for GK and FGK Kepler stars.  They found, for example, $\eta_{\oplus} = 0.40^{+0.54}_{-0.23}$ for GK stars and $\eta_{\oplus} = 0.63^{+1.18}_{-0.39}$ for FGK stars, less than a $1 \sigma$ effect.  The choice of stellar population may have an impact, but this is likely not a large effect.  


\subsubsection{Completeness} \label{section:completeness}
The completeness of the DR25 exoplanet catalog is well characterized via the injection of transit signals as described in \S\ref{section:doingOurBest} for exoplanets with a period $< 500$ days.  For longer period exoplanets the completeness is, strictly speaking, unknown in DR25, but we can use results in  \citet{Bryson2021} to bound the impact of unknown completeness beyond 500 days. In a possible excess of ideological zeal, \citet{Bryson2021} computed completeness using completeness contours based on \citet{BurkeJCat2017} that are functions of period, expected S/N and host effective temperature. Beyond 500 days \citet{Bryson2021} made the unrealistic assumption that the completeness beyond 500 days is independent of period for two cases: {\it constant extrapolation} assumed that the completeness beyond 500 days is equal to that at 500 days, and {\it zero extrapolation} assumed that the completeness dropped to zero beyond 500 days.  In zero extrapolation there is no dependence on S/N or host effective temperature beyond 500 days.  Generally speaking, completeness is averaged over the parent stellar population.  Example values for the resulting average completeness using zero-extrapolation from \citet{Bryson2021} can be found in contours in the bottom panel of Figure~\ref{figure:populations}, and the difference between the constant and zero completeness extrapolations can be found in Figure 6 of \citet{Bryson2021}.  While these two extrapolation cases are unrealistic, the resulting values of $\eta_\oplus$ for the optimistic habitable zone, shown in Figure~\ref{fig:etaEarthHistory}, for the constant ($\eta_\oplus = 0.58^{+0.73}_{-0.33}$) and zero ($\eta_\oplus = 0.88^{+1.28}_{-0.51}$) extrapolations provide a bound for the sensitivity of $\eta_\oplus$ on the unknown completeness beyond 500 days.  While $\eta_\oplus$ resulting from these two extrapolations are within each other's error bars, this difference in $\eta_\oplus$ is a bias impacting accuracy.  This bias range for the optimistic habitable zone is $< 0.3$, or $<41\%$ of the mean value.  For the conservative habitable zone, constant extrapolation gives ($\eta_\oplus = 0.37^{+0.48}_{-0.21}$) while zero extrapolation gives ($\eta_\oplus = 0.60^{+0.90}_{-0.36}$), for a bias range of  $<47\%$ of the mean value.  Therefore uncertainty in long-period completeness is a significant, though bound, contributor to inaccurate estimates of $\eta_\oplus$.

The actual completeness is likely closer to the zero-extrapolation case: a more realistic analysis would assume a linear ramp from the completeness at a period of 500 days, where three transits will likely be captured, to zero completeness at 710 days, where three transits cannot be captured.  Beyond 710 days completeness is zero because of the three-transit detection requirement.  The resulting completeness contours would be closer to the zero extrapolation case, so $\eta_\oplus$ is likely closer (though not close) to the zero-extrapolation values.  While this approach is more realistic than the bounding cases in \citet{Bryson2021}, a new injection study of completeness out to 710 days is required to eliminate inaccuracy due to unknown completeness.

\subsubsection{Reliability} \label{section:reliability}

As described in \S\ref{section:currentSituation}, Kepler exoplanet candidate catalogs are likely contaminated by unidentified astrophysical false positives and false alarms.  Therefore, just as for completeness, these catalogs should not be used without correcting for this contamination, particularly in the $\eta_\oplus$ regime. 

Because reliability correction removes exoplanets (sometimes in a probabilistic way), it will reduce estimates of $\eta_\oplus$.  For example, using a power law population model in period and radius, \citet{Bryson2020} finds that reliability correction drops the SAG13 definition of $\eta_\oplus$ from $0.302^{+0.181}_{-0.113}$ to $0.126^{+0.095}_{-0.055}$.

\begin{figure}[ht]
  \centering
  \includegraphics[width=0.95\linewidth]{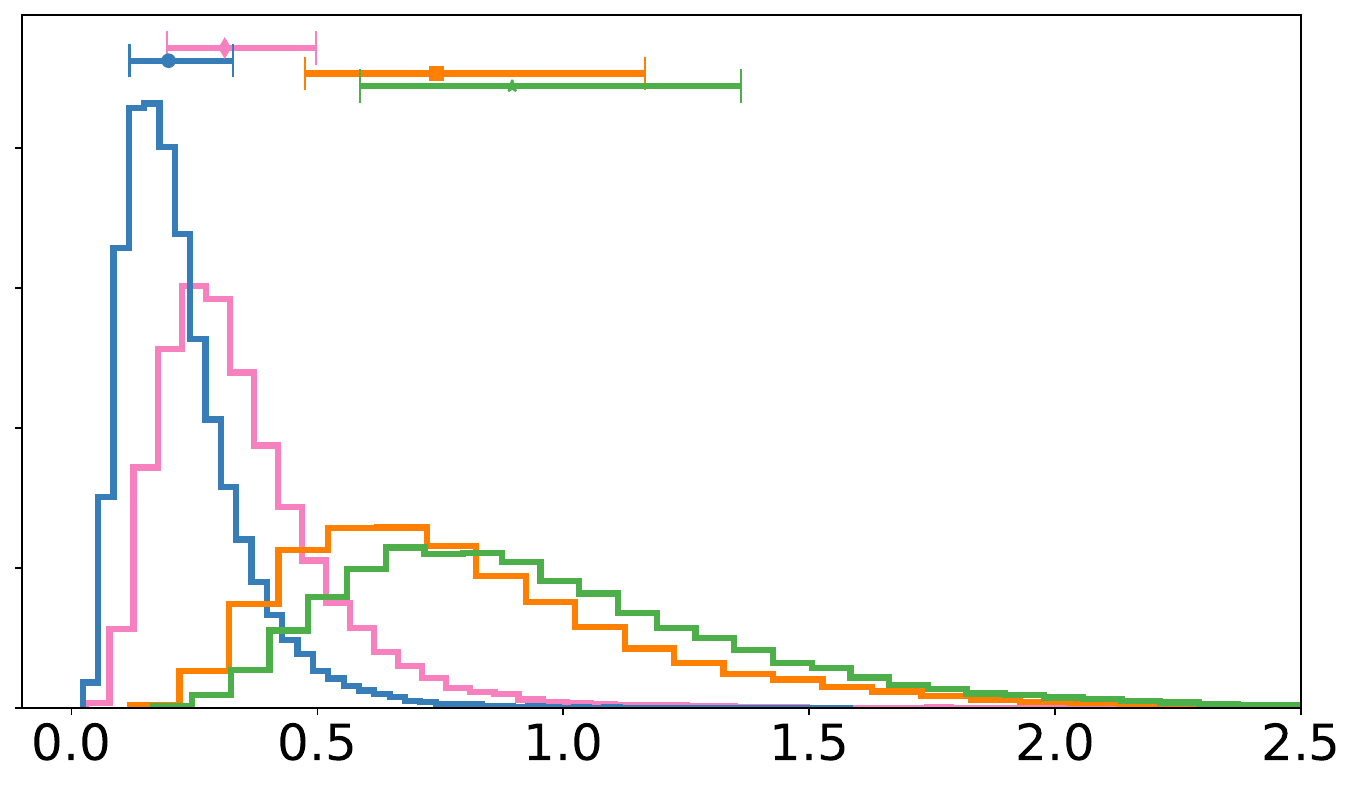} \\
  \includegraphics[width=0.95\linewidth]{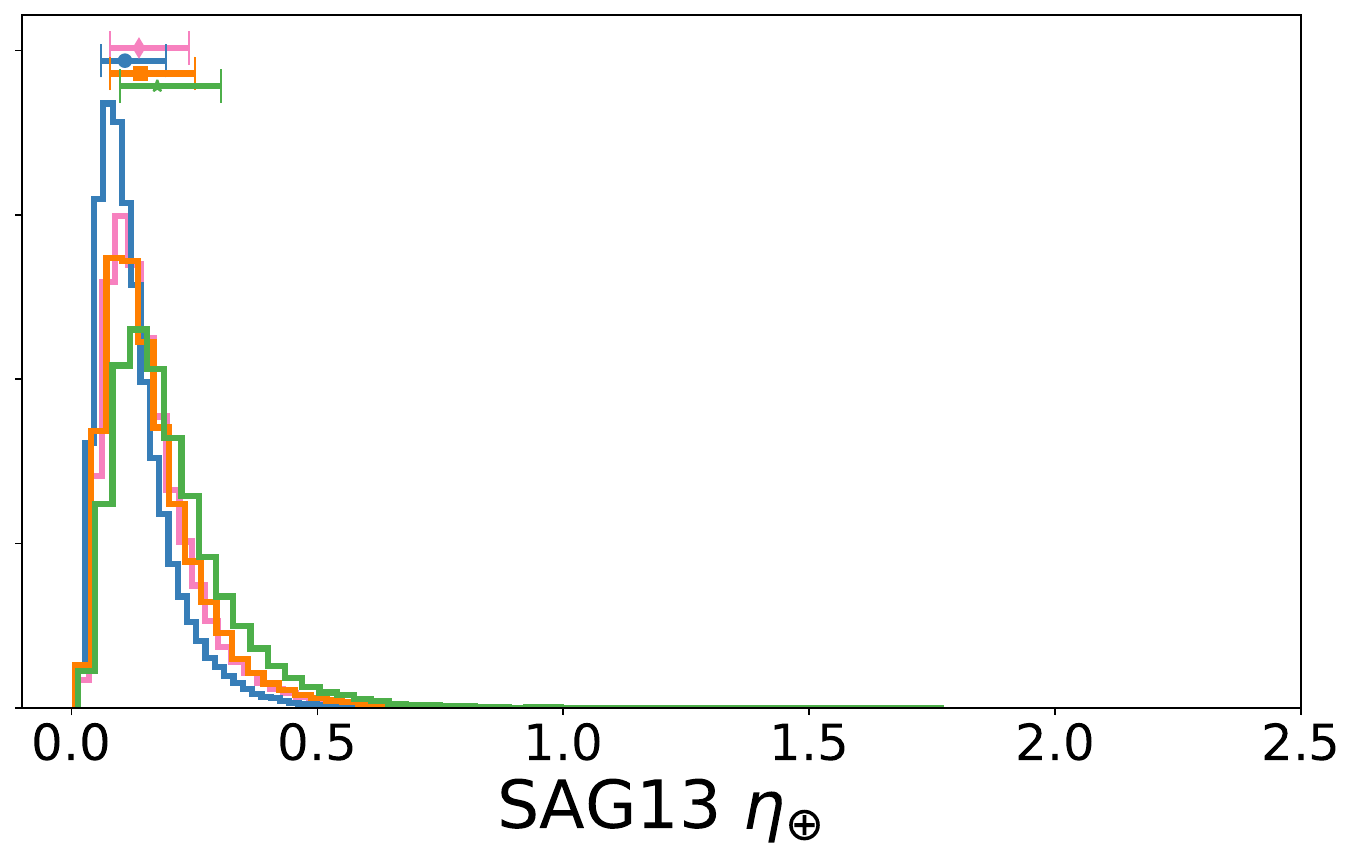} 
\caption{An example of the impact of including reliability in an $\eta_\oplus$ estimate, comparing the distribution in SAG13 $\eta_\oplus$ values using different vetting thresholds.  Top: $\eta_\oplus$ values without accounting for reliability.  Bottom: $\eta_\oplus$ values accounting for reliability.  The colors indicate the choice of thresholds: DR25, which balances completeness and reliability (pink), tuned for high-reliability (blue), tuned for high-completeness (orange), and tuned to best match the certified false positive table (green).  From \citet{Bryson2020b}.} \label{figure:compareReliability}
\end{figure}

\citet{Bryson2020b} studied the impact of correcting catalogs for reliability.  Four catalogs were used for comparison, each with vetting thresholds tuned to different goals.  Three of the catalogs are from the DR25 baseline thresholds, tuned to balance completeness and reliability, ``high-reliability'' thresholds, tuned for fewer, more reliable exoplanets and ``high-completeness'' thresholds, tuned for more, less reliable exoplanets as described in \citep{Thompson2018}.  The fourth catalog, ``FPWG'', was tuned to best match the certified false positive table generated by the False Positive Working Group \citep{Bryson2015}.  These catalogs were created from the same set of observations with different vetting criteria, so we expect that they describe the same underlying population statistics.  \citet{Bryson2020b} found that the estimate of the SAG13 $\eta_\oplus$ from these catalogs, based on a power law population model in period and radius, gives notably different distributions when not correcting for reliability, with median values varying by up to a factor of 4.5 (the ratio of the highest to the lowest value).  When correcting for reliability, median $\eta_\oplus$ from the four catalogs varies by a factor of 1.6.  The distributions of $\eta_\oplus$ from these catalogs is shown in Figure~\ref{figure:compareReliability}, where we see that without reliability correction the $\eta_\oplus$ distributions are quite different, while with reliability correction the distributions are much more alike.  Because these distributions are from the measurement of the same underlying population, the lack of consistency in the distributions when not accounting for reliability compared with the much more consistent distributions when accounting for reliability illustrates the importance of reliability correction for a robust estimate of $\eta_\oplus$ that is independent of how the exoplanet catalog is created.

\subsection{The Hard Problem: Fundamental Limitations in $\eta_\oplus$ Estimates Using Kepler Data}\label{section:limitations}

It is not surprising that $\eta_\oplus$ studies using different exoplanet catalogs, different stellar properties and different definitions of $\eta_\oplus$ (as described in \S\ref{section:underControl}) result in different estimates of $\eta_\oplus$.  But the studies shown in Figure~\ref{fig:etaEarthHistory} from 2020 to the present used similar (or similarly constructed) catalogs, definitions, reliability and completeness corrections, yet yield very different estimates of $\eta_\oplus$. In this section we discuss aspects of Kepler data that create difficulties in estimating $\eta_\oplus$ that are much more difficult to control than those described in \S\ref{section:underControl}.  Different studies (sometimes by the same authors, as in the case of \citet{Bryson2020} and \citet{Bryson2021}) take different approaches to these difficulties, yielding different values of $\eta_\oplus$.




\subsubsection{The Lack of Detections} \label{section:lackDetections}
As described in \S\ref{section:lackDetections}, there is a dearth of detections of Earth-size exoplanets in the habitable zones of Sun-like stars.  This lack of detections is made worse when reliability is taken into account as described in \S\ref{section:reliability}, because, as shown in Figure~\ref{figure:populations}, habitable-zone Earth-like exoplanets have typically low reliability.  

An immediate impact of a small number of detections is large uncertainties due to Poisson statistics.  We can gain insight on the impact of small numbers of detection from Table 5 of \citet{Bryson2021}, where $\eta_\oplus$ is computed for the exoplanet radius range of $0.5$ -- $1.5~R_\oplus$ where there are few detections, and $1.5$ -- $2.5~R_\oplus$ where there are significantly more detections, though these larger exoplanets are unlikely to be rocky (see the lower panel of Figure~\ref{figure:populations}). For the zero-extrapolation completeness case (see \S\ref{section:completeness}), for exoplanet sizes $0.5$ -- $1.5~R_\oplus$, where we find 9 PCs, $\eta_\oplus = 0.60^{+0.90}_{-0.36} = 0.60^{+150\%}_{-60\%}$ while for $1.5$ -- $2.5~R_\oplus$, where we have 38 PCs, $\eta_\oplus = 0.24^{+0.14}_{-0.08} = 0.24^{+58\%}_{-33\%}$.  So a decrease in the number of exoplanets by about a factor of 4 results in an increase in the uncertainty by a factor of 2 to 3, consistent with the expectations from Poisson statistics.

\subsubsection{The Need to Extrapolate} \label{section:needExtrapolate}
Beyond the increase in uncertainty from Poisson statistics due to small numbers of detections discussed in \S\ref{section:lackDetections}, extrapolation to cover the habitable zone can induce further severe uncertainties and biases.
Extrapolation is necessary to determine $\eta_\oplus$ using Kepler data: as shown in Figures \ref{figure:populationPerRad} and \ref{figure:populations}, Kepler data does not cover the entirety of the habitable zone for stars hotter than $T_\mathrm{eff} \sim 5500\mathrm{K}$.  An estimate of the number of exoplanets in the habitable zones of these hotter stars necessarily involves the integration of extrapolations into regions that contain no data. For example, several authors \citep[e.g.,][]{ Burke2015,Mulders2018,Bryson2020,Bryson2021,Bergsten2022} extrapolate power laws that are dominated by more abundant exoplanet detections: either from larger exoplanets to smaller exoplanets at long orbital period, and/or from short-period to long-period small exoplanets. But the population of larger, likely not rocky, exoplanets may not represent the population of rocky exoplanets.

Similarly, the occurrence of small close-in exoplanets may not reflect the rocky population in the HZ because the radius distribution at short orbital periods is strongly affected by atmospheric loss. \citet{Lopez2018} and \citet{Pascucci2019} showed that including the numerous small ($<1.8,R_\oplus$), short-period ($<25$,days) exoplanets -- many likely evolved sub-Neptunes stripped of their atmospheres -- in $\eta_\oplus$ extrapolations biases the estimates to high values. Using the same method and HZ definition as \citet{Mulders2018}, they found that excluding these close-in exoplanets lowers $\eta_\oplus$ by factors of $\sim4$–$8$ (Figure~\ref{fig:etaEarthHistory}). Recent studies of young clusters and associations find more sub-Neptunes than around Gyr-old stars \citep{Christiansen2023,Vach2024,Fernandes2025}, highlighting the role of evaporation in reshaping exoplanet radii for the close-in population and the importance of excluding this population from $\eta_\oplus$ extrapolations as pointed out in \citet{Pascucci2019}.  In addition, different authors use different parametric population models (compare, for example, \citet{Bryson2021} and \citet{Bergsten2022}), whose extrapolations can differ considerably even when extrapolating the same data.

So extrapolation is unavoidable when etimating $\eta_\oplus$ using Keler data, and it can be difficult to determine what population model should be extrapolated.  But extrapolation presents dangers beyond the ignorance of the model: even when a parametric population model has been determined, it will have unknown parameters that need to be fitted to the observed data.  Extrapolating such a fitted model introduces two important sources of error and uncertainty:
\begin{itemize}
    \item Extrapolation of a function introduces an error proportional to that function's derivatives: functions with large second and higher derivatives may vary rapidly as we extrapolate beyond the domain where data are available. Because these derivatives typically depend on the fitted model parameters, it is very difficult to predict the resulting extrapolation error, which acts as an unknown bias.
    \item When there are few observed data points near the edge of the observed data domain, extrapolation beyond that domain is sensitively dependent on those few data points.  Changes in those data points -- for example, small differences between two different author's exoplanet catalogs -- can have a significant impact on the extrapolation.  This is a particularly pernicious problem when estimating $\eta_\oplus$ because small, long-period/low insolation exoplanets are at Kepler's detection limit, so small changes in the catalog generation process may or may not include the same exoplanets.
\end{itemize}

\begin{figure}[ht]
  \centering
  \includegraphics[width=0.95\linewidth]{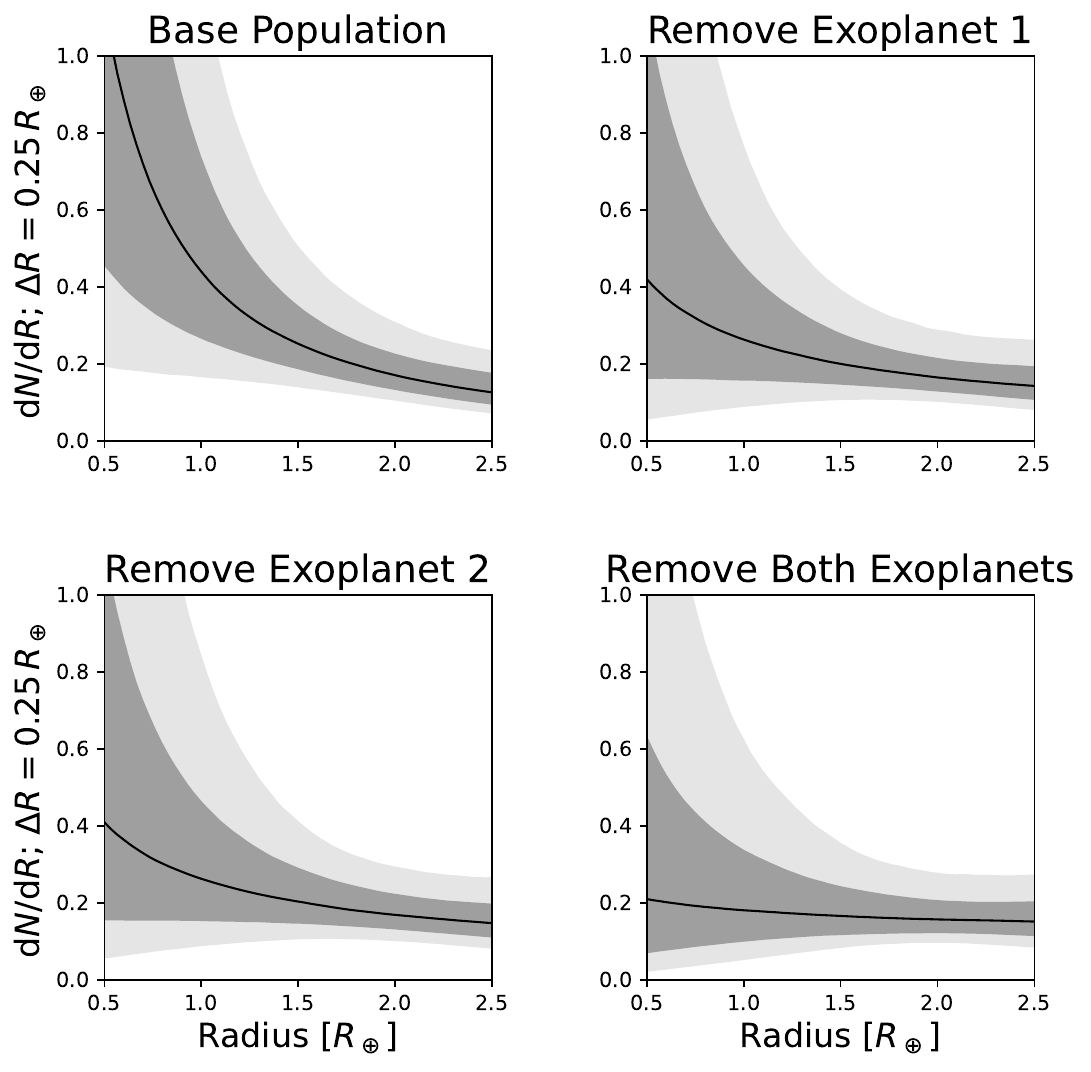} \\
\caption{The impact of removing the exoplanets labeled ``1'' and ``2'' in the bottom panel of Figure~\ref{figure:populations} on the marginalized dependence of exoplanet occurrence on exoplanet radius.  Upper left: including all the exoplanets in Figure~\ref{figure:populations}.  Upper right: removing exoplanet 1. Lower left: removing exoplanet 2.  Lower right: removing exoplanets 1 and 2.} \label{figure:fitDependence}
\end{figure}

\begin{figure}[ht]
  \centering
  \includegraphics[width=0.95\linewidth]{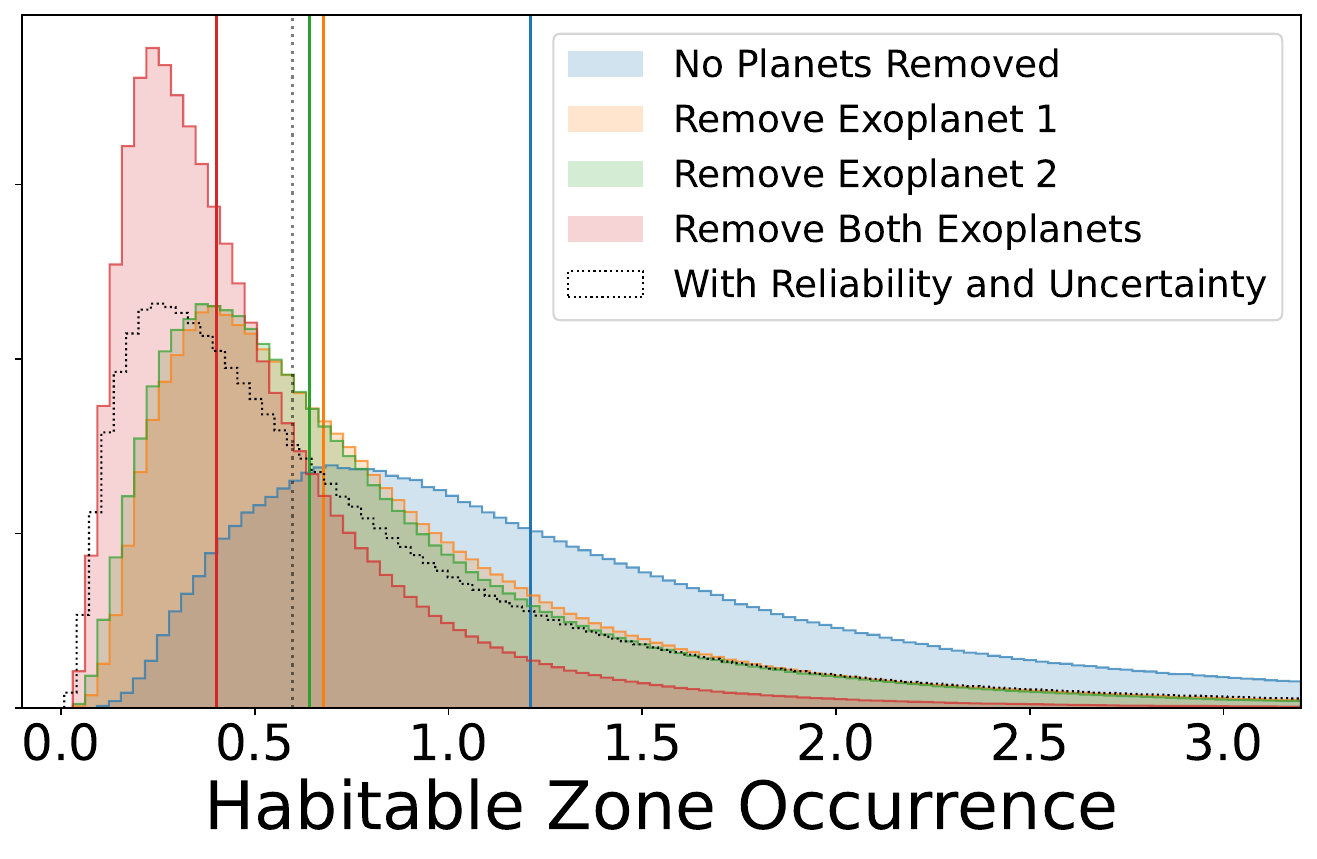} \\
\caption{The impact of removing the exoplanets labeled ``1'' and ``2'' in the bottom panel of Figure~\ref{figure:populations}, corresponding to the four cases shown in Figure~\ref{figure:fitDependence}, on $\eta_\oplus$.  The distributions when removing either exoplanet 1 or 2 are very similar.  The result when accounting for reliability and uncertainty from \citet{Bryson2021} is included for comparison.} \label{figure:etaEatrhFitDependence}
\end{figure}

We present a simple example of the second point, the sensitivity of extrapolation to small changes at the edge of the data set.  Specifically, we examine the impact of removing one or two exoplanets (labeled ``1'' and ``2'' in the bottom panel of Figure~\ref{figure:populations}) on $\eta_\oplus$. These exoplanets were selected because they are at the edge of the data domain beyond which extrapolation is required to compute $\eta_\oplus$.
Figures \ref{figure:fitDependence} and \ref{figure:etaEatrhFitDependence}  show how $\eta_\oplus$ changes with the exclusion of these exoplanets, computed using the methods and software of \citet{Bryson2021}; the final $\eta_\oplus$ distribution from \citet{Bryson2021} is also shown in Figure~\ref{figure:etaEatrhFitDependence} for comparison. Briefly, we modeled the exoplanet population as a power law in radius, instellation flux and host star effective temperature, using the Poisson likelihood of \citet{Bryson2021}, for the case of zero completeness extrapolation, without accounting for reliability.  Figure~\ref{figure:fitDependence} shows the resulting marginalized power law dependence of exoplanet occurrence on exoplanet radius.  We see that the inferred power law changes dramatically at small radii when we remove one of the exoplanets, and very dramatically when we remove both exoplanets. 
Figure~\ref{figure:etaEatrhFitDependence} shows the resulting value of $\eta_\oplus$ in each case.  We see that removing one exoplanet at the edge of the observed data can reduce $\eta_\oplus$ by a factor of 2, while removing both exoplanets reduces $\eta_\oplus$ by a factor of 3.

The exoplanets we remove in this example are in a regime of very few, low-reliability detections (see Figure~\ref{figure:populations}) so different studies may or may not include them. Their inclusion (or exclusion) in different exoplanet catalogs can go a long way towards accounting for the differences we see in Figure~\ref{fig:etaEarthHistory}.    

The above example illustrating the impact of data near the edge of the data domain did not consider the reliability of these exoplanets.  When combined with considerations of exoplanet reliability, that impact manifests as increased uncertainty in the $\eta_\oplus$ estimate.  The $\eta_\oplus$ estimate from \cite{Bryson2021} accounted for reliability by combining $\eta_\oplus$ posteriors for computations that included exoplanets with a probability equal to their reliability.  Thus, their final $\eta_\oplus$ posteriors (included as the black line in Figure~\ref{figure:etaEatrhFitDependence}) combined cases with and without those (and other) low-reliability exoplanets, which accounts for the wider $\eta_\oplus$ distribution when considering reliability (and uncertainty in exoplanet properties).

To summarize, the need to extrapolate population models that are poorly constrained by theory or data to cover the habitable zone leads to unknown and possibly large biases in the resulting value of $\eta_\oplus$.  We believe that these biases are a dominant source of uncertainty in $\eta_\oplus$ estimates.  We stress that these biases are unknown and typically are not included in the error estimates of $\eta_\oplus$.  These biases may be a significant contributor to the spread of values in Figure~\ref{fig:etaEarthHistory}.

\begin{table*}[ht]
\caption{Summary of Issues and Their Impacts}\label{table:summary}
\begin{tabular}{p{0.40\textwidth}|p{0.43\textwidth}}
\multicolumn{2}{c}{Choices} \\
\hline
Definition of $\eta_\oplus$ (\S\ref{section:habZone}) & Large impact on accuracy, sensitive to extrapolation.  \\
\hline
Data vintage (\S\ref{section:catalogs}) &  Accuracy and precision of the input data impacts accuracy and precision of the result.  \\
\hline
Correct for catalog reliability  (\S\ref{section:reliability}) &  Impacts accuracy (by multiple factors) and precision (by falsely reducing uncertainty)\\
\hline
\multicolumn{2}{c}{Intrinsic Limitations} \\
\hline
Extrapolation of unknown populations (\S\ref{section:needExtrapolate}) & Large, unbounded impact on accuracy. \\
\hline
Lack of detections  (\S\ref{section:lackDetections}) &  Impacts precision through small-number statistics \\
\hline
Limited Completeness measurement  (\S\ref{section:completeness}) &  Impacts accuracy, bounded by about $50\%$ of the value.  \\
\hline
\end{tabular}
\end{table*}





\section{What Can Be Done?} \label{section:whatToDo}
In \S\ref{section:spread} we present what we believe to be the issues contributing to the spread of $\eta_\oplus$ values in Figure~\ref{fig:etaEarthHistory}.  These issues are summarized in Table~\ref{table:summary}. In this table we use the language of {\it accuracy}, the extent to which the central value of a measurement distribution is correct, and {\it precision}, the width of the measurement distribution around the central value.  Both accuracy and precision contribute to the uncertainty of a measurement, but in importantly different ways: a high-precision measurement can have low accuracy.  Some of the issues discussed in this work primarily impact the accuracy of an $\eta_\oplus$ estimate, leading to a biased result, while others primarily impact precision.

Our main conclusion is that, while differences in $\eta_\oplus$ due to varying choices can be reconciled (the easy problem of \S\ref{section:underControl}), there are intrinsic limitations with no easy or obvious solution (the hard problem of \S\ref{section:limitations}).  Different studies can take different, equally legitimate approaches to these intrinsic limitations that will result in possibly very different values of $\eta_\oplus$.  

\subsection{Can We Obtain More Data?}

The best way to improve our ability to estimate $\eta_\oplus$ is to reliably detect more Earth-sized exoplanets in the entirety of the habitable zone of Sun-like stars.  There are two space telescopes moving towards launch that are specifically designed to detect Habitable-zone Earth-size exoplanets around Sun-like stars via transits: PLATO \citep{Rauer2014, Rauer2024} and Earth 2.0 \citep{Ge2022, Zhang2022}.  Both of these missions use multiple overlapping small telescopes to cover a wide field of view and reduce systematic noise.  In addition, the Nancy Grace Roman space telescope will allow transit detection of small exoplanets in the habitable zone of M-dwarf stars \citep{Wilson2023}.

PLATO, planned for launch in late 2026, is predicted to detect ``about a dozen'' small habitable zone exoplanets around FGK stars \citep{Heller2022}, requiring two transits for detection.  As of this writing, PLATO's baseline observational plan is four years of operations, observing two fields for two years each.  In this baseline plan, even accepting two transits as a detection, the longest detectable orbital period will be 730 days, similar to Kepler's 710-day three-transit limit.  Therefore, PLATO's design and baseline observational strategy does not cover the entirety of the habitable zones of early G and late F stars  \citep[see also Figure~2 of ][]{Rauer2014}. Because an Earth-Sun analog is near PLATO's detection limit, we expect that it will be challenging to distinguish true exoplanets with two transits from noise fluctuations.   The baseline plan of two-year observations per field is provisional, and other possibilities, including a three-year observation of a single field, are being considered.  Further, the PLATO spacecraft is designed for over 8 years of operations, so there is the possibility of mission extensions allowing further long-term observations of single fields.  PLATO currently plans a partial overlap with the Kepler prime mission field, allowing the possibility of detections with longer orbital periods by combining Kepler and PLATO data. Plato's detection of Kepler exoplanet candidates in the $\eta_\oplus$ regime would significantly increase the reliability of these candidates.

Earth 2.0, planned for launch in 2028, is predicted to detect 10 to 20 small habitable zone exoplanets around FGK stars \citep{Ge2022}, requiring three transits for detection.  Earth 2.0's operational plan is to observe a large field that entirely covers the Kepler prime mission field for four years.  Combining Earth 2.0 with Kepler prime mission data can potentially identify small exoplanets throughout the habitable zone of early K through late F stars. 

The Nancy Grace Roman telescope, set to launch in 2026, is expected to detect over a thousand exoplanets via microlensing, a few tens of which are in the habitable zone of Sun-like stars \citep{Penny2019}. In addition, Roman can detect small habitable zone exoplanets orbiting M dwarfs using other methods: a few tens such exoplanets via transit observations \citep{Wilson2023}, and about 13 via astrometry \citep{tamburo2023}.  

\subsection{Further Exploitation of the Kepler Data}
While we look forward to observations from these space telescopes, it will take several years for these observations to impact $\eta_\oplus$ studies.  In the mean time, we believe that there are opportunities to further exploit existing Kepler prime mission data, creating new exoplanet candidate catalogs that have higher completeness {\it and} higher reliability.  Just a few higher reliability exoplanet in the $\eta_\oplus$ regime could significantly improve constraints on $\eta_\oplus$.

\begin{figure}[ht]
  \centering
  \includegraphics[width=0.95\linewidth]{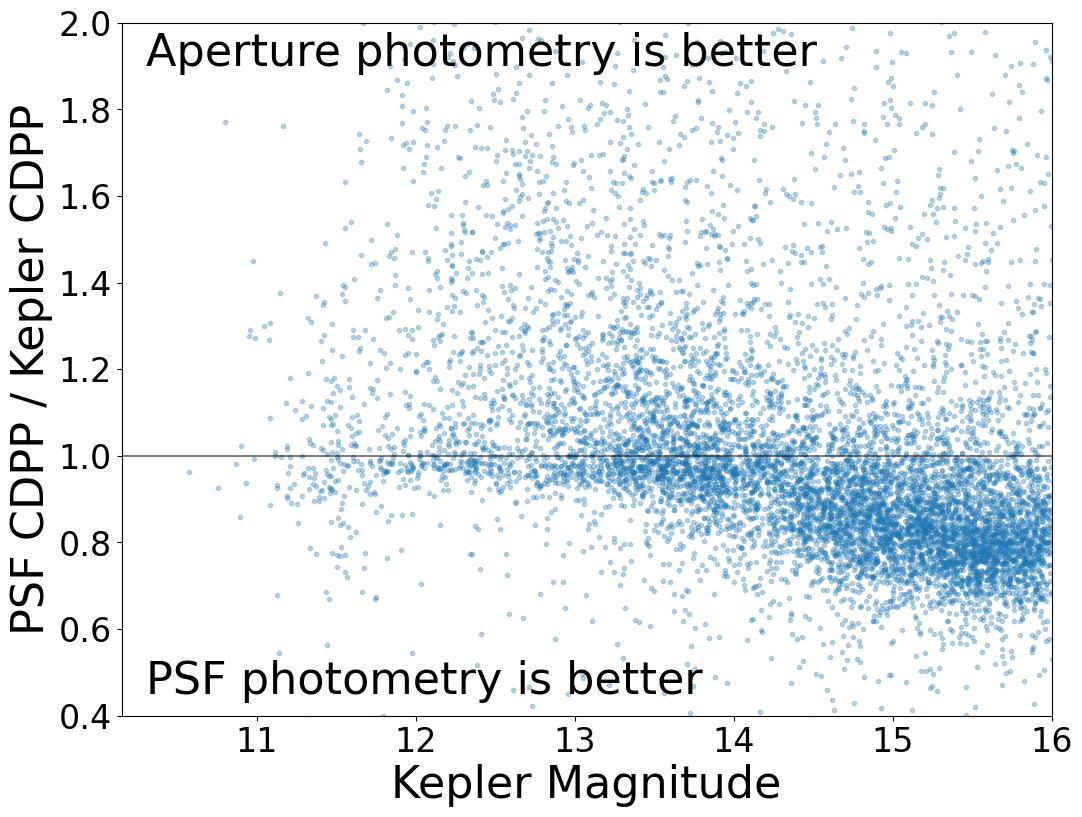} \\
\caption{The ratio of the combined differential photometric precision (CDPP) of light curves from Kepler DR25 aperture photometry to the CDPP of light curves from the PSF photometry of \citet{Martinez2023}.} \label{figure:keplerBonus}
\end{figure}

As described in \S\ref{section:doingOurBest}, the aggressive vetting required to remove the false alarms very likely removed true exoplanets detected in the $\eta_\oplus$ regime. Developing improved detection algorithms that pass fewer false alarms would allow less aggressive vetting, possibly passing more true exoplanets.  One example of an improvement in the detection algorithms is the use of pixel level data to check that a transit signal in a light curve is caused by pixel-level changes consistent with a change in star brightness.  Analysis of the systematic noise properties of light curves such as the statistical bootstrap used in \citet{Jenkins2015} and methods developed in \citet{Matesic2024} or \citet{Robnik2025} provide a candidate-specific analysis of the false alarm probability that is likely more accurate than the population-based reliability analysis of DR25.  Such a pixel-level check would weed out cosmic-ray or electronic artifact events that can mimic small-exoplanet transits in the light curve.  Improved methods of creating higher-precision light curves, such as the PSF photometry technique developed for Kepler in \citet{Hedges2021} and implemented in \citet{Martinez2023} would improve our ability to detect small exoplanets in Kepler data.  We compared the light curves from Kepler DR25 aperture photometry with those from \citet{Martinez2023}, comparing their combined 6.5 hour differential photometric precision (CDPP) \citep{Christiansen2012}, computed via the lightkurve package \citep{lightkurve2018}.  Lower CDPP indicates better precision.  Figure~\ref{figure:keplerBonus} shows the CDPP ratio of the Kepler aperture photometry to the PSF photometry for Kepler target stars brighter than 16th magnitude.  For target stars between magnitude 14 and 16, the PSF photometry for most target has about 20\% better precision. \citet{Martinez2023} also computes photometry for neighboring stars that happened to be captured in the pixels for Kepler target stars, which may reveal more exoplanets. But we caution that these neighboring light curves are for dimmer stars that are not likely to have the precision required to detect small exoplanets in their habitable zones. Lessons learned from the analysis of TESS data may provide other opportunities.  An effort to reprocess Kepler data starting from calibrated pixels that incorporates such improvements is underway \citep{Bryson2025}.  

Machine learning methods have been developed to detect (for example, \citet{Shallue2018}) and vet (for example, \citet{Armstrong2021, Valizadegan2022, Valizadegan2023}) exoplanets in Kepler data.  While we encourage further development of such machine learning approaches, they have not demonstrated effectiveness in detection or vetting for exoplanets in the $\eta_\oplus$ regime.  In particular, we are concerned about how machine learning approaches are typically trained on exoplanet populations such as the Kepler TCE table \citep{Twicken2016} or the NASA confirmed exoplanet table \citep{Christiansen2025}, where there are many exoplanets at short orbital period and few exoplanets in the $\eta_\oplus$ regime.  Therefore the characteristics of Kepler data for long orbital periods described in \S\ref{section:KeplerProblems}, which are very different from those at shorter orbital period, are possibly not well represented in such training data.  While we believe that there may be future opportunities for machine learning in distinguishing exoplanet detections from false positives, we feel that the detection of exoplanets in the $\eta_\oplus$ regime via machine learning is an open, unsolved research area.

Finding more exoplanets in Kepler data is not enough to constrain $\eta_\oplus$, however.  As discussed in \S\ref{section:needExtrapolate}, there are orbital periods in the habitable zone of hotter stars that are not detectable by Kepler (with the three-transit detection requirement), so extrapolation of population models is required to fully cover those habitable zones.  At this time, there is no consensus on what population models to extrapolate: simple broken power laws (with few breaks) do not seem to adequately fit the data.  Accepting two-transit detections would cover the habitable zones of all but the hottest stars, but the issues with systematic noise in Kepler data described in \S\ref{section:currentSituation} would become much worse when requiring only two transits.  It is not clear whether requiring two rather than three transits in the Kepler data would reduce uncertainties in $\eta_\oplus$.  


Extrapolating through the habitable zone can be accomplished with a number of semi-empirical and theoretical approaches.  \citet{Fernandes2019} explored the demographics of more massive Jovian exoplanets discovered with the radial velocity technique, uncovering an occurrence rate turnover at larger semi-major axes.  Such demographics could be extrapolated to lower exoplanet masses by combining the demographics of radial velocity and Kepler derived-demographics, as was done in \citet{Dulz2020} utilizing older SAG12 estimates of $\eta_\oplus$.  However, given that Jovian, Neptune and terrestrial worlds are distinct exoplanet populations, and could have very different demographics trends with semi-major axis, such extrapolation in exoplanet size may not reflect reality. 


An unexplored approach to extrapolating over the habitable zone is to understand the underlying exoplanet population with the help of formation and evolution theories. Several competing theories currently aim to explain the formation and evolution of exoplanets, many of which successfully reproduce the observed ``radius valley'' at $\sim$1.8–2~$R_\oplus$ in the radius histogram of short-period ($\lesssim$100 days) Kepler exoplanets. For example, models involving atmospheric evolution, such as photoevaporation models \citep{Owen2017, Rogers2021} and core-powered mass loss models \citep{Ginzburg2018, Gupta2019, Gupta2020, Misener2021}, are effective at explaining the short-period population and may be useful to de-bias extrapolations of $\eta_\oplus$ that have made use of the entire population of close-in exoplanets as discussed in \S\ref{section:needExtrapolate}.

Formation and evolution models more directly relevant to the $\eta_\oplus$ regime are available.  For example, models that invoke late-stage gas accretion from a depleted disk near the end of its lifetime \citep{Lee2016, Lee2021, Lee2022} can account for both the radius valley and the existence of long-period terrestrial exoplanets. Such models can provide the relative occurrence of rocky versus gas-rich , so fitting these models to the abundant larger observed exoplanet population can constrain the rocky exoplanet population.  These models often rely on parametric assumptions for the initial distributions of mass, period, and composition. Constraints on these initial distributions can come from formation model grids derived from \textit{N}-body simulations \citep{Hansen2013, Dawson2016, Mulders2020, Chakrabarty2024}. 

Current observational data does not provide enough information to select the ``best'' formation and evolution models from the several competing available theories, so it is likely premature to rely on such models for an accurate estimate of $\eta_\oplus$. However, the use of formation-and-evolution-theory-based population models may provide a more physically motivated basis for extrapolation compared with parametric population models (such as power laws) imposed by fiat. Further, it's possible that $\eta_\oplus$ estimates using a specific model will provide opportunities to constrain that model's parameters.  Such an approach could be explored for different formation and evolution models, and reports of $\eta_\oplus$ could include which formation and evolution model was used for extrapolation.

An important aspect of this consideration is the extent to which planetary system architectures shape the composition of exoplanets in the liquid--water zone.  For example, having a gas giant form near the iceline (e.g. Jupiter in our Solar System) may separate volatile--rich  from volatile--poor exoplanet--forming reservoirs, inhibiting the formation of water worlds, with several percent water mass fraction near the habitable zone.  Such exoplanets may not be suitable for the biochemical origins of life.  The Earth contains $< 10^{-3}$ mass fraction of water for comparison.  If, like the liquid-water zone, the water mass fraction of an exoplanet has a limited range needed for life, and this range depends on system architectures, we then wonder if the probability of having a rocky exoplanet in the liquid--water zone is independent of the probability of having an iceline gas giant.  In that case, the probability of having a habitable exoplanet could be the product of eta Earth and eta iceline-gas-giant.  However, recent work suggests possible correlation between the presence of inner sub-Neptunes and the presence of outer gas giants (\citep{2024ApJ...968L..25B}).


In tandem with the extrapolation of population models to the $\eta_\oplus$ regime, the role of exoplanet multiplicity is another important, though largely unexplored, consideration that can potentially impact calculations of $\eta_\oplus$. The Kepler mission revealed that multi-exoplanet systems are ubiquitous. About $\sim 40\%$ of all transiting exoplanets observed by Kepler occur in systems with more than one transiting exoplanet \citep{Lissauer2011, Fabrycky2014, Lissauer2024}. Correcting for completeness, this implies that the underlying frequency of multi-exoplanet systems (within the period range Kepler was sensitive to, $P \lesssim 1$ yr) is even higher, with calculations ranging from $\sim$50-90\% (e.g., \citealt{FangMargot2012, Ballard2016, Zhu2018, Sandford2019, He2019, He2020}).  Therefore multi-exoplanet systems are very common, so multiplicity should be accounted for when computing $\eta_\oplus$.  Observed correlations in the structure of multiple-exoplanet systems can provide another way of extrapolating short-period populations to longer-period habitable zones

Such an approach is challenging: while systems often host multiple small exoplanets within $\sim 1$ AU, it remains unclear how well observed patterns of these exoplanets extrapolate to longer periods. To make matters worse, there is some evidence for an outer ``edge" in the pattern of orderly, multi-exoplanet systems based on a detailed study of the Kepler statistics and its detection efficiency drop-off \citep{Millholland2022}. Despite these difficulties, developing techniques to leverage off of multi-exoplanet correlations can provide useful constraints on $\eta_\oplus$.

Dynamical packing and stability simulations provide additional constraints for additional hypothetical exoplanets in constraining $\eta_\oplus$. Dynamical packing was considered in \citet{Dulz2020} as well, and they concluded that their approximate dynamical stability criteria inferred from simulations and compact multiples could limit $\eta_\oplus$, and was increasingly relevant for more massive exoplanets, and exoplanets at larger semi-major axes, consistent with the turnover reported in \citet{Fernandes2019}.  

Exoplanet multiplicity raises the problem of how many habitable-zone exoplanets orbit the same star.  Understanding the impact of exoplanet multiplicity on $\eta_\oplus$ would require modeling the number of exoplanets that can exist in the habitable zones of individual stars. The width of the habitable zone generally increases towards hotter stars and is on the order of one to a few AU for Sun-like stars \citep{Bryson2021}. Thus, the ratio of semi-major axes of the outer to inner habitable zone boundaries is a factor of several. Considering that the Kepler-observed period-ratio distribution peaks at $\lesssim 2$, this back-of-the-envelope calculation suggests that the habitable zones of Sun-like stars are wide enough to potentially host multiple Earth-mass exoplanets on stable orbits.

Modeling exoplanet multiplicity in the $\eta_\oplus$ regime enables the distinction between the mean number of rocky habitable-zone exoplanets per star and the fraction of stars with such exoplanets. Essentially all Kepler studies to date have computed the former quantity. Discerning between these two quantities offers an understanding of not only how frequently rocky exoplanets exist in the habitable zones of their host stars, but also how these exoplanets are distributed across their host stars.  The fraction of stars with habitable exoplanets is more directly relevant to, for example, HWO target selection than $\eta_\oplus$, the average number of habitable exoplanets per star.

\section{Summary} \label{section:summary}

In this work we have identified various possible reasons for the community not reaching consensus on the value of $\eta_\oplus$ using Kepler data, as shown in Figure~\ref{fig:etaEarthHistory}.  Some of these reasons, described in \S\ref{section:underControl}, are simple matters of choice and convention, but we have identified fundamental issues, discussed in \S\ref{section:limitations}, that defy consensus and cannot be resolved by analysis of currently existing Kepler data releases alone. These fundamental issues are driven by, in the $\eta_\oplus$ regime, a) a lack of detections of exoplanets, and b) our inability to fill in that lack of detections due to our ignorance of the exoplanet population structure.  Addressing these issues requires either more data from future telescopes, or deeper analysis of Kepler data than in current data releases combined with further advances in understanding exoplanet population structure.

We reach these conclusions by telling the story of the Kepler mission and how that mission and the exoplanet community addressed the unexpected limitations encountered once Kepler was on orbit.  
The Kepler space telescope was designed to accomplish the ambitious goal of estimating $\eta_\oplus$, but for a variety of reasons (described in \S\ref{section:currentSituation} it fell short, leaving us tantalizingly close to that goal.  Many studies have attempted to bridge the gap between the data we have and estimates of $\eta_\oplus$, but, as shown in Figure~\ref{fig:etaEarthHistory}, there has been little agreement about the value of $\eta_\oplus$, though most estimates agree that there are very large uncertainties.  In this paper we describe what we believe are the dominant contributions to this lack of agreement.  Beyond the simple disagreement described in \S\ref{section:underControl} about exactly what is being estimated, in \S\ref{section:limitations} we identify two important problems that every attempt to estimate $\eta_\oplus$ using Kepler data will have to grapple with: the lack of detections leading to large uncertainties, and possibly large unknown biases due to the need to extrapolate unknown population models to cover the habitable zone.  We conclude in \S\ref{section:whatToDo} with thoughts on how to improve the situation with improved analysis of Kepler data and, ultimately, with more data from upcoming space telescopes.  

We hope this work will help our colleagues tackle the subtle, difficult, and exciting challenge of estimating $\eta_\oplus$. 


\section{Acknowledgments} 
We thank Eric Petigura, Eric Mamajec, Chris Stark, and Susan Thompson for helpful discussions.  We thank the anonymous reviewer for several excellent suggestions resulting in an improved paper. A.C. acknowledges support from the NASA Postdoctoral Program at NASA Ames Research Center, administered by Oak Ridge Associated Universities under contract with NASA.
G.D.M. acknowledges support from FONDECYT project 1252141 and the ANID BASAL project FB210003. I.P. acknowledges partial support by the National Aeronautics and Space Administration under agreement No. 80NSSC21K0593 for the program
“Alien Earths.” M.K. acknowledges the support of the Natural Sciences and Engineering Research Council of Canada (NSERC), RGPIN-2024-06452. Cette recherche a été financée par le Conseil de recherches en sciences naturelles et en génie du Canada (CRSNG), RGPIN-2024-06452.  The opinions expressed in this paper are those of the authors, and do not reflect the position of NASA, the United States Government, or any institution. 
This research made use of Lightkurve, a Python package for Kepler and TESS data analysis \citep{lightkurve2018}.

\facilities{Kepler}
\software{\texttt{lightkurve} \citep{lightkurve2018}, \texttt{matplotlib} \citep{Hunter2007}, \texttt{numpy} \citep{Harris2020}, \texttt{pandas} \citep{reback2020pandas, mckinney-proc-scipy-2010},  \texttt{scipy} \citep{Virtanen2020}}

\bibliography{refs}{}

\begin{thebibliography}{}
\expandafter\ifx\csname natexlab\endcsname\relax\def\natexlab#1{#1}\fi
\providecommand{\url}[1]{\href{#1}{#1}}
\providecommand{\dodoi}[1]{doi:~\href{http://doi.org/#1}{\nolinkurl{#1}}}
\providecommand{\doeprint}[1]{\href{http://ascl.net/#1}{\nolinkurl{http://ascl.net/#1}}}
\providecommand{\doarXiv}[1]{\href{https://arxiv.org/abs/#1}{\nolinkurl{https://arxiv.org/abs/#1}}}

\bibitem[{D.~J. {Armstrong} {et~al.}(2021){Armstrong}, {Gamper}, \&
  {Damoulas}}]{Armstrong2021}
{Armstrong}, D.~J., {Gamper}, J., \& {Damoulas}, T. 2021,
  \bibinfo{title}{{Exoplanet validation with machine learning: 50 new validated
  Kepler planets},} \mnras, 504, 5327, \dodoi{10.1093/mnras/staa2498}

\bibitem[{S. {Ballard} \& J.~A. {Johnson}(2016){Ballard} \&
  {Johnson}}]{Ballard2016}
{Ballard}, S., \& {Johnson}, J.~A. 2016, \bibinfo{title}{{The Kepler Dichotomy
  among the M Dwarfs: Half of Systems Contain Five or More Coplanar Planets},}
  \apj, 816, 66, \dodoi{10.3847/0004-637X/816/2/66}

\bibitem[{N.~M. {Batalha} {et~al.}(2013){Batalha}, {Rowe}, {Bryson}, {Barclay},
  {Burke}, {Caldwell}, {Christiansen}, {Mullally}, {Thompson}, {Brown},
  {Dupree}, {Fabrycky}, {Ford}, {Fortney}, {Gilliland}, {Isaacson}, {Latham},
  {Marcy}, {Quinn}, {Ragozzine}, {Shporer}, {Borucki}, {Ciardi}, {Gautier},
  {Haas}, {Jenkins}, {Koch}, {Lissauer}, {Rapin}, {Basri}, {Boss}, {Buchhave},
  {Carter}, {Charbonneau}, {Christensen-Dalsgaard}, {Clarke}, {Cochran},
  {Demory}, {Desert}, {Devore}, {Doyle}, {Esquerdo}, {Everett}, {Fressin},
  {Geary}, {Girouard}, {Gould}, {Hall}, {Holman}, {Howard}, {Howell},
  {Ibrahim}, {Kinemuchi}, {Kjeldsen}, {Klaus}, {Li}, {Lucas}, {Meibom},
  {Morris}, {Pr{\v s}a}, {Quintana}, {Sanderfer}, {Sasselov}, {Seader},
  {Smith}, {Steffen}, {Still}, {Stumpe}, {Tarter}, {Tenenbaum}, {Torres},
  {Twicken}, {Uddin}, {Van Cleve}, {Walkowicz}, \& {Welsh}}]{Batalha2013}
{Batalha}, N.~M., {Rowe}, J.~F., {Bryson}, S.~T., {et~al.} 2013,
  \bibinfo{title}{{Planetary Candidates Observed by Kepler. III. Analysis of
  the First 16 Months of Data},} \apjs, 204, 24,
  \dodoi{10.1088/0067-0049/204/2/24}

\bibitem[{T.~A. {Berger} {et~al.}(2018){Berger}, {Huber}, {Gaidos}, \& {van
  Saders}}]{berger18}
{Berger}, T.~A., {Huber}, D., {Gaidos}, E., \& {van Saders}, J.~L. 2018,
  \bibinfo{title}{{Revised Radii of Kepler Stars and Planets Using Gaia Data
  Release 2},} \apj, 866, 99, \dodoi{10.3847/1538-4357/aada83}

\bibitem[{T.~A. {Berger} {et~al.}(2020{\natexlab{a}}){Berger}, {Huber},
  {Gaidos}, {van Saders}, \& {Weiss}}]{Berger2020b}
{Berger}, T.~A., {Huber}, D., {Gaidos}, E., {van Saders}, J.~L., \& {Weiss},
  L.~M. 2020{\natexlab{a}}, \bibinfo{title}{{The Gaia-Kepler Stellar Properties
  Catalog. II. Planet Radius Demographics as a Function of Stellar Mass and
  Age},} \aj, 160, 108, \dodoi{10.3847/1538-3881/aba18a}

\bibitem[{T.~A. {Berger} {et~al.}(2020{\natexlab{b}}){Berger}, {Huber}, {van
  Saders}, {Gaidos}, {Tayar}, \& {Kraus}}]{Berger2020a}
{Berger}, T.~A., {Huber}, D., {van Saders}, J.~L., {et~al.} 2020{\natexlab{b}},
  \bibinfo{title}{{The Gaia-Kepler Stellar Properties Catalog. I. Homogeneous
  Fundamental Properties for 186,301 Kepler Stars},} \aj, 159, 280,
  \dodoi{10.3847/1538-3881/159/6/280}

\bibitem[{G.~J. {Bergsten} {et~al.}(2022){Bergsten}, {Pascucci}, {Mulders},
  {Fernandes}, \& {Koskinen}}]{Bergsten2022}
{Bergsten}, G.~J., {Pascucci}, I., {Mulders}, G.~D., {Fernandes}, R.~B., \&
  {Koskinen}, T.~T. 2022, \bibinfo{title}{{The Demographics of Kepler's Earths
  and Super-Earths into the Habitable Zone},} \aj, 164, 190,
  \dodoi{10.3847/1538-3881/ac8fea}

\bibitem[{J. {Bin} {et~al.}(2018){Bin}, {Tian}, \& {Liu}}]{Bin2018}
{Bin}, J., {Tian}, F., \& {Liu}, L. 2018, \bibinfo{title}{{New inner boundaries
  of the habitable zones around M dwarfs},} E\&PSL, 492, 121,
  \dodoi{10.1016/j.epsl.2018.04.003}

\bibitem[{W. {Borucki} {et~al.}(2008){Borucki}, {Koch}, {Basri}, {Batalha},
  {Brown}, {Caldwell}, {Christensen-Dalsgaard}, {Cochran}, {Dunham}, {Gautier},
  {Geary}, {Gilliland}, {Jenkins}, {Kondo}, {Latham}, {Lissauer}, \&
  {Monet}}]{Borucki2008}
{Borucki}, W., {Koch}, D., {Basri}, G., {et~al.} 2008, in IAU Symposium, Vol.
  249, Exoplanets: Detection, Formation and Dynamics, ed. Y.-S. {Sun},
  S.~{Ferraz-Mello}, \& J.-L. {Zhou}, 17--24, \dodoi{10.1017/S174392130801630X}

\bibitem[{W.~J. {Borucki}(2016){Borucki}}]{Borucki2016}
{Borucki}, W.~J. 2016, \bibinfo{title}{{KEPLER Mission: development and
  overview},} RPPh, 79, 036901, \dodoi{10.1088/0034-4885/79/3/036901}

\bibitem[{W.~J. {Borucki} {et~al.}(2010){Borucki}, {Koch}, {Basri}, {Batalha},
  {Brown}, {Caldwell}, {Caldwell}, {Christensen-Dalsgaard}, {Cochran}, \&
  {DeVore}}]{Borucki2010}
{Borucki}, W.~J., {Koch}, D., {Basri}, G., {et~al.} 2010,
  \bibinfo{title}{{Kepler Planet-Detection Mission: Introduction and First
  Results},} Sci, 327, 977, \dodoi{10.1126/science.1185402}

\bibitem[{W.~J. {Borucki} {et~al.}(2011){Borucki}, {Koch}, {Basri}, {Batalha},
  {Brown}, {Bryson}, {Caldwell}, {Christensen-Dalsgaard}, {Cochran}, {DeVore},
  {Dunham}, {Gautier}, {Geary}, {Gilliland}, {Gould}, {Howell}, {Jenkins},
  {Latham}, {Lissauer}, {Marcy}, {Rowe}, {Sasselov}, {Boss}, {Charbonneau},
  {Ciardi}, {Doyle}, {Dupree}, {Ford}, {Fortney}, {Holman}, {Seager},
  {Steffen}, {Tarter}, {Welsh}, {Allen}, {Buchhave}, {Christiansen}, {Clarke},
  {Das}, {D{\'e}sert}, {Endl}, {Fabrycky}, {Fressin}, {Haas}, {Horch},
  {Howard}, {Isaacson}, {Kjeldsen}, {Kolodziejczak}, {Kulesa}, {Li}, {Lucas},
  {Machalek}, {McCarthy}, {MacQueen}, {Meibom}, {Miquel}, {Prsa}, {Quinn},
  {Quintana}, {Ragozzine}, {Sherry}, {Shporer}, {Tenenbaum}, {Torres},
  {Twicken}, {Van Cleve}, {Walkowicz}, {Witteborn}, \& {Still}}]{Borucki2011}
{Borucki}, W.~J., {Koch}, D.~G., {Basri}, G., {et~al.} 2011,
  \bibinfo{title}{{Characteristics of Planetary Candidates Observed by Kepler.
  II. Analysis of the First Four Months of Data},} \apj, 736, 19,
  \dodoi{10.1088/0004-637X/736/1/19}

\bibitem[{T.~M. {Brown} {et~al.}(2011){Brown}, {Latham}, {Everett}, \&
  {Esquerdo}}]{Brown2011}
{Brown}, T.~M., {Latham}, D.~W., {Everett}, M.~E., \& {Esquerdo}, G.~A. 2011,
  \bibinfo{title}{{Kepler Input Catalog: Photometric Calibration and Stellar
  Classification},} \aj, 142, 112, \dodoi{10.1088/0004-6256/142/4/112}

\bibitem[{M.~L. {Bryan} \& E.~J. {Lee}(2024){Bryan} \&
  {Lee}}]{2024ApJ...968L..25B}
{Bryan}, M.~L., \& {Lee}, E.~J. 2024, \bibinfo{title}{{Friends Not Foes: Strong
  Correlation between Inner Super-Earths and Outer Gas Giants},} \apjl, 968,
  L25, \dodoi{10.3847/2041-8213/ad5013}

\bibitem[{S. {Bryson}(2025){Bryson}}]{Bryson2025b}
{Bryson}, S. 2025, \bibinfo{title}{{How does
  {\ensuremath{\Gamma}}$_{{\ensuremath{\oplus}}}$ Relate to
  {\ensuremath{\eta}}$_{{\ensuremath{\oplus}}}$?},} Research Notes of the
  American Astronomical Society, 9, 255, \dodoi{10.3847/2515-5172/ae0d7f}

\bibitem[{S. {Bryson} {et~al.}(2020{\natexlab{a}}){Bryson}, {Coughlin},
  {Batalha}, {Berger}, {Huber}, {Burke}, {Dotson}, \& {Mullally}}]{Bryson2020}
{Bryson}, S., {Coughlin}, J., {Batalha}, N.~M., {et~al.} 2020{\natexlab{a}},
  \bibinfo{title}{{A Probabilistic Approach to Kepler Completeness and
  Reliability for Exoplanet Occurrence Rates},} \aj, 159, 279,
  \dodoi{10.3847/1538-3881/ab8a30}

\bibitem[{S. {Bryson} {et~al.}(2020{\natexlab{b}}){Bryson}, {Coughlin},
  {Kunimoto}, \& {Mullally}}]{Bryson2020b}
{Bryson}, S., {Coughlin}, J., {Kunimoto}, M., \& {Mullally}, S.~E.
  2020{\natexlab{b}}, \bibinfo{title}{Reliability Correction is Key for Robust
  Kepler Occurrence Rates,} \aj, 160, 200, \dodoi{10.3847/1538-3881/abb316}

\bibitem[{S. {Bryson} {et~al.}(2025){Bryson}, {Jenkins}, {Caldwell}, {Twicken},
  {Chakrabarty}, {He}, {Tenenbaum}, \& {Wohler}}]{Bryson2025}
{Bryson}, S., {Jenkins}, J., {Caldwell}, D., {et~al.} 2025, in American
  Astronomical Society Meeting Abstracts, Vol. 245, American Astronomical
  Society Meeting Abstracts, 207.06

\bibitem[{S. {Bryson} {et~al.}(2021){Bryson}, {Kunimoto}, {Kopparapu},
  {Coughlin}, {Borucki}, {Koch}, {Aguirre}, {Allen}, {Barentsen}, {Batalha},
  {Berger}, {Boss}, {Buchhave}, {Burke}, {Caldwell}, {Campbell}, {Catanzarite},
  {Chandrasekaran}, {Chaplin}, {Christiansen}, {Christensen-Dalsgaard},
  {Ciardi}, {Clarke}, {Cochran}, {Dotson}, {Doyle}, {Duarte}, {Dunham},
  {Dupree}, {Endl}, {Fanson}, {Ford}, {Fujieh}, {Gautier}, {Geary},
  {Gilliland}, {Girouard}, {Gould}, {Haas}, {Henze}, {Holman}, {Howard},
  {Howell}, {Huber}, {Hunter}, {Jenkins}, {Kjeldsen}, {Kolodziejczak},
  {Larson}, {Latham}, {Li}, {Mathur}, {Meibom}, {Middour}, {Morris}, {Morton},
  {Mullally}, {Mullally}, {Pletcher}, {Prsa}, {Quinn}, {Quintana}, {Ragozzine},
  {Ramirez}, {Sanderfer}, {Sasselov}, {Seader}, {Shabram}, {Shporer}, {Smith},
  {Steffen}, {Still}, {Torres}, {Troeltzsch}, {Twicken}, {Uddin}, {Van Cleve},
  {Voss}, {Weiss}, {Welsh}, {Wohler}, \& {Zamudio}}]{Bryson2021}
{Bryson}, S., {Kunimoto}, M., {Kopparapu}, R.~K., {et~al.} 2021,
  \bibinfo{title}{{The Occurrence of Rocky Habitable-zone Planets around
  Solar-like Stars from Kepler Data},} \aj, 161, 36,
  \dodoi{10.3847/1538-3881/abc418}

\bibitem[{S.~T. {Bryson} \& T.~D. {Morton}(2017){Bryson} \&
  {Morton}}]{Bryson2017}
{Bryson}, S.~T., \& {Morton}, T.~D. 2017, \bibinfo{title}{{Planet Reliability
  Metrics: Astrophysical Positional Probabilities for Data Release 25},} NExScI
  Exoplanet Archive, KSCI-19108-001

\bibitem[{S.~T. {Bryson} {et~al.}(2015){Bryson}, {Abdul-Masih}, {Batalha},
  {Burke}, {Caldwell}, {Colon}, {Coughlin}, {Esquerdo}, {Haas}, {Henze},
  {Huber}, {Latham}, {Morton}, {Romine}, {Rowe}, {Thompson}, \&
  {Wolfgang}}]{Bryson2015}
{Bryson}, S.~T., {Abdul-Masih}, M., {Batalha}, N., {et~al.} 2015,
  \bibinfo{title}{{The Kepler Certified False Positive Table},} NExScI
  Exoplanet Archive, KSCI-19093-002

\bibitem[{C.~J. {Burke} \& J. {Catanzarite}(2017){Burke} \&
  {Catanzarite}}]{BurkeJCat2017}
{Burke}, C.~J., \& {Catanzarite}, J. 2017, \bibinfo{title}{{Planet Detection
  Metrics: Per-Target Detection Contours for Data Release 25},} NExScI
  Exoplanet Archive, KSCI-19111-002

\bibitem[{C.~J. {Burke} {et~al.}(2014){Burke}, {Bryson}, {Mullally}, {Rowe},
  {Christiansen}, {Thompson}, {Coughlin}, {Haas}, {Batalha}, {Caldwell},
  {Jenkins}, {Still}, {Barclay}, {Borucki}, {Chaplin}, {Ciardi}, {Clarke},
  {Cochran}, {Demory}, {Esquerdo}, {Gautier}, {Gilliland}, {Girouard}, {Havel},
  {Henze}, {Howell}, {Huber}, {Latham}, {Li}, {Morehead}, {Morton}, {Pepper},
  {Quintana}, {Ragozzine}, {Seader}, {Shah}, {Shporer}, {Tenenbaum}, {Twicken},
  \& {Wolfgang}}]{Burke2014}
{Burke}, C.~J., {Bryson}, S.~T., {Mullally}, F., {et~al.} 2014,
  \bibinfo{title}{{Planetary Candidates Observed by Kepler IV: Planet Sample
  from Q1-Q8 (22 Months)},} \apjs, 210, 19, \dodoi{10.1088/0067-0049/210/2/19}

\bibitem[{C.~J. {Burke} {et~al.}(2015){Burke}, {Christiansen}, {Mullally},
  {Seader}, {Huber}, {Rowe}, {Coughlin}, {Thompson}, {Catanzarite}, {Clarke},
  {Morton}, {Caldwell}, {Bryson}, {Haas}, {Batalha}, {Jenkins}, {Tenenbaum},
  {Twicken}, {Li}, {Quintana}, {Barclay}, {Henze}, {Borucki}, {Howell}, \&
  {Still}}]{Burke2015}
{Burke}, C.~J., {Christiansen}, J.~L., {Mullally}, F., {et~al.} 2015,
  \bibinfo{title}{{Terrestrial Planet Occurrence Rates for the Kepler GK Dwarf
  Sample},} \apj, 809, 8, \dodoi{10.1088/0004-637X/809/1/8}

\bibitem[{J. {Catanzarite} \& M. {Shao}(2011){Catanzarite} \&
  {Shao}}]{CatanzariteShao2011}
{Catanzarite}, J., \& {Shao}, M. 2011, \bibinfo{title}{{The Occurrence Rate of
  Earth Analog Planets Orbiting Sun-like Stars},} \apj, 738, 151,
  \dodoi{10.1088/0004-637X/738/2/151}

\bibitem[{A. {Chakrabarty} \& G.~D. {Mulders}(2024){Chakrabarty} \&
  {Mulders}}]{Chakrabarty2024}
{Chakrabarty}, A., \& {Mulders}, G.~D. 2024, \bibinfo{title}{{Where Are the
  Water Worlds? Identifying Exo-water-worlds Using Models of Planet Formation
  and Atmospheric Evolution},} \apj, 966, 185, \dodoi{10.3847/1538-4357/ad3802}

\bibitem[{J.~L. {Christiansen}(2017){Christiansen}}]{Christiansen2017}
{Christiansen}, J.~L. 2017, \bibinfo{title}{{Planet Detection Metrics:
  Pixel-Level Transit Injection Tests of Pipeline Detection Efficiency for Data
  Release 25},} NExScI Exoplanet Archive, KSCI-19110-001

\bibitem[{J.~L. {Christiansen} {et~al.}(2012){Christiansen}, {Jenkins},
  {Caldwell}, {Burke}, {Tenenbaum}, {Seader}, {Thompson}, {Barclay}, {Clarke},
  {Li}, {Smith}, {Stumpe}, {Twicken}, \& {Van Cleve}}]{Christiansen2012}
{Christiansen}, J.~L., {Jenkins}, J.~M., {Caldwell}, D.~A., {et~al.} 2012,
  \bibinfo{title}{{The Derivation, Properties, and Value of
  Kepler{\textquoteright}s Combined Differential Photometric Precision},}
  \pasp, 124, 1279, \dodoi{10.1086/668847}

\bibitem[{J.~L. {Christiansen} {et~al.}(2013){Christiansen}, {Clarke}, {Burke},
  {Jenkins}, {Barclay}, {Ford}, {Haas}, {Sabale}, {Seader}, {Claiborne Smith},
  {Tenenbaum}, {Twicken}, {Kamal Uddin}, \& {Thompson}}]{Christiansen2013}
{Christiansen}, J.~L., {Clarke}, B.~D., {Burke}, C.~J., {et~al.} 2013,
  \bibinfo{title}{{Measuring Transit Signal Recovery in the Kepler Pipeline. I.
  Individual Events},} \apjs, 207, 35, \dodoi{10.1088/0067-0049/207/2/35}

\bibitem[{J.~L. {Christiansen} {et~al.}(2015){Christiansen}, {Clarke}, {Burke},
  {Seader}, {Jenkins}, {Twicken}, {Catanzarite}, {Smith}, {Batalha}, {Haas},
  {Thompson}, {Campbell}, {Sabale}, \& {Kamal Uddin}}]{Christiansen2015}
{Christiansen}, J.~L., {Clarke}, B.~D., {Burke}, C.~J., {et~al.} 2015,
  \bibinfo{title}{{Measuring Transit Signal Recovery in the Kepler Pipeline II:
  Detection Efficiency as Calculated in One Year of Data},} \apj, 810, 95,
  \dodoi{10.1088/0004-637X/810/2/95}

\bibitem[{J.~L. {Christiansen} {et~al.}(2016){Christiansen}, {Clarke}, {Burke},
  {Jenkins}, {Bryson}, {Coughlin}, {Mullally}, {Thompson}, {Twicken},
  {Batalha}, {Haas}, {Catanzarite}, {Campbell}, {Kamal Uddin}, {Zamudio},
  {Smith}, \& {Henze}}]{Christiansen2016}
{Christiansen}, J.~L., {Clarke}, B.~D., {Burke}, C.~J., {et~al.} 2016,
  \bibinfo{title}{{Measuring Transit Signal Recovery in the Kepler Pipeline.
  III. Completeness of the Q1-Q17 DR24 Planet Candidate Catalogue with
  Important Caveats for Occurrence Rate Calculations},} \apj, 828, 99,
  \dodoi{10.3847/0004-637X/828/2/99}

\bibitem[{J.~L. {Christiansen} {et~al.}(2020){Christiansen}, {Clarke}, {Burke},
  {Jenkins}, {Bryson}, {Coughlin}, {Mullally}, {Twicken}, {Batalha},
  {Catanzarite}, {Uddin}, {Zamudio}, {Smith}, {Henze}, \&
  {Campbell}}]{Christiansen2020}
{Christiansen}, J.~L., {Clarke}, B.~D., {Burke}, C.~J., {et~al.} 2020,
  \bibinfo{title}{{Measuring Transit Signal Recovery in the Kepler Pipeline.
  IV. Completeness of the DR25 Planet Candidate Catalog},} \aj, 160, 159,
  \dodoi{10.3847/1538-3881/abab0b}

\bibitem[{J.~L. {Christiansen} {et~al.}(2023){Christiansen}, {Zink},
  {Hardegree-Ullman}, {Fernandes}, {Hopkins}, {Rebull}, {Boley}, {Bergsten}, \&
  {Bhure}}]{Christiansen2023}
{Christiansen}, J.~L., {Zink}, J.~K., {Hardegree-Ullman}, K.~K., {et~al.} 2023,
  \bibinfo{title}{{Scaling K2. VII. Evidence For a High Occurrence Rate of Hot
  Sub-Neptunes at Intermediate Ages},} \aj, 166, 248,
  \dodoi{10.3847/1538-3881/acf9f9}

\bibitem[{J.~L. {Christiansen} {et~al.}(2025){Christiansen}, {McElroy},
  {Harbut}, {Ciardi}, {Crane}, {Good}, {Hardegree-Ullman}, {Kesseli}, {Lund},
  {Lynn}, {Muthiar}, {Nilsson}, {Oluyide}, {Papin}, {Rivera}, {Swain},
  {Susemiehl}, {Tam}, {van Eyken}, \& {Beichman}}]{Christiansen2025}
{Christiansen}, J.~L., {McElroy}, D.~L., {Harbut}, M., {et~al.} 2025,
  \bibinfo{title}{{The NASA Exoplanet Archive and Exoplanet Follow-up Observing
  Program: Data, Tools, and Usage},} arXiv e-prints, arXiv:2506.03299,
  \dodoi{10.48550/arXiv.2506.03299}

\bibitem[{J.~L. {Coughlin}(2017){Coughlin}}]{Coughlin2017}
{Coughlin}, J.~L. 2017, \bibinfo{title}{{Planet Detection Metrics: Robovetter
  Completeness and Effectiveness for Data Release 25},} NExScI Exoplanet
  Archive, KSCI-19114-002

\bibitem[{J.~L. Coughlin {et~al.}(2016)Coughlin, Mullally, Thompson, Rowe,
  Burke, Latham, Batalha, Ofir, Quarles, Henze, Wolfgang, Caldwell, Bryson,
  Shporer, Catanzarite, Akeson, Barclay, Borucki, Boyajian, Campbell,
  Christiansen, Girouard, Haas, Howell, Huber, Jenkins, Li, Patil-Sabale,
  Quintana, Ramirez, Seader, Smith, Tenenbaum, Twicken, \&
  Zamudio}]{Coughlin2016}
Coughlin, J.~L., Mullally, F., Thompson, S.~E., {et~al.} 2016,
  \bibinfo{title}{Planetary Candidates Observed by Kepler. VII. The First Fully
  Uniform Catalog Based on the Entire 48-month Data Set (Q1–Q17 DR24),}
  \apjs, 224, 12.
\newblock \url{http://stacks.iop.org/0067-0049/224/i=1/a=12}

\bibitem[{R.~I. {Dawson} {et~al.}(2016){Dawson}, {Lee}, \&
  {Chiang}}]{Dawson2016}
{Dawson}, R.~I., {Lee}, E.~J., \& {Chiang}, E. 2016,
  \bibinfo{title}{{Correlations between Compositions and Orbits Established by
  the Giant Impact Era of Planet Formation},} \apj, 822, 54,
  \dodoi{10.3847/0004-637X/822/1/54}

\bibitem[{S.~D. {Dulz} {et~al.}(2020){Dulz}, {Plavchan}, {Crepp}, {Stark},
  {Morgan}, {Kane}, {Newman}, {Matzko}, \& {Mulders}}]{Dulz2020}
{Dulz}, S.~D., {Plavchan}, P., {Crepp}, J.~R., {et~al.} 2020,
  \bibinfo{title}{{Joint Radial Velocity and Direct Imaging Planet Yield
  Calculations. I. Self-consistent Planet Populations},} \apj, 893, 122,
  \dodoi{10.3847/1538-4357/ab7b73}

\bibitem[{D.~C. {Fabrycky} {et~al.}(2014){Fabrycky}, {Lissauer}, {Ragozzine},
  {Rowe}, {Steffen}, {Agol}, {Barclay}, {Batalha}, {Borucki}, {Ciardi}, {Ford},
  {Gautier}, {Geary}, {Holman}, {Jenkins}, {Li}, {Morehead}, {Morris},
  {Shporer}, {Smith}, {Still}, \& {Van Cleve}}]{Fabrycky2014}
{Fabrycky}, D.~C., {Lissauer}, J.~J., {Ragozzine}, D., {et~al.} 2014,
  \bibinfo{title}{{Architecture of Kepler's Multi-transiting Systems. II. New
  Investigations with Twice as Many Candidates},} \apj, 790, 146,
  \dodoi{10.1088/0004-637X/790/2/146}

\bibitem[{J. {Fang} \& J.-L. {Margot}(2012){Fang} \& {Margot}}]{FangMargot2012}
{Fang}, J., \& {Margot}, J.-L. 2012, \bibinfo{title}{{Architecture of Planetary
  Systems Based on Kepler Data: Number of Planets and Coplanarity},} \apj, 761,
  92, \dodoi{10.1088/0004-637X/761/2/92}

\bibitem[{L. {Feinberg} {et~al.}(2024){Feinberg}, {Ziemer}, {Ansdell},
  {Crooke}, {Dressing}, {Mennesson}, {O'Meara}, {Pepper}, \&
  {Roberge}}]{Feinberg2024}
{Feinberg}, L., {Ziemer}, J., {Ansdell}, M., {et~al.} 2024, in Society of
  Photo-Optical Instrumentation Engineers (SPIE) Conference Series, Vol. 13092,
  Space Telescopes and Instrumentation 2024: Optical, Infrared, and Millimeter
  Wave, ed. L.~E. {Coyle}, S.~{Matsuura}, \& M.~D. {Perrin}, 130921N,
  \dodoi{10.1117/12.3018328}

\bibitem[{R.~B. {Fernandes} {et~al.}(2019){Fernandes}, {Mulders}, {Pascucci},
  {Mordasini}, \& {Emsenhuber}}]{Fernandes2019}
{Fernandes}, R.~B., {Mulders}, G.~D., {Pascucci}, I., {Mordasini}, C., \&
  {Emsenhuber}, A. 2019, \bibinfo{title}{{Hints for a Turnover at the Snow Line
  in the Giant Planet Occurrence Rate},} \apj, 874, 81,
  \dodoi{10.3847/1538-4357/ab0300}

\bibitem[{R.~B. {Fernandes} {et~al.}(2025{\natexlab{a}}){Fernandes}, {Johnson},
  {Bergsten}, {Bhure}, {Boley}, {Boss}, {Bryson}, {Dietrich}, {Duck},
  {Giacalone}, {Gupta}, {He}, {Kunimoto}, {Ment}, {Sagear}, {Silverstein},
  {Sullivan}, {Vrijmoet}, {Wagner}, {Wilson}, {Brefka}, {Belikov},
  {Chakrabarty}, {Christiansen}, {Ciardi}, {Dattilo}, {DeRocco}, {Fitzmaurice},
  {Ford}, {Hotnisky}, {Jones}, {Kar}, {Kopparapu}, {Lowson}, {Mamajek},
  {Mennesson}, {Meyer}, {Millholland}, {Mulders}, {Mullally}, {Murlidhar},
  {Pascucci}, {Ragozzine}, {Robertson}, {Stapelfeldt}, \&
  {Wright}}]{Fernandes2025b}
{Fernandes}, R.~B., {Johnson}, S., {Bergsten}, G.~J., {et~al.}
  2025{\natexlab{a}}, \bibinfo{title}{Are We There Yet? Challenges in
  Quantifying the Frequency of Earth Analogs in the Habitable Zone,} submitted
  to PASP, tbd, tbd, \dodoi{tbd}

\bibitem[{R.~B. {Fernandes} {et~al.}(2025{\natexlab{b}}){Fernandes},
  {Bergsten}, {Mulders}, {Pascucci}, {Hardegree-Ullman}, {Giacalone},
  {Christiansen}, {Rogers}, {Gupta}, {Dawson}, {Koskinen}, {Boley}, {Curtis},
  {Cunha}, {Mamajek}, {Sagynbayeva}, {Bhure}, {Ciardi}, {Karpoor}, {Pearson},
  {Zink}, \& {Feiden}}]{Fernandes2025}
{Fernandes}, R.~B., {Bergsten}, G.~J., {Mulders}, G.~D., {et~al.}
  2025{\natexlab{b}}, \bibinfo{title}{{Signatures of Atmospheric Mass Loss and
  Planet Migration in the Time Evolution of Short-period Transiting
  Exoplanets},} \aj, 169, 208, \dodoi{10.3847/1538-3881/adb97e}

\bibitem[{D. {Foreman-Mackey} {et~al.}(2014){Foreman-Mackey}, {Hogg}, \&
  {Morton}}]{ForemanMackey2014}
{Foreman-Mackey}, D., {Hogg}, D.~W., \& {Morton}, T.~D. 2014,
  \bibinfo{title}{{Exoplanet Population Inference and the Abundance of Earth
  Analogs from Noisy, Incomplete Catalogs},} \apj, 795, 64,
  \dodoi{10.1088/0004-637X/795/1/64}

\bibitem[{B.~J. {Fulton} {et~al.}(2017){Fulton}, {Petigura}, {Howard},
  {Isaacson}, {Marcy}, {Cargile}, {Hebb}, {Weiss}, {Johnson}, {Morton},
  {Sinukoff}, {Crossfield}, \& {Hirsch}}]{Fulton2017}
{Fulton}, B.~J., {Petigura}, E.~A., {Howard}, A.~W., {et~al.} 2017,
  \bibinfo{title}{{The California-Kepler Survey. III. A Gap in the Radius
  Distribution of Small Planets},} \aj, 154, 109,
  \dodoi{10.3847/1538-3881/aa80eb}

\bibitem[{E. {Furlan} \& S.~B. {Howell}(2020){Furlan} \& {Howell}}]{Furlan2020}
{Furlan}, E., \& {Howell}, S.~B. 2020, \bibinfo{title}{{Unresolved Binary
  Exoplanet Host Stars Fit as Single Stars: Effects on the Stellar
  Parameters},} \apj, 898, 47, \dodoi{10.3847/1538-4357/ab9c9c}

\bibitem[{J. {Ge} {et~al.}(2022){Ge}, {Zhang}, {Zang}, {Deng}, {Mao}, {Xie},
  {Liu}, {Zhou}, {Willis}, {Huang}, {Howell}, {Feng}, {Zhu}, {Yao}, {Liu},
  {Aizawa}, {Zhu}, {Li}, {Ma}, {Ye}, {Yu}, {Xiang}, {Yu}, {Liu}, {Yang},
  {Wang}, {Shi}, {Fang}, {Zong}, {Liu}, {Zhang}, {Zhang}, {El-Badry}, {Shen},
  {Tam}, {Hu}, {Yang}, {Zou}, {Wu}, {Lei}, {Wei}, {Wu}, {Sun}, {Wang}, {Zhang},
  {Xu}, {Yang}, {Li}, {Xiang}, {Wang}, {Wang}, {Zhang}, {Jia}, {Yuan}, {Zhang},
  {Xuesong Wang}, {Gan}, {Wang}, {Zhao}, {Liu}, {Wei}, {Kang}, {Yang}, {Qi},
  {Liu}, {Zhang}, {Zhu}, {Zhou}, {Zhang}, {Yu}, {Zhang}, {Li}, {Tang}, {Wang},
  {Wang}, {Li}, {Cheng}, {Shen}, {Li}, {Pan}, {Yang}, {Gao}, {Song}, {Wang},
  {Zhang}, {Chen}, {Wang}, {Zhang}, {Wang}, {Zeng}, {Zheng}, {Zhu}, {Guo},
  {Zhang}, {Li}, {Wen}, {Feng}, {Chen}, {Chen}, {Han}, {Yang}, {Wang}, {Duan},
  {Huang}, {Liang}, {Bi}, {Gai}, {Ge}, {Guo}, {Huang}, {Li}, {Li}, {Li},
  {Yuxi}, {Lu}, {Rix}, {Shi}, {Song}, {Tang}, {Ting}, {Wu}, {Wu}, {Yang},
  {Yin}, {Gould}, {Lee}, {Dong}, {Yee}, {Shvartzvald}, {Yang}, {Kuang},
  {Zhang}, {Liao}, {Qi}, {Yang}, {Zhang}, {Jiang}, {Ou}, {Li}, {Beck},
  {Bedding}, {Campante}, {Chaplin}, {Christensen-Dalsgaard}, {Garc{\'\i}a},
  {Gaulme}, {Gizon}, {Hekker}, {Huber}, {Khanna}, {Li}, {Mathur}, {Miglio},
  {Mosser}, {Ong}, {Santos}, {Stello}, {Bowman}, {Lares-Martiz}, {Murphy},
  {Niu}, {Ma}, {Moln{\'a}r}, {Fu}, {De Cat}, {Su}, \& {consortium}}]{Ge2022}
{Ge}, J., {Zhang}, H., {Zang}, W., {et~al.} 2022, \bibinfo{title}{{ET White
  Paper: To Find the First Earth 2.0},} arXiv e-prints, arXiv:2206.06693,
  \dodoi{10.48550/arXiv.2206.06693}

\bibitem[{R.~L. {Gilliland} {et~al.}(2015){Gilliland}, {Chaplin}, {Jenkins},
  {Ramsey}, \& {Smith}}]{Gilliland2015}
{Gilliland}, R.~L., {Chaplin}, W.~J., {Jenkins}, J.~M., {Ramsey}, L.~W., \&
  {Smith}, J.~C. 2015, \bibinfo{title}{{Kepler Mission Stellar and Instrument
  Noise Properties Revisited},} \aj, 150, 133,
  \dodoi{10.1088/0004-6256/150/4/133}

\bibitem[{R.~L. {Gilliland} {et~al.}(2011){Gilliland}, {Chaplin}, {Dunham},
  {Argabright}, {Borucki}, {Basri}, {Bryson}, {Buzasi}, {Caldwell}, {Elsworth},
  {Jenkins}, {Koch}, {Kolodziejczak}, {Miglio}, {van Cleve}, {Walkowicz}, \&
  {Welsh}}]{gilliland2011}
{Gilliland}, R.~L., {Chaplin}, W.~J., {Dunham}, E.~W., {et~al.} 2011,
  \bibinfo{title}{{Kepler Mission Stellar and Instrument Noise Properties},}
  \apjs, 197, 6, \dodoi{10.1088/0067-0049/197/1/6}

\bibitem[{S. {Ginzburg} {et~al.}(2018){Ginzburg}, {Schlichting}, \&
  {Sari}}]{Ginzburg2018}
{Ginzburg}, S., {Schlichting}, H.~E., \& {Sari}, R. 2018,
  \bibinfo{title}{{Core-powered mass-loss and the radius distribution of small
  exoplanets},} \mnras, 476, 759, \dodoi{10.1093/mnras/sty290}

\bibitem[{M. {Godolt} {et~al.}(2015){Godolt}, {Grenfell}, {Hamann-Reinus},
  {Kitzmann}, {Kunze}, {Langematz}, {von Paris}, {Patzer}, {Rauer}, \&
  {Stracke}}]{Godolt2015}
{Godolt}, M., {Grenfell}, J.~L., {Hamann-Reinus}, A., {et~al.} 2015,
  \bibinfo{title}{{3D climate modeling of Earth-like extrasolar planets
  orbiting different types of host stars},} P\&SS, 111, 62,
  \dodoi{10.1016/j.pss.2015.03.010}

\bibitem[{A. {Gupta} \& H.~E. {Schlichting}(2019){Gupta} \&
  {Schlichting}}]{Gupta2019}
{Gupta}, A., \& {Schlichting}, H.~E. 2019, \bibinfo{title}{{Sculpting the
  valley in the radius distribution of small exoplanets as a by-product of
  planet formation: the core-powered mass-loss mechanism},} \mnras, 487, 24,
  \dodoi{10.1093/mnras/stz1230}

\bibitem[{A. {Gupta} \& H.~E. {Schlichting}(2020){Gupta} \&
  {Schlichting}}]{Gupta2020}
{Gupta}, A., \& {Schlichting}, H.~E. 2020, \bibinfo{title}{{Signatures of the
  core-powered mass-loss mechanism in the exoplanet population: dependence on
  stellar properties and observational predictions},} \mnras, 493, 792,
  \dodoi{10.1093/mnras/staa315}

\bibitem[{B.~M.~S. {Hansen} \& N. {Murray}(2013){Hansen} \&
  {Murray}}]{Hansen2013}
{Hansen}, B. M.~S., \& {Murray}, N. 2013, \bibinfo{title}{{Testing in Situ
  Assembly with the Kepler Planet Candidate Sample},} \apj, 775, 53,
  \dodoi{10.1088/0004-637X/775/1/53}

\bibitem[{C.~R. Harris {et~al.}(2020)Harris, Millman, van~der Walt, Gommers,
  Virtanen, Cournapeau, Wieser, Taylor, Berg, Smith, Kern, Picus, Hoyer, van
  Kerkwijk, Brett, Haldane, del R{\'{i}}o, Wiebe, Peterson,
  G{\'{e}}rard-Marchant, Sheppard, Reddy, Weckesser, Abbasi, Gohlke, \&
  Oliphant}]{Harris2020}
Harris, C.~R., Millman, K.~J., van~der Walt, S.~J., {et~al.} 2020,
  \bibinfo{title}{Array programming with {NumPy},} Nature, 585, 357,
  \dodoi{10.1038/s41586-020-2649-2}

\bibitem[{M.~Y. {He} {et~al.}(2019){He}, {Ford}, \& {Ragozzine}}]{He2019}
{He}, M.~Y., {Ford}, E.~B., \& {Ragozzine}, D. 2019,
  \bibinfo{title}{{Architectures of exoplanetary systems - I. A clustered
  forward model for exoplanetary systems around Kepler's FGK stars},} \mnras,
  490, 4575, \dodoi{10.1093/mnras/stz2869}

\bibitem[{M.~Y. {He} {et~al.}(2020){He}, {Ford}, {Ragozzine}, \&
  {Carrera}}]{He2020}
{He}, M.~Y., {Ford}, E.~B., {Ragozzine}, D., \& {Carrera}, D. 2020,
  \bibinfo{title}{{Architectures of Exoplanetary Systems. III. Eccentricity and
  Mutual Inclination Distributions of AMD-stable Planetary Systems},} \aj, 160,
  276, \dodoi{10.3847/1538-3881/abba18}

\bibitem[{C. {Hedges} {et~al.}(2021){Hedges}, {Luger}, {Martinez-Palomera},
  {Dotson}, \& {Barentsen}}]{Hedges2021}
{Hedges}, C., {Luger}, R., {Martinez-Palomera}, J., {Dotson}, J., \&
  {Barentsen}, G. 2021, \bibinfo{title}{{Linearized Field Deblending:
  Point-spread Function Photometry for Impatient Astronomers},} \aj, 162, 107,
  \dodoi{10.3847/1538-3881/ac0825}

\bibitem[{R. {Heller} {et~al.}(2022){Heller}, {Harre}, \&
  {Samadi}}]{Heller2022}
{Heller}, R., {Harre}, J.-V., \& {Samadi}, R. 2022, \bibinfo{title}{{Transit
  least-squares survey. IV. Earth-like transiting planets expected from the
  PLATO mission},} \aap, 665, A11, \dodoi{10.1051/0004-6361/202141640}

\bibitem[{A.~W. {Howard} {et~al.}(2012){Howard}, {Marcy}, {Bryson}, {Jenkins},
  {Rowe}, {Batalha}, {Borucki}, {Koch}, {Dunham}, {Gautier}, {Van Cleve},
  {Cochran}, {Latham}, {Lissauer}, {Torres}, {Brown}, {Gilliland}, {Buchhave},
  {Caldwell}, {Christensen-Dalsgaard}, {Ciardi}, {Fressin}, {Haas}, {Howell},
  {Kjeldsen}, {Seager}, {Rogers}, {Sasselov}, {Steffen}, {Basri},
  {Charbonneau}, {Christiansen}, {Clarke}, {Dupree}, {Fabrycky}, {Fischer},
  {Ford}, {Fortney}, {Tarter}, {Girouard}, {Holman}, {Johnson}, {Klaus},
  {Machalek}, {Moorhead}, {Morehead}, {Ragozzine}, {Tenenbaum}, {Twicken},
  {Quinn}, {Isaacson}, {Shporer}, {Lucas}, {Walkowicz}, {Welsh}, {Boss},
  {Devore}, {Gould}, {Smith}, {Morris}, {Prsa}, {Morton}, {Still}, {Thompson},
  {Mullally}, {Endl}, \& {MacQueen}}]{Howard2012}
{Howard}, A.~W., {Marcy}, G.~W., {Bryson}, S.~T., {et~al.} 2012,
  \bibinfo{title}{{Planet Occurrence within 0.25 AU of Solar-type Stars from
  Kepler},} \apjs, 201, 15, \dodoi{10.1088/0067-0049/201/2/15}

\bibitem[{S.~B. {Howell} {et~al.}(2014){Howell}, {Sobeck}, {Haas}, {Still},
  {Barclay}, {Mullally}, {Troeltzsch}, {Aigrain}, {Bryson}, {Caldwell},
  {Chaplin}, {Cochran}, {Huber}, {Marcy}, {Miglio}, {Najita}, {Smith},
  {Twicken}, \& {Fortney}}]{howell2014}
{Howell}, S.~B., {Sobeck}, C., {Haas}, M., {et~al.} 2014, \bibinfo{title}{{The
  K2 Mission: Characterization and Early Results},} \pasp, 126, 398,
  \dodoi{10.1086/676406}

\bibitem[{D.~C. {Hsu} {et~al.}(2019){Hsu}, {Ford}, {Ragozzine}, \&
  {Ashby}}]{Hsu2019}
{Hsu}, D.~C., {Ford}, E.~B., {Ragozzine}, D., \& {Ashby}, K. 2019,
  \bibinfo{title}{{Occurrence Rates of Planets Orbiting FGK Stars: Combining
  Kepler DR25, Gaia DR2, and Bayesian Inference},} \aj, 158, 109,
  \dodoi{10.3847/1538-3881/ab31ab}

\bibitem[{D.~C. {Hsu} {et~al.}(2018){Hsu}, {Ford}, {Ragozzine}, \&
  {Morehead}}]{Hsu2018}
{Hsu}, D.~C., {Ford}, E.~B., {Ragozzine}, D., \& {Morehead}, R.~C. 2018,
  \bibinfo{title}{{Improving the Accuracy of Planet Occurrence Rates from
  Kepler Using Approximate Bayesian Computation},} \aj, 155, 205,
  \dodoi{10.3847/1538-3881/aab9a8}

\bibitem[{D. {Huber} {et~al.}(2014){Huber}, {Silva Aguirre}, {Matthews},
  {Pinsonneault}, {Gaidos}, {Garc{\'\i}a}, {Hekker}, {Mathur}, {Mosser},
  {Torres}, {Bastien}, {Basu}, {Bedding}, {Chaplin}, {Demory}, {Fleming},
  {Guo}, {Mann}, {Rowe}, {Serenelli}, {Smith}, \& {Stello}}]{Huber2014}
{Huber}, D., {Silva Aguirre}, V., {Matthews}, J.~M., {et~al.} 2014,
  \bibinfo{title}{{Revised Stellar Properties of Kepler Targets for the Quarter
  1-16 Transit Detection Run},} \apjs, 211, 2,
  \dodoi{10.1088/0067-0049/211/1/2}

\bibitem[{J.~D. Hunter(2007)Hunter}]{Hunter2007}
Hunter, J.~D. 2007, \bibinfo{title}{Matplotlib: A 2D graphics environment,}
  Computing in Science \& Engineering, 9, 90, \dodoi{10.1109/MCSE.2007.55}

\bibitem[{J.~M. {Jenkins}(2002){Jenkins}}]{Jenkins2002}
{Jenkins}, J.~M. 2002, \bibinfo{title}{{The Impact of Solar-like Variability on
  the Detectability of Transiting Terrestrial Planets},} \apj, 575, 493,
  \dodoi{10.1086/341136}

\bibitem[{J.~M. {Jenkins}(2012){Jenkins}}]{Jenkins2012}
{Jenkins}, J.~M. 2012, in American Astronomical Society Meeting Abstracts, Vol.
  220, American Astronomical Society Meeting Abstracts \#220, 318.05

\bibitem[{J.~M. {Jenkins} \&  {et al.}(2020){Jenkins} \& {et
  al.}}]{Jenkins2020KDPH}
{Jenkins}, J.~M., \& {et al.} 2020, \bibinfo{title}{{Kepler Science Document
  KSCI-19081-003},}, Kepler Science Document KSCI-19081-003, Edited by Jon M.
  Jenkins.

\bibitem[{J.~M. {Jenkins} {et~al.}(2010){Jenkins}, {Caldwell},
  {Chandrasekaran}, {Twicken}, {Bryson}, {Quintana}, {Clarke}, {Li}, {Allen},
  {Tenenbaum}, {Wu}, {Klaus}, {Van Cleve}, {Dotson}, {Haas}, {Gilliland},
  {Koch}, \& {Borucki}}]{KSOCoverview2010}
{Jenkins}, J.~M., {Caldwell}, D.~A., {Chandrasekaran}, H., {et~al.} 2010,
  \bibinfo{title}{{Initial Characteristics of Kepler Long Cadence Data for
  Detecting Transiting Planets},} \apjl, 713, L120,
  \dodoi{10.1088/2041-8205/713/2/L120}

\bibitem[{J.~M. {Jenkins} {et~al.}(2015){Jenkins}, {Twicken}, {Batalha},
  {Caldwell}, {Cochran}, {Endl}, {Latham}, {Esquerdo}, {Seader}, {Bieryla},
  {Petigura}, {Ciardi}, {Marcy}, {Isaacson}, {Huber}, {Rowe}, {Torres},
  {Bryson}, {Buchhave}, {Ramirez}, {Wolfgang}, {Li}, {Campbell}, {Tenenbaum},
  {Sanderfer}, {Henze}, {Catanzarite}, {Gilliland}, \& {Borucki}}]{Jenkins2015}
{Jenkins}, J.~M., {Twicken}, J.~D., {Batalha}, N.~M., {et~al.} 2015,
  \bibinfo{title}{{Discovery and Validation of Kepler-452b: A 1.6 R$_{e}$ Super
  Earth Exoplanet in the Habitable Zone of a G2 Star},} \aj, 150, 56,
  \dodoi{10.1088/0004-6256/150/2/56}

\bibitem[{J.~A. {Johnson} {et~al.}(2017){Johnson}, {Petigura}, {Fulton},
  {Marcy}, {Howard}, {Isaacson}, {Hebb}, {Cargile}, {Morton}, {Weiss}, {Winn},
  {Rogers}, {Sinukoff}, \& {Hirsch}}]{Johnson2017}
{Johnson}, J.~A., {Petigura}, E.~A., {Fulton}, B.~J., {et~al.} 2017,
  \bibinfo{title}{{The California-Kepler Survey. II. Precise Physical
  Properties of 2025 Kepler Planets and Their Host Stars},} \aj, 154, 108,
  \dodoi{10.3847/1538-3881/aa80e7}

\bibitem[{D.~G. {Koch} {et~al.}(2010){Koch}, {Borucki}, {Basri}, {Batalha},
  {Brown}, {Caldwell}, {Christensen-Dalsgaard}, {Cochran}, {DeVore}, \&
  {Dunham}}]{Koch2010}
{Koch}, D.~G., {Borucki}, W.~J., {Basri}, G., {et~al.} 2010,
  \bibinfo{title}{{Kepler Mission Design, Realized Photometric Performance, and
  Early Science},} \apj, 713, L79, \dodoi{10.1088/2041-8205/713/2/L79}

\bibitem[{R.~K. {Kopparapu} {et~al.}(2014){Kopparapu}, {Ramirez},
  {SchottelKotte}, {Kasting}, {Domagal-Goldman}, \& {Eymet}}]{Kopparapu2014}
{Kopparapu}, R.~K., {Ramirez}, R.~M., {SchottelKotte}, J., {et~al.} 2014,
  \bibinfo{title}{{Habitable Zones around Main-sequence Stars: Dependence on
  Planetary Mass},} \apjl, 787, L29, \dodoi{10.1088/2041-8205/787/2/L29}

\bibitem[{R.~k. {Kopparapu} {et~al.}(2017){Kopparapu}, {Wolf}, {Arney},
  {Batalha}, {Haqq-Misra}, {Grimm}, \& {Heng}}]{Kopparapu2017}
{Kopparapu}, R.~k., {Wolf}, E.~T., {Arney}, G., {et~al.} 2017,
  \bibinfo{title}{{Habitable Moist Atmospheres on Terrestrial Planets near the
  Inner Edge of the Habitable Zone around M Dwarfs},} \apj, 845, 5,
  \dodoi{10.3847/1538-4357/aa7cf9}

\bibitem[{R.~K. Kopparapu {et~al.}(2016)Kopparapu, Wolf, Haqq-Misra, Yang,
  Kasting, Meadows, Terrien, \& Mahadevan}]{Kopparapu2016}
Kopparapu, R.~K., Wolf, E.~T., Haqq-Misra, J., {et~al.} 2016,
  \bibinfo{title}{The inner edge of the habitable zone for synchronously
  rotating planets around low-mass stars using general circulation models,} The
  \apj, 819, 84

\bibitem[{R.~K. {Kopparapu} {et~al.}(2013){Kopparapu}, {Ramirez}, {Kasting},
  {Eymet}, {Robinson}, {Mahadevan}, {Terrien}, {Domagal-Goldman}, {Meadows}, \&
  {Deshpande}}]{Kopparapu2013}
{Kopparapu}, R.~K., {Ramirez}, R., {Kasting}, J.~F., {et~al.} 2013,
  \bibinfo{title}{{Habitable Zones around Main-sequence Stars: New Estimates},}
  \apj, 765, 131, \dodoi{10.1088/0004-637X/765/2/131}

\bibitem[{M. {Kunimoto} \& J.~M. {Matthews}(2020){Kunimoto} \&
  {Matthews}}]{Kunimoto2020a}
{Kunimoto}, M., \& {Matthews}, J.~M. 2020, \bibinfo{title}{{Searching the
  Entirety of Kepler Data. II. Occurrence Rate Estimates for FGK Stars},} \aj,
  159, 248, \dodoi{10.3847/1538-3881/ab88b0}

\bibitem[{E.~J. {Lee} \& E. {Chiang}(2016){Lee} \& {Chiang}}]{Lee2016}
{Lee}, E.~J., \& {Chiang}, E. 2016, \bibinfo{title}{{Breeding Super-Earths and
  Birthing Super-puffs in Transitional Disks},} \apj, 817, 90,
  \dodoi{10.3847/0004-637X/817/2/90}

\bibitem[{E.~J. {Lee} \& N.~J. {Connors}(2021){Lee} \& {Connors}}]{Lee2021}
{Lee}, E.~J., \& {Connors}, N.~J. 2021, \bibinfo{title}{{Primordial Radius Gap
  and Potentially Broad Core Mass Distributions of Super-Earths and
  Sub-Neptunes},} \apj, 908, 32, \dodoi{10.3847/1538-4357/abd6c7}

\bibitem[{E.~J. {Lee} {et~al.}(2022){Lee}, {Karalis}, \& {Thorngren}}]{Lee2022}
{Lee}, E.~J., {Karalis}, A., \& {Thorngren}, D.~P. 2022,
  \bibinfo{title}{{Creating the Radius Gap without Mass Loss},} \apj, 941, 186,
  \dodoi{10.3847/1538-4357/ac9c66}

\bibitem[{ {Lightkurve Collaboration} {et~al.}(2018){Lightkurve Collaboration},
  {Cardoso}, {Hedges}, {Gully-Santiago}, {Saunders}, {Cody}, {Barclay}, {Hall},
  {Sagear}, {Turtelboom}, {Zhang}, {Tzanidakis}, {Mighell}, {Coughlin}, {Bell},
  {Berta-Thompson}, {Williams}, {Dotson}, \& {Barentsen}}]{lightkurve2018}
{Lightkurve Collaboration}, {Cardoso}, J.~V.~d.~M., {Hedges}, C., {et~al.}
  2018, \bibinfo{title}{{Lightkurve: Kepler and TESS time series analysis in
  Python},}, Astrophysics Source Code Library \doeprint{1812.013}

\bibitem[{L. Lindegren(2018)Lindegren}]{gaiaRuwe2018}
Lindegren, L. 2018.
\newblock \url{http://www.rssd.esa.int/doc_fetch.php?id=3757412}

\bibitem[{J.~J. {Lissauer} {et~al.}(2024){Lissauer}, {Rowe}, {Jontof-Hutter},
  {Fabrycky}, {Ford}, {Ragozzine}, {Steffen}, \& {Nizam}}]{Lissauer2024}
{Lissauer}, J.~J., {Rowe}, J.~F., {Jontof-Hutter}, D., {et~al.} 2024,
  \bibinfo{title}{{Updated Catalog of Kepler Planet Candidates: Focus on
  Accuracy and Orbital Periods},} PSJ, 5, 152, \dodoi{10.3847/PSJ/ad0e6e}

\bibitem[{J.~J. {Lissauer} {et~al.}(2011){Lissauer}, {Ragozzine}, {Fabrycky},
  {Steffen}, {Ford}, {Jenkins}, {Shporer}, {Holman}, {Rowe}, {Quintana},
  {Batalha}, {Borucki}, {Bryson}, {Caldwell}, {Carter}, {Ciardi}, {Dunham},
  {Fortney}, {Gautier}, {Howell}, {Koch}, {Latham}, {Marcy}, {Morehead}, \&
  {Sasselov}}]{Lissauer2011}
{Lissauer}, J.~J., {Ragozzine}, D., {Fabrycky}, D.~C., {et~al.} 2011,
  \bibinfo{title}{{Architecture and Dynamics of Kepler's Candidate Multiple
  Transiting Planet Systems},} \apjs, 197, 8, \dodoi{10.1088/0067-0049/197/1/8}

\bibitem[{E.~D. {Lopez} \& K. {Rice}(2018){Lopez} \& {Rice}}]{Lopez2018}
{Lopez}, E.~D., \& {Rice}, K. 2018, \bibinfo{title}{{How formation time-scales
  affect the period dependence of the transition between rocky super-Earths and
  gaseous sub-Neptunesand implications for
  {\ensuremath{\eta}}$_{{\ensuremath{\oplus}}}$},} \mnras, 479, 5303,
  \dodoi{10.1093/mnras/sty1707}

\bibitem[{J. {Mart{\'\i}nez-Palomera} {et~al.}(2023){Mart{\'\i}nez-Palomera},
  {Hedges}, \& {Dotson}}]{Martinez2023}
{Mart{\'\i}nez-Palomera}, J., {Hedges}, C., \& {Dotson}, J. 2023,
  \bibinfo{title}{{Kepler Bonus: Light Curves of Kepler Background Sources},}
  \aj, 166, 265, \dodoi{10.3847/1538-3881/ad0727}

\bibitem[{M.~R.~B. {Matesic} {et~al.}(2024){Matesic}, {Rowe}, {Livingston},
  {Dholakia}, {Jontof-Hutter}, \& {Lissauer}}]{Matesic2024}
{Matesic}, M. R.~B., {Rowe}, J.~F., {Livingston}, J.~H., {et~al.} 2024,
  \bibinfo{title}{{Gaussian Processes and Nested Sampling Applied to Kepler's
  Small Long-period Exoplanet Candidates},} \aj, 167, 68,
  \dodoi{10.3847/1538-3881/ad0fe9}

\bibitem[{S. {Mathur} {et~al.}(2017){Mathur}, {Huber}, {Batalha}, {Ciardi},
  {Bastien}, {Bieryla}, {Buchhave}, {Cochran}, {Endl}, {Esquerdo}, {Furlan},
  {Howard}, {Howell}, {Isaacson}, {Latham}, {MacQueen}, \&
  {Silva}}]{Mathur2017ApJS}
{Mathur}, S., {Huber}, D., {Batalha}, N.~M., {et~al.} 2017,
  \bibinfo{title}{{Revised Stellar Properties of Kepler Targets for the Q1-17
  (DR25) Transit Detection Run},} \apjs, 229, 30,
  \dodoi{10.3847/1538-4365/229/2/30}

\bibitem[{S.~C. {Millholland} {et~al.}(2022){Millholland}, {He}, \&
  {Zink}}]{Millholland2022}
{Millholland}, S.~C., {He}, M.~Y., \& {Zink}, J.~K. 2022,
  \bibinfo{title}{{Edge-of-the-Multis: Evidence for a Transition in the Outer
  Architectures of Compact Multiplanet Systems},} \aj, 164, 72,
  \dodoi{10.3847/1538-3881/ac7c67}

\bibitem[{W. {Misener} \& H.~E. {Schlichting}(2021){Misener} \&
  {Schlichting}}]{Misener2021}
{Misener}, W., \& {Schlichting}, H.~E. 2021, \bibinfo{title}{{To cool is to
  keep: residual H/He atmospheres of super-Earths and sub-Neptunes},} \mnras,
  503, 5658, \dodoi{10.1093/mnras/stab895}

\bibitem[{T.~D. {Morton}(2012){Morton}}]{Morton2012}
{Morton}, T.~D. 2012, \bibinfo{title}{{An Efficient Automated Validation
  Procedure for Exoplanet Transit Candidates},} \apj, 761, 6,
  \dodoi{10.1088/0004-637X/761/1/6}

\bibitem[{T.~D. {Morton} {et~al.}(2016){Morton}, {Bryson}, {Coughlin}, {Rowe},
  {Ravichandran}, {Petigura}, {Haas}, \& {Batalha}}]{Morton2016}
{Morton}, T.~D., {Bryson}, S.~T., {Coughlin}, J.~L., {et~al.} 2016,
  \bibinfo{title}{{False Positive Probabilities for all Kepler Objects of
  Interest: 1284 Newly Validated Planets and 428 Likely False Positives},}
  \apj, 822, 86, \dodoi{10.3847/0004-637X/822/2/86}

\bibitem[{T.~D. {Morton} \& J.~A. {Johnson}(2011){Morton} \&
  {Johnson}}]{Morton2011}
{Morton}, T.~D., \& {Johnson}, J.~A. 2011, \bibinfo{title}{{On the Low False
  Positive Probabilities of Kepler Planet Candidates},} \apj, 738, 170,
  \dodoi{10.1088/0004-637X/738/2/170}

\bibitem[{G.~D. {Mulders} {et~al.}(2020){Mulders}, {O'Brien}, {Ciesla}, {Apai},
  \& {Pascucci}}]{Mulders2020}
{Mulders}, G.~D., {O'Brien}, D.~P., {Ciesla}, F.~J., {Apai}, D., \& {Pascucci},
  I. 2020, \bibinfo{title}{{Earths in Other Solar Systems' N-body Simulations:
  The Role of Orbital Damping in Reproducing the Kepler Planetary Systems},}
  \apj, 897, 72, \dodoi{10.3847/1538-4357/ab9806}

\bibitem[{G.~D. {Mulders} {et~al.}(2018){Mulders}, {Pascucci}, {Apai}, \&
  {Ciesla}}]{Mulders2018}
{Mulders}, G.~D., {Pascucci}, I., {Apai}, D., \& {Ciesla}, F.~J. 2018,
  \bibinfo{title}{{The Exoplanet Population Observation Simulator. I. The Inner
  Edges of Planetary Systems},} \aj, 156, 24, \dodoi{10.3847/1538-3881/aac5ea}

\bibitem[{F. {Mullally} {et~al.}(2015){Mullally}, {Coughlin}, {Thompson},
  {Rowe}, {Burke}, {Latham}, {Batalha}, {Bryson}, {Christiansen}, {Henze},
  {Ofir}, {Quarles}, {Shporer}, {Van Eylen}, {Van Laerhoven}, {Shah},
  {Wolfgang}, {Chaplin}, {Xie}, {Akeson}, {Argabright}, {Bachtell}, {Barclay},
  {Borucki}, {Caldwell}, {Campbell}, {Catanzarite}, {Cochran}, {Duren},
  {Fleming}, {Fraquelli}, {Girouard}, {Haas}, {He{\l}miniak}, {Howell},
  {Huber}, {Larson}, {Gautier}, {Jenkins}, {Li}, {Lissauer}, {McArthur},
  {Miller}, {Morris}, {Patil-Sabale}, {Plavchan}, {Putnam}, {Quintana},
  {Ramirez}, {Silva Aguirre}, {Seader}, {Smith}, {Steffen}, {Stewart},
  {Stober}, {Still}, {Tenenbaum}, {Troeltzsch}, {Twicken}, \&
  {Zamudio}}]{Mullally2015}
{Mullally}, F., {Coughlin}, J.~L., {Thompson}, S.~E., {et~al.} 2015,
  \bibinfo{title}{{Planetary Candidates Observed by Kepler. VI. Planet Sample
  from Q1--Q16 (47 Months)},} \apjs, 217, 31,
  \dodoi{10.1088/0067-0049/217/2/31}

\bibitem[{S. {M{\"u}ller} {et~al.}(2024){M{\"u}ller}, {Baron}, {Helled},
  {Bouchy}, \& {Parc}}]{Muller2024}
{M{\"u}ller}, S., {Baron}, J., {Helled}, R., {Bouchy}, F., \& {Parc}, L. 2024,
  \bibinfo{title}{{The mass-radius relation of exoplanets revisited},} \aap,
  686, A296, \dodoi{10.1051/0004-6361/202348690}

\bibitem[{J.~F. {Otegi} {et~al.}(2020){Otegi}, {Bouchy}, \&
  {Helled}}]{Otegi2020}
{Otegi}, J.~F., {Bouchy}, F., \& {Helled}, R. 2020, \bibinfo{title}{{Revisited
  mass-radius relations for exoplanets below 120 M$_{{\ensuremath{\oplus}}}$},}
  \aap, 634, A43, \dodoi{10.1051/0004-6361/201936482}

\bibitem[{J.~E. {Owen} \& Y. {Wu}(2017){Owen} \& {Wu}}]{Owen2017}
{Owen}, J.~E., \& {Wu}, Y. 2017, \bibinfo{title}{{The Evaporation Valley in the
  Kepler Planets},} \apj, 847, 29, \dodoi{10.3847/1538-4357/aa890a}

\bibitem[{I. {Pascucci} {et~al.}(2019){Pascucci}, {Mulders}, \&
  {Lopez}}]{Pascucci2019}
{Pascucci}, I., {Mulders}, G.~D., \& {Lopez}, E. 2019, \bibinfo{title}{{The
  Impact of Stripped Cores on the Frequency of Earth-size Planets in the
  Habitable Zone},} \apjl, 883, L15, \dodoi{10.3847/2041-8213/ab3dac}

\bibitem[{M.~T. {Penny} {et~al.}(2019){Penny}, {Gaudi}, {Kerins}, {Rattenbury},
  {Mao}, {Robin}, \& {Calchi Novati}}]{Penny2019}
{Penny}, M.~T., {Gaudi}, B.~S., {Kerins}, E., {et~al.} 2019,
  \bibinfo{title}{{Predictions of the WFIRST Microlensing Survey. I. Bound
  Planet Detection Rates},} \apjs, 241, 3, \dodoi{10.3847/1538-4365/aafb69}

\bibitem[{E.~A. {Petigura} {et~al.}(2013){Petigura}, {Howard}, \&
  {Marcy}}]{Petigura2013}
{Petigura}, E.~A., {Howard}, A.~W., \& {Marcy}, G.~W. 2013,
  \bibinfo{title}{{Prevalence of Earth-size planets orbiting Sun-like stars},}
  PNAS, 110, 19273, \dodoi{10.1073/pnas.1319909110}

\bibitem[{E.~A. {Petigura} {et~al.}(2017){Petigura}, {Howard}, {Marcy},
  {Johnson}, {Isaacson}, {Cargile}, {Hebb}, {Fulton}, {Weiss}, {Morton},
  {Winn}, {Rogers}, {Sinukoff}, {Hirsch}, \& {Crossfield}}]{Petigura2017}
{Petigura}, E.~A., {Howard}, A.~W., {Marcy}, G.~W., {et~al.} 2017,
  \bibinfo{title}{{The California-Kepler Survey. I. High-resolution
  Spectroscopy of 1305 Stars Hosting Kepler Transiting Planets},} \aj, 154,
  107, \dodoi{10.3847/1538-3881/aa80de}

\bibitem[{H. {Rauer} {et~al.}(2014){Rauer}, {Catala}, {Aerts}, {Appourchaux},
  {Benz}, {Brandeker}, {Christensen-Dalsgaard}, {Deleuil}, {Gizon}, {Goupil},
  {G{\"u}del}, {Janot-Pacheco}, {Mas-Hesse}, {Pagano}, {Piotto}, {Pollacco},
  {Santos}, {Smith}, {Su{\'a}rez}, {Szab{\'o}}, {Udry}, {Adibekyan}, {Alibert},
  {Almenara}, {Amaro-Seoane}, {Eiff}, {Asplund}, {Antonello}, {Barnes},
  {Baudin}, {Belkacem}, {Bergemann}, {Bihain}, {Birch}, {Bonfils}, {Boisse},
  {Bonomo}, {Borsa}, {Brand{\~a}o}, {Brocato}, {Brun}, {Burleigh}, {Burston},
  {Cabrera}, {Cassisi}, {Chaplin}, {Charpinet}, {Chiappini}, {Church},
  {Csizmadia}, {Cunha}, {Damasso}, {Davies}, {Deeg}, {D{\'\i}az}, {Dreizler},
  {Dreyer}, {Eggenberger}, {Ehrenreich}, {Eigm{\"u}ller}, {Erikson}, {Farmer},
  {Feltzing}, {de Oliveira Fialho}, {Figueira}, {Forveille}, {Fridlund},
  {Garc{\'\i}a}, {Giommi}, {Giuffrida}, {Godolt}, {Gomes da Silva}, {Granzer},
  {Grenfell}, {Grotsch-Noels}, {G{\"u}nther}, {Haswell}, {Hatzes},
  {H{\'e}brard}, {Hekker}, {Helled}, {Heng}, {Jenkins}, {Johansen},
  {Khodachenko}, {Kislyakova}, {Kley}, {Kolb}, {Krivova}, {Kupka}, {Lammer},
  {Lanza}, {Lebreton}, {Magrin}, {Marcos-Arenal}, {Marrese}, {Marques},
  {Martins}, {Mathis}, {Mathur}, {Messina}, {Miglio}, {Montalban}, {Montalto},
  {Monteiro}, {Moradi}, {Moravveji}, {Mordasini}, {Morel}, {Mortier},
  {Nascimbeni}, {Nelson}, {Nielsen}, {Noack}, {Norton}, {Ofir}, {Oshagh},
  {Ouazzani}, {P{\'a}pics}, {Parro}, {Petit}, {Plez}, {Poretti}, {Quirrenbach},
  {Ragazzoni}, {Raimondo}, {Rainer}, {Reese}, {Redmer}, {Reffert},
  {Rojas-Ayala}, {Roxburgh}, {Salmon}, {Santerne}, {Schneider}, {Schou},
  {Schuh}, {Schunker}, {Silva-Valio}, {Silvotti}, {Skillen}, {Snellen}, {Sohl},
  {Sousa}, {Sozzetti}, {Stello}, {Strassmeier}, {{\v{S}}vanda}, {Szab{\'o}},
  {Tkachenko}, {Valencia}, {Van Grootel}, {Vauclair}, {Ventura}, {Wagner},
  {Walton}, {Weingrill}, {Werner}, {Wheatley}, \& {Zwintz}}]{Rauer2014}
{Rauer}, H., {Catala}, C., {Aerts}, C., {et~al.} 2014, \bibinfo{title}{{The
  PLATO 2.0 mission},} Experimental Astronomy, 38, 249,
  \dodoi{10.1007/s10686-014-9383-4}

\bibitem[{H. {Rauer} {et~al.}(2024){Rauer}, {Aerts}, {Cabrera}, {Deleuil},
  {Erikson}, {Gizon}, {Goupil}, {Heras}, {Lorenzo-Alvarez}, {Marliani},
  {Martin-Garcia}, {Mas-Hesse}, {O'Rourke}, {Osborn}, {Pagano}, {Piotto},
  {Pollacco}, {Ragazzoni}, {Ramsay}, {Udry}, {Appourchaux}, {Benz},
  {Brandeker}, {G{\"u}del}, {Janot-Pacheco}, {Kabath}, {Kjeldsen}, {Min},
  {Santos}, {Smith}, {Suarez}, {Werner}, {Aboudan}, {Abreu}, {Acu a}, {Adams},
  {Adibekyan}, {Affer}, {Agneray}, {Agnor}, {Aguirre B{\o}rsen-Koch}, {Ahmed},
  {Aigrain}, {Al-Bahlawan}, {Alcacera Gil}, {Alei}, {Alencar}, {Alexander},
  {Alfonso-Garz{\'o}n}, {Alibert}, {Allende Prieto}, {Almeida}, {Alonso
  Sobrino}, {Altavilla}, {Althaus}, {Alonso Alvarez Trujillo}, {Amarsi},
  {Ammler-von Eiff}, {Am{\^o}res}, {Andrade}, {Antoniadis-Karnavas},
  {Ant{\'o}nio}, {Aparicio del Moral}, {Appolloni}, {Arena}, {Armstrong},
  {Aroca Aliaga}, {Asplund}, {Audenaert}, {Auricchio}, {Avelino}, {Baeke},
  {Bailli{\'e}}, {Balado}, {Ballber Balaguer{\'o}}, {Balestra}, {Ball},
  {Ballans}, {Ballot}, {Barban}, {Barbary}, {Barbieri}, {Barcel{\'o} Forteza},
  {Barker}, {Barklem}, {Barnes}, {Barrado Navascues}, {Barragan}, {Baruteau},
  {Basu}, {Baudin}, {Baumeister}, {Bayliss}, {Bazot}, {Beck}, {Bedding},
  {Belkacem}, {Bellinger}, {Benatti}, {Benomar}, {B{\'e}rard}, {Bergemann},
  {Bergomi}, {Bernardo}, {Biazzo}, {Bignamini}, {Bigot}, {Billot}, {Binet},
  {Biondi}, {Biondi}, {Birch}, {Bitsch}, {Bluhm Ceballos}, {B{\'o}di},
  {Bogn{\'a}r}, {Boisse}, {Bolmont}, {Bonanno}, {Bonavita}, {Bonfanti},
  {Bonfils}, {Bonito}, {Bonomo}, {B{\"o}rner}, {Boro Saikia}, {Borreguero
  Mart{\'\i}n}, {Borsa}, {Borsato}, {Bossini}, {Bouchy}, {Bou{\'e}},
  {Boufleur}, {Boumier}, {Bourrier}, {Bowman}, {Bozzo}, {Bradley}, {Bray},
  {Bressan}, {Breton}, {Brienza}, {Brito}, {Brogi}, {Brown}, {Brown}, {Brun},
  {Bruno}, {Bruns}, {Buchhave}, {Bugnet}, {Buldgen}, {Burgess}, {Busatta},
  {Busso}, {Buzasi}, {Caballero}, {Cabral}, {Cabrero Gomez}, {Calderone},
  {Cameron}, {Cameron}, {Campante}, {Campos Gestal}, {Canto Martins}, {Cara},
  {Carone}, {Carrasco}, {Casagrande}, {Casewell}, {Cassisi}, {Castellani},
  {Castro}, {Catala}, {Catal{\'a}n Fern{\'a}ndez}, {Catelan}, {Cegla},
  {Cerruti}, {Cessa}, {Chadid}, {Chaplin}, {Charpinet}, {Chiappini},
  {Chiarucci}, {Chiavassa}, {Chinellato}, {Chirulli}, {Christensen-Dalsgaard},
  {Church}, {Claret}, {Clarke}, {Claudi}, {Clermont}, {Coelho}, {Coelho},
  {Cogato}, {Colom{\'e}}, {Condamin}, {Conde Garc{\'\i}a}, \&
  {Conseil}}]{Rauer2024}
{Rauer}, H., {Aerts}, C., {Cabrera}, J., {et~al.} 2024, \bibinfo{title}{{The
  PLATO Mission},} arXiv e-prints, arXiv:2406.05447,
  \dodoi{10.48550/arXiv.2406.05447}

\bibitem[{J. {Robnik} \& U. {Seljak}(2025){Robnik} \& {Seljak}}]{Robnik2025}
{Robnik}, J., \& {Seljak}, U. 2025, \bibinfo{title}{{Reassessment of Kepler's
  habitable zone Earth-like exoplanets with data-driven null-signal
  templates},} arXiv e-prints, arXiv:2509.07409,
  \dodoi{10.48550/arXiv.2509.07409}

\bibitem[{J.~G. {Rogers} \& J.~E. {Owen}(2021){Rogers} \& {Owen}}]{Rogers2021}
{Rogers}, J.~G., \& {Owen}, J.~E. 2021, \bibinfo{title}{{Unveiling the planet
  population at birth},} \mnras, 503, 1526, \dodoi{10.1093/mnras/stab529}

\bibitem[{L.~A. {Rogers}(2015){Rogers}}]{Rogers2015}
{Rogers}, L.~A. 2015, \bibinfo{title}{{Most 1.6 Earth-radius Planets are Not
  Rocky},} \apj, 801, 41, \dodoi{10.1088/0004-637X/801/1/41}

\bibitem[{E. {Sandford} {et~al.}(2019){Sandford}, {Kipping}, \&
  {Collins}}]{Sandford2019}
{Sandford}, E., {Kipping}, D., \& {Collins}, M. 2019, \bibinfo{title}{{The
  multiplicity distribution of Kepler's exoplanets},} \mnras, 489, 3162,
  \dodoi{10.1093/mnras/stz2350}

\bibitem[{A.~B. {Savel} {et~al.}(2020){Savel}, {Dressing}, {Hirsch}, {Ciardi},
  {Fleming}, {Giacalone}, {Mayo}, \& {Christiansen}}]{Savel2020}
{Savel}, A.~B., {Dressing}, C.~D., {Hirsch}, L.~A., {et~al.} 2020,
  \bibinfo{title}{{A Closer Look at Exoplanet Occurrence Rates: Considering the
  Multiplicity of Stars without Detected Planets},} \aj, 160, 287,
  \dodoi{10.3847/1538-3881/abc47d}

\bibitem[{C.~J. {Shallue} \& A. {Vanderburg}(2018){Shallue} \&
  {Vanderburg}}]{Shallue2018}
{Shallue}, C.~J., \& {Vanderburg}, A. 2018, \bibinfo{title}{{Identifying
  Exoplanets with Deep Learning: A Five-planet Resonant Chain around Kepler-80
  and an Eighth Planet around Kepler-90},} \aj, 155, 94,
  \dodoi{10.3847/1538-3881/aa9e09}

\bibitem[{A. {Silburt} {et~al.}(2015){Silburt}, {Gaidos}, \&
  {Wu}}]{Silburt2015}
{Silburt}, A., {Gaidos}, E., \& {Wu}, Y. 2015, \bibinfo{title}{{A Statistical
  Reconstruction of the Planet Population around Kepler Solar-type Stars},}
  \apj, 799, 180, \dodoi{10.1088/0004-637X/799/2/180}

\bibitem[{C.~C. {Stark} {et~al.}(2024){Stark}, {Mennesson}, {Bryson}, {Ford},
  {Robinson}, {Belikov}, {Bolcar}, {Feinberg}, {Guyon}, {Latouf}, {Mandell},
  {Rauscher}, {Sirbu}, \& {Tuchow}}]{Stark2024}
{Stark}, C.~C., {Mennesson}, B., {Bryson}, S., {et~al.} 2024,
  \bibinfo{title}{{Paths to robust exoplanet science yield margin for the
  Habitable Worlds Observatory},} Journal of Astronomical Telescopes,
  Instruments, and Systems, 10, 034006, \dodoi{10.1117/1.JATIS.10.3.034006}

\bibitem[{K. {Sullivan} \& A.~L. {Kraus}(2022){Sullivan} \&
  {Kraus}}]{Sullivan2022}
{Sullivan}, K., \& {Kraus}, A.~L. 2022, \bibinfo{title}{{Revising Properties of
  Planet-Host Binary Systems. II. Apparent Near-Earth-analog Planets in
  Binaries Are Often Sub-Neptunes},} \aj, 164, 138,
  \dodoi{10.3847/1538-3881/ac89ed}

\bibitem[{P. {Tamburo} {et~al.}(2023){Tamburo}, {Muirhead}, \&
  {Dressing}}]{tamburo2023}
{Tamburo}, P., {Muirhead}, P.~S., \& {Dressing}, C.~D. 2023,
  \bibinfo{title}{{Predicting the Yield of Small Transiting Exoplanets around
  Mid-M and Ultracool Dwarfs in the Nancy Grace Roman Space Telescope Galactic
  Bulge Time Domain Survey},} \aj, 165, 251, \dodoi{10.3847/1538-3881/acd1de}

\bibitem[{T.~P.~D. Team(2020)Team}]{reback2020pandas}
Team, T. P.~D. 2020, \bibinfo{title}{pandas-dev/pandas: Pandas,}, latest
  Zenodo, \dodoi{10.5281/zenodo.3509134}

\bibitem[{S.~E. {Thompson} {et~al.}(2018){Thompson}, {Coughlin}, {Hoffman},
  {Mullally}, {Christiansen}, {Burke}, {Bryson}, {Batalha}, {Haas},
  {Catanzarite}, {Rowe}, {Barentsen}, {Caldwell}, {Clarke}, {Jenkins}, {Li},
  {Latham}, {Lissauer}, {Mathur}, {Morris}, {Seader}, {Smith}, {Klaus},
  {Twicken}, {Van Cleve}, {Wohler}, {Akeson}, {Ciardi}, {Cochran}, {Henze},
  {Howell}, {Huber}, {Pr{\v s}a}, {Ram{\'{\i}}rez}, {Morton}, {Barclay},
  {Campbell}, {Chaplin}, {Charbonneau}, {Christensen-Dalsgaard}, {Dotson},
  {Doyle}, {Dunham}, {Dupree}, {Ford}, {Geary}, {Girouard}, {Isaacson},
  {Kjeldsen}, {Quintana}, {Ragozzine}, {Shabram}, {Shporer}, {Silva Aguirre},
  {Steffen}, {Still}, {Tenenbaum}, {Welsh}, {Wolfgang}, {Zamudio}, {Koch}, \&
  {Borucki}}]{Thompson2018}
{Thompson}, S.~E., {Coughlin}, J.~L., {Hoffman}, K., {et~al.} 2018,
  \bibinfo{title}{{Planetary Candidates Observed by Kepler. VIII. A Fully
  Automated Catalog with Measured Completeness and Reliability Based on Data
  Release 25},} \apjs, 235, 38, \dodoi{10.3847/1538-4365/aab4f9}

\bibitem[{J.~D. {Twicken} {et~al.}(2016){Twicken}, {Jenkins}, {Seader},
  {Tenenbaum}, {Smith}, {Brownston}, {Burke}, {Catanzarite}, {Clarke}, {Cote},
  {Girouard}, {Klaus}, {Li}, {McCauliff}, {Morris}, {Wohler}, {Campbell},
  {Kamal Uddin}, {Zamudio}, {Sabale}, {Bryson}, {Caldwell}, {Christiansen},
  {Coughlin}, {Haas}, {Henze}, {Sanderfer}, \& {Thompson}}]{Twicken2016}
{Twicken}, J.~D., {Jenkins}, J.~M., {Seader}, S.~E., {et~al.} 2016,
  \bibinfo{title}{{Detection of Potential Transit Signals in 17 Quarters of
  Kepler Data: Results of the Final Kepler Mission Transiting Planet Search
  (DR25)},} \apj, 152, 158, \dodoi{10.3847/0004-6256/152/6/158}

\bibitem[{S. {Vach} {et~al.}(2024){Vach}, {Zhou}, {Huang}, {Rogers}, {Bouma},
  {Douglas}, {Kunimoto}, {Mann}, {Barber}, {Quinn}, {Latham}, {Bieryla}, \&
  {Collins}}]{Vach2024}
{Vach}, S., {Zhou}, G., {Huang}, C.~X., {et~al.} 2024, \bibinfo{title}{{The
  Occurrence of Small, Short-period Planets Younger than 200 Myr with TESS},}
  \aj, 167, 210, \dodoi{10.3847/1538-3881/ad3108}

\bibitem[{H. {Valizadegan} {et~al.}(2023){Valizadegan}, {Martinho}, {Jenkins},
  {Caldwell}, {Twicken}, \& {Bryson}}]{Valizadegan2023}
{Valizadegan}, H., {Martinho}, M. J.~S., {Jenkins}, J.~M., {et~al.} 2023,
  \bibinfo{title}{{Multiplicity Boost of Transit Signal Classifiers: Validation
  of 69 New Exoplanets using the Multiplicity Boost of ExoMiner},} \aj, 166,
  28, \dodoi{10.3847/1538-3881/acd344}

\bibitem[{H. {Valizadegan} {et~al.}(2022){Valizadegan}, {Martinho}, {Wilkens},
  {Jenkins}, {Smith}, {Caldwell}, {Twicken}, {Gerum}, {Walia}, {Hausknecht},
  {Lubin}, {Bryson}, \& {Oza}}]{Valizadegan2022}
{Valizadegan}, H., {Martinho}, M. J.~S., {Wilkens}, L.~S., {et~al.} 2022,
  \bibinfo{title}{{ExoMiner: A Highly Accurate and Explainable Deep Learning
  Classifier That Validates 301 New Exoplanets},} \apj, 926, 120,
  \dodoi{10.3847/1538-4357/ac4399}

\bibitem[{J.~E. {Van Cleve} \& D.~A. {Caldwell}(2009){Van Cleve} \&
  {Caldwell}}]{VanCleve2009}
{Van Cleve}, J.~E., \& {Caldwell}, D.~A. 2009, \bibinfo{title}{{Kepler
  Instrument Handbook},} MAST Archive, KSCI-19033-001

\bibitem[{P. Virtanen {et~al.}(2020)Virtanen, Gommers, Oliphant, Haberland,
  Reddy, Cournapeau, Burovski, Peterson, Weckesser, Bright, {van der Walt},
  Brett, Wilson, Millman, Mayorov, Nelson, Jones, Kern, Larson, Carey, Polat,
  Feng, Moore, {VanderPlas}, Laxalde, Perktold, Cimrman, Henriksen, Quintero,
  Harris, Archibald, Ribeiro, Pedregosa, {van Mulbregt}, \& {SciPy 1.0
  Contributors}}]{Virtanen2020}
Virtanen, P., Gommers, R., Oliphant, T.~E., {et~al.} 2020,
  \bibinfo{title}{{{SciPy} 1.0: Fundamental Algorithms for Scientific Computing
  in Python},} Nature Methods, 17, 261, \dodoi{10.1038/s41592-019-0686-2}

\bibitem[{M.~J. {Way} {et~al.}(2015){Way}, {Del Genio}, {Kelley}, {Aleinov}, \&
  {Clune}}]{Way2015}
{Way}, M.~J., {Del Genio}, A.~D., {Kelley}, M., {Aleinov}, I., \& {Clune}, T.
  2015, \bibinfo{title}{{Exploring the Inner Edge of the Habitable Zone with
  Fully Coupled Oceans},} arXiv e-prints, arXiv:1511.07283.
\newblock \doarXiv{1511.07283}

\bibitem[{ {W}es {M}c{K}inney(2010){W}es
  {M}c{K}inney}]{mckinney-proc-scipy-2010}
{W}es {M}c{K}inney. 2010, in {P}roceedings of the 9th {P}ython in {S}cience
  {C}onference, ed. {S}t\'efan van~der {W}alt \& {J}arrod {M}illman, 56 -- 61,
  \dodoi{10.25080/Majora-92bf1922-00a}

\bibitem[{R.~F. {Wilson} {et~al.}(2023){Wilson}, {Barclay}, {Powell},
  {Schlieder}, {Hedges}, {Montet}, {Quintana}, {Mcdonald}, {Penny}, {Espinoza},
  \& {Kerins}}]{Wilson2023}
{Wilson}, R.~F., {Barclay}, T., {Powell}, B.~P., {et~al.} 2023,
  \bibinfo{title}{{Transiting Exoplanet Yields for the Roman Galactic Bulge
  Time Domain Survey Predicted from Pixel-level Simulations},} \apjs, 269, 5,
  \dodoi{10.3847/1538-4365/acf3df}

\bibitem[{E.~T. {Wolf} \& O.~B. {Toon}(2015){Wolf} \& {Toon}}]{Wolf2015a}
{Wolf}, E.~T., \& {Toon}, O.~B. 2015, \bibinfo{title}{{The evolution of
  habitable climates under the brightening Sun},} JRGD, 120, 5775,
  \dodoi{10.1002/2015JD023302}

\bibitem[{J. Yang {et~al.}(2014)Yang, Bou{\'e}, Fabrycky, \& Abbot}]{Yang2014b}
Yang, J., Bou{\'e}, G., Fabrycky, D.~C., \& Abbot, D.~S. 2014,
  \bibinfo{title}{Strong dependence of the inner edge of the habitable zone on
  planetary rotation rate,} \apjl, 787, L2

\bibitem[{J. {Yang} {et~al.}(2013){Yang}, {Cowan}, \& {Abbot}}]{Yang2013}
{Yang}, J., {Cowan}, N.~B., \& {Abbot}, D.~S. 2013,
  \bibinfo{title}{{Stabilizing Cloud Feedback Dramatically Expands the
  Habitable Zone of Tidally Locked Planets},} \apjl, 771, L45,
  \dodoi{10.1088/2041-8205/771/2/L45}

\bibitem[{A.~N. {Youdin}(2011){Youdin}}]{Youdin2011}
{Youdin}, A.~N. 2011, \bibinfo{title}{{The Exoplanet Census: A General Method
  Applied to Kepler},} \apj, 742, 38, \dodoi{10.1088/0004-637X/742/1/38}

\bibitem[{H. {Zhang} {et~al.}(2022){Zhang}, {Ge}, {Deng}, {Yao}, {Zhu}, {Zang},
  {Mao}, {Zhu}, {Wang}, {Xie}, {Yang}, {Jiang}, {Chen}, {Wang}, {Tang}, {Sun},
  {Willis}, {Huang}, {Ma}, {Wang}, {Shen}, {Tam}, {Hu}, {Yang}, {Feng}, {Liu},
  {Ye}, {Xiang}, {Yu}, {Zhang}, {Wu}, {Zong}, {Yuan}, {Li}, {Zhao}, {Zou}, \&
  {Liu}}]{Zhang2022}
{Zhang}, H., {Ge}, J., {Deng}, H., {et~al.} 2022, in Society of Photo-Optical
  Instrumentation Engineers (SPIE) Conference Series, Vol. 12180, Space
  Telescopes and Instrumentation 2022: Optical, Infrared, and Millimeter Wave,
  ed. L.~E. {Coyle}, S.~{Matsuura}, \& M.~D. {Perrin}, 1218016,
  \dodoi{10.1117/12.2630151}

\bibitem[{W. {Zhu} {et~al.}(2018){Zhu}, {Petrovich}, {Wu}, {Dong}, \&
  {Xie}}]{Zhu2018}
{Zhu}, W., {Petrovich}, C., {Wu}, Y., {Dong}, S., \& {Xie}, J. 2018,
  \bibinfo{title}{{About 30\% of Sun-like Stars Have Kepler-like Planetary
  Systems: A Study of Their Intrinsic Architecture},} \apj, 860, 101,
  \dodoi{10.3847/1538-4357/aac6d5}

\bibitem[{J.~K. {Zink} {et~al.}(2019){Zink}, {Christiansen}, \&
  {Hansen}}]{zinkChristiansen2019}
{Zink}, J.~K., {Christiansen}, J.~L., \& {Hansen}, B. M.~S. 2019,
  \bibinfo{title}{{Accounting for incompleteness due to transit multiplicity in
  Kepler planet occurrence rates},} \mnras, 483, 4479,
  \dodoi{10.1093/mnras/sty3463}

\bibitem[{J.~K. {Zink} \& B.~M.~S. {Hansen}(2019){Zink} \& {Hansen}}]{zink2019}
{Zink}, J.~K., \& {Hansen}, B. M.~S. 2019, \bibinfo{title}{{Accounting for
  multiplicity in calculating eta Earth},} \mnras, 487, 246,
  \dodoi{10.1093/mnras/stz1246}

\end{thebibliography}
\bibliographystyle{aasjournalV7}

\end{document}